\documentclass[pdflatex,sn-mathphys-num]{sn-jnl}% Math and Physical Sciences Numbered Reference Style
\usepackage{graphicx}%
\usepackage{multirow}%
\usepackage{amsmath,amssymb,amsfonts}%
\usepackage{amsthm}%
\usepackage{mathrsfs}%
\usepackage[title]{appendix}%
\usepackage{xcolor}%
\usepackage{textcomp}%
\usepackage{manyfoot}%
\usepackage{booktabs}%
\usepackage{algorithm}%
\usepackage{algorithmicx}%
\usepackage{algpseudocode}%
\usepackage{listings}%

\usepackage{amsmath}
\usepackage{dsfont}
\usepackage{pifont}
\usepackage{booktabs}
\usepackage{tabularx}
\usepackage{siunitx}
\usepackage{epstopdf}
\usepackage{epsfig}
\usepackage{framed}
\usepackage{makeidx}
\usepackage{simplewick}
\usepackage{hyperref}
\usepackage{placeins}
\usepackage{bbold}
\usepackage{listofitems} % for \readlist to create arrays
\usepackage[font=small,labelfont=bf]{caption}
\usepackage{tikz}
\usepackage[framemethod=TikZ]{mdframed}
\usepackage{braket}
\usepackage[printsolution=true]{exercises}
\usepackage[absolute, overlay]{textpos}

\usetikzlibrary{backgrounds,snakes}
\usetikzlibrary{decorations.markings}
\usetikzlibrary{decorations.pathmorphing,decorations.markings}
\usetikzlibrary{positioning}% To get more advances positioning options
\usetikzlibrary{arrows}
\usetikzlibrary{colorbrewer}

\TPGrid[0pt, 0pt]{1}{1}

\usetikzlibrary{arrows.meta} % for arrow size

\usetikzlibrary{shadows}

\tikzset{>=latex} % for LaTeX arrow head
\colorlet{myred}{red!80!black}
\colorlet{myblue}{blue!80!black}
\colorlet{mygreen}{green!60!black}
\colorlet{myorange}{orange!70!red!60!black}
\colorlet{mydarkred}{red!30!black}
\colorlet{mydarkblue}{blue!40!black}
\colorlet{mydarkgreen}{green!30!black}
\tikzstyle{node}=[thick,circle,draw=myblue,minimum size=22,inner sep=0.5,outer sep=0.6]
\tikzstyle{node in}=[node,green!20!black,draw=mygreen!30!black,fill=mygreen!25]
\tikzstyle{node hidden}=[node,blue!20!black,draw=myblue!30!black,fill=myblue!20]
\tikzstyle{node convol}=[node,orange!20!black,draw=myorange!30!black,fill=myorange!20]
\tikzstyle{node out}=[node,red!20!black,draw=myred!30!black,fill=myred!20]
\tikzstyle{connect}=[thick,mydarkblue] %,line cap=round
\tikzstyle{connect arrow}=[-{Latex[length=4,width=3.5]},thick,mydarkblue,shorten <=0.5,shorten >=1]
\tikzset{ % node styles, numbered for easy mapping with \nstyle
  node 1/.style={node in},
  node 2/.style={node hidden},
  node 3/.style={node out},
}
\def\nstyle{int(\lay<\Nnodlen?min(2,\lay):3)} % map layer number onto 1, 2, or 3

\DeclareSIUnit{\barn}{b}

\newcommand{\nmodel}{N_{\rm model}}

\newcommand{\nreps}{N_{\rm rep}}

\newcommand{\ie}{{\it i.e.}}
\newcommand{\eg}{{\it e.g.}}

\newcommand{\ndat}{N_{\mathrm{dat}}}
\newcommand{\ngrid}{N_{\mathrm{grid}}}

\newcommand{\cov}{\mathrm{Cov}}
\newcommand{\FKtab}{(\mathrm{FK})}
\newcommand{\FKtabT}{(\mathrm{FK})^T}

\newcommand{\ddt}{\frac{d}{dt}}

\newcommand{\bpmat}{\begin{pmatrix}}
\newcommand{\epmat}{\end{pmatrix}}

\graphicspath{{./figs/}}

\newcommand{\fin}{{f_0}}
\newcommand{\finperp}{{f_0^{\perp}}}
\newcommand{\finpar}{{f_0^{\parallel}}}

\theoremstyle{thmstyleone}%
\theoremstyle{thmstyletwo}%
\newtheorem{example}{Example}%

\theoremstyle{thmstylethree}%

\begin{document}

\title[Some Inverse Problems in Particle Physics]{Some Inverse Problems in Particle Physics}

%%=============================================================%%
%% GivenName	-> \fnm{Joergen W.}
%% Particle	-> \spfx{van der} -> surname prefix
%% FamilyName	-> \sur{Ploeg}
%% Suffix	-> \sfx{IV}
%% \author*[1,2]{\fnm{Joergen W.} \spfx{van der} \sur{Ploeg} 
%%  \sfx{IV}}\email{iauthor@gmail.com}
%%=============================================================%%

\author*[1]{\fnm{Luigi} \sur{Del Debbio}}\email{luigi.del.debbio@ed.ac.uk}

% \author[2,3]{\fnm{Second} \sur{Author}}\email{iiauthor@gmail.com}
% \equalcont{These authors contributed equally to this work.}

% \author[1,2]{\fnm{Third} \sur{Author}}\email{iiiauthor@gmail.com}
% \equalcont{These authors contributed equally to this work.}

\affil*[1]{\orgdiv{School of Physics and Astronomy}, \orgname{The University of Edinburgh}, 
\orgaddress{\street{Peter Guthrie Tait Road}, \city{Edinburgh}, \postcode{EH9 3FD}, 
%\state{State}, 
\country{United Kingdom}}}

% \affil[2]{\orgdiv{Department}, \orgname{Organization}, \orgaddress{\street{Street}, \city{City}, \postcode{10587}, \state{State}, \country{Country}}}

% \affil[3]{\orgdiv{Department}, \orgname{Organization}, \orgaddress{\street{Street}, \city{City}, \postcode{610101}, \state{State}, \country{Country}}}

%%==================================%%
%% Sample for unstructured abstract %%
%%==================================%%

\abstract{
        Inverse problems play a central role in current areas of research in particle phenomenology. In these 
    lectures we focus on two examples, the extraction of Parton Distribution Functions (PDFs) from 
    experimental data (or, equivalently, from pseudo- and quasi-PDFs computed in lattice QCD), and the 
    extraction of spectral functions from lattice Euclidean time correlators. We investigate in 
    detail three different approaches, namely Backus-Gilbert, Gaussian Processes and fits based on 
    Neural Network parametrizations. \\

    \bigskip
    \noindent
    Lectures given at CERN, {\em School on Continuum Foundations of Lattice Gauge Theories}, 
    in July 2024.

    }


%%\pacs[JEL Classification]{D8, H51}

%%\pacs[MSC Classification]{35A01, 65L10, 65L12, 65L20, 65L70}

\maketitle

\newpage

\tableofcontents

\newpage

\section{Introduction}
\label{sec:intro}

Inverse problems appear in many areas of physics and, more widely, of computational science. The solution of 
these problems requires a mathematically robust framework that allows the extraction of information 
about theoretical models from (increasingly large) finite sets of data. It is often the case that solutions
are plagued by the ill-posed nature of the problem and require some form of regularization.

In these lectures we focus on applications to particle phenomenology, a subject area that is increasingly
driven towards precision physics as a way to search for new physics beyond the Standard Model. 
A crucial requirement for precision physics is the possibility of defining statistically robust error 
bars, and 
controlling the systematic errors in the solution. Solutions need to be both precise and accurate, 
calling for rigorous uncertainty quantification.

More specifically, we consider here examples where we aim to determine a function defined on a one-dimensional real 
interval, which we denote $\mathcal{I}$, using a finite set of data points. In the absence of further 
hypotheses, these functions are elements of infinite-dimensional spaces, so that the inverse 
problem is necessarily ill-posed. In the absence of infinitely many data points,
further assumptions are needed in order to get a solution. 

We frame our presentation from a probabilistic point of view, where the quantities that need 
to be determined are promoted
to stochastic variables and the solution to the inverse problem is provided in terms of probability distributions. 
The probabilistic approach has, in our opinion, two main advantages: 
\begin{enumerate}
    \item it spells out explicitly all the assumptions about the model in the {\em prior}\ 
        probability distribution;
    \item it yields a characterization of 
        the possible solutions in terms of {\em posterior}\ probabilities.
\end{enumerate}
The division between prior and posterior distributions is an appealing feature 
of the probabilistic approach, which clearly distinguishes the impact of any model assumption from
the impact of the available data. We will see examples where we consider the impact of 
larger data sets, of more precise data, and of different choices of priors. 

Building upon the mathematical literature, we address the solution of two inverse problems that are relevant
for current particle phenomenology, namely 
\begin{enumerate}
    \item the determination of Parton Distribution Functions (PDFs) from 
    experimental (or lattice) data,
    \item the inversion of the Laplace transform that allows to extract 
    spectral densities from Euclidean lattice correlators.     
\end{enumerate}

The lectures are organised as follows. The notation and the mathematical framework are introduced in 
Sect.~\ref{sec:prelim}. We then focus on three different methods to address the solution of the 
inverse problem that are widely in use in the particle phenomenology community. Sect.~\ref{sec:bg} is 
devoted to the Backus-Gilbert method; we discuss the basic idea behind the method and its applications, 
in particular for the extraction of spectral densities from lattice QCD correlator. 
Sect.~\ref{sec:gp} introduces the usage of Gaussian Processes for inverse problems. Gaussian Processes 
are a non-parametric method, which relies on Bayesian inference in order to deduce the posterior distribution
for the solution of the inverse problem. Finally, Sects.~\ref{sec:nn}, \ref{sec:training} and 
\ref{sec:LazyTraining} are devoted to the solution of the inverse problem using Neural Networks. The solution
based on Neural Networks is divided into three parts: in Sect.~\ref{sec:nn} we introduce the notation used 
and discuss the prior distribution of the output of the Neural Networks at initialization; 
Sect.~\ref{sec:training} is devoted to the analysis of the training dynamics of Neural Networks, using the
Neural Tangent Kernel; finally, in Sect.~\ref{sec:LazyTraining} we discuss the so-called lazy training regime, 
where an analytical solution can be obtained. Understanding in detail the training dynamics is a key ingredient
in order to assess the robustness of the solution and to relate the solution obtained using 
Neural Networks to the ones obtained using other methods, like Backus-Gilbert or Gaussian Processes. 
In Sect.~\ref{sec:Synth} we attempt a synthesis of the results obtained using the different methods, and 
we discuss the relations between them. 

In writing these lectures, we had to make a number of choices about the topics to be covered and the level of detail. 
We have tried to provide a self-contained presentation, focussing on methods that are more familiar 
to us and that we have extensively used. In a limited amount of hours, it was impossible to provide an 
exhaustive overview of the field, and we apologize for the omissions~\footnote{
    For instance, an interesting review~\cite{Rothkopf:2026wdj} has appeared as we were finishing 
    writing these lectures, which covers a number of topics that we have not discussed here.
}. We hope nonetheless 
that these lectures will be useful for students and researchers interested in the solution of inverse problems 
in particle phenomenology, and more widely in computational science.

\newpage

\section{Preliminaries}
\label{sec:prelim}

\subsection{Notation and Definitions}
\label{sec:notation}

Let us begin by introducing the notation used in these lectures. 

\paragraph{Input Functions.} 
The space of {\it input functions}\ is a Banach space, $X$. 
A Banach space is a {\em complete, normed}\ vector space. The actual space of functions that we want to explore depends on the 
problem at hand, but we always assume the existence of a metric that allows us to define the distance between two functions. 
The completeness of the space is important to prove mathematical theorems, but is not needed in practice for the problems 
considered here.

\begin{example}
    The space of parton distribution functions (PDFs) is the space of functions 
    \begin{equation}
        \label{eq:pdfspace}
        f : [0,1] \to \mathbb{R}\, ,
    \end{equation}
    sometimes with the additional requirement that $f$ is integrable so that the sum rules can be satisfied. 
    In {\em all}\ current determinations of the PDFs, see \eg\ ~\cite{NNPDF:2021njg,Bailey:2020ooq,Hou:2019efy}, 
    a choice of functional form is made, which is defined by a set of $\nmodel$ parameters. Therefore the set of input functions
    is effectively reduced to a finite-dimensional space spanned by these parameters. For example, one could use a basis of 
    orthogonal polynomials or a neural network parametrization to represent the functions. These choices introduce a model 
    dependence in the determination of the PDFs.
\end{example}

\begin{example}
    The space of spectral densities is the space of functions 
    \begin{equation}
        \label{eq:spectralspace}
        \rho : [0,\infty) \to \mathbb{R}^+\, ,
    \end{equation}
    with the additional requirement that $\rho$ is integrable, and more specifically that
    \begin{equation}
        \label{eq:spectralnorm}
        \int_0^\infty d\omega\, \rho(\omega) = 1\, .   
    \end{equation}
    Note that single particle stable states correspond to delta functions in the spectral density, which is therefore a 
    distribution, not a function. We will come back to this point later. 
\end{example}

\begin{example}
    When considering the scattering of two particles, the amplitude can be decomposed in 
    partial waves, so that the physical process is described by a set of functions
    $\mathcal{M}_{\ell}(\sqrt{s})$. It is customary to rewrite
    \begin{equation}
        \label{eq:AmplDec}
        \mathcal{M}_{\ell}^{-1} = \mathcal{K}_{\ell}^{-1} - i \rho\, ,
    \end{equation}
    where $\rho$ is the phase space factor and the K-matrix $\mathcal{K}_{\ell}$,
    which is a meromorphic function of $\sqrt{s}$. The space of input functions in this case
    is the space of meromorphic functions defined on the complex $\sqrt{s}$ plane. 
\end{example}

\paragraph{Forward Map.} 
The {\it forward map}\ maps the input functions to some output. The forward map 
is sometimes called the {\it model}\, and is denoted by $\mathcal{F}$, 
\begin{align}
    \label{eq:FwdMap}
    \mathcal{F} : X &\to R\, , \\
    f &\mapsto r = \mathcal{F}(f)\, . \nonumber
\end{align}
Given a model $\mathcal{F}$ and an input function $f$, we can compute the output $r$. 
We assume that the space of output functions $R$ is also a Banach space. The output functions are the 
{\em theoretical predictions}\ for the observables that are measured in the experiments.

\begin{example}
    In the case of PDFs, the forward map that yields the theoretical prediction for the 
    structure functions measured in a DIS experiment is 
    \begin{equation}
        \label{eq:DISFwdMap}
        F(x,Q^2) = \mathcal{F}(f) = \int_0^1 \frac{d\xi}{\xi}\, C(x/\xi, Q^2)\, f(\xi)\, ,
    \end{equation}
    where $C(x,Q^2)$ is the coefficient function that is known from perturbation theory. The coefficient functions
    can be computed at different orders in perturbation theory, yielding different (NLO, NNLO, ...) theoretical
    predictions.  
\end{example}

\begin{example}
    The Laplace transform of the spectral densities yields the theoretical prediction for the 
    Euclidean correlators
    \begin{equation}
        \label{eq:EuclCorr}
        C(t) = \int_0^\infty d\omega\, e^{-\omega t}\, \rho(\omega)\, , \quad \text{for } t > 0\, .
    \end{equation}
    The lattice correlators are computed from Monte Carlo simulations of the Euclidean path integral. 
    They are the data used to infer $\rho(\omega)$, just like the measurements of structure functions 
    in the case of PDFs. The use of the letter $C$ to denote the correlators is standard in the lattice 
    community; they should not be confused with the coefficient functions $C(x,Q^2)$ in the case of PDFs
    discussed in the example above. 
    Rather than changing a consolidated notation, we hope that the context will make it clear which of the 
    two is being referred to.
\end{example}

\begin{example}
    The scattering amplitudes determine the discrete levels of the energy spectrum of 
    the theory in a finite volume. The forward map in this case is given by the L\"uscher 
    formula, which states that the energy levels are the solutions of the equation
    \begin{equation}
        \label{eq:Luscher}
        \det \left[ \mathcal{K}^{-1}(E) + F(E,\mathbf{P},L)\right] = 0\, ,
    \end{equation}
    where $F(E,\mathbf{P},L)$ is a known function of the energy, the total momentum and the 
    volume.
\end{example}

\paragraph{Observables.}
The {\it observables}\ are the quantities that are measured in the experiments. The central values quoted
by the experiments are denoted by $Y$. Experiments
only ever measure a finite number of observables, $\ndat$, hence 
\begin{equation}
    \label{eq:ObsSpace}
    Y = \{ Y_I, \, I = 1, \ldots, \ndat \} \in \mathbb{R}^{\ndat}\, .
\end{equation}
Observables will be labelled by capital roman indices, $I, J, \ldots$. At times we will write all indices explicitly, 
however, in general, we suppress the indices and use a compact notation, where $Y$ is a vector in $\mathbb{R}^{\ndat}$.

The corresponding theoretical predictions are denoted by $T$ and are computed by applying the forward map at the 
points corresponding to the observables, 
\begin{align}
    \label{eq:ObsPred}
    T : X &\to \mathbb{R}^{\ndat}\, , \\
    f &\mapsto T[f]\, . \nonumber
\end{align}
There is a theoretical prediction for each observable, \ie\ $T_I[f] \in \mathbb{R}\, , I = 1, \ldots, \ndat$.
\begin{example}
    In the case of PDFs, the observables are the structure functions measured at various kinematical 
    points in a DIS experiment. The experimental data points are identified by the values of the 
    Bjorken $x$ and the virtuality $Q^2$. The observables are then defined as
    \begin{equation}
        \label{eq:DISObs}
        Y_I = F(x_I, Q^2_I)\, , \quad I = 1, \ldots, \ndat\, .
    \end{equation}
    The theoretical predictions are computed from the forward map as
    \begin{equation}
        T_I = T_I[f] = \int_0^1 \frac{d\xi}{\xi}\, C(x_I/\xi, Q_I^2)\, f(\xi)\, .
    \end{equation}
    In order to emphasize the linear dependence of the obervables on the PDFs, we can rewrite the equation above as
    \begin{equation}
        \label{eq:DISObsLinear}
        T_I = T_I[f] = \int_0^1 d\xi\, C_{I}(\xi)\, f(\xi)\, ,
    \end{equation}
    where $C_{I}(\xi) = C(x_I/\xi, Q_I^2)/\xi$ is the coefficient function evaluated at the kinematical point 
    corresponding to the observable $Y_I$.
\end{example}

\begin{example}
    When extracting spectral densities, we can only compute the Euclidean correlators for a finite 
    number of time separations on a discretized spacetime lattice. Hence
    \begin{equation}
        \label{eq:SpecDenObs}
        Y_I = C(t_I)\, , \quad I = 1, \ldots, \ndat\, ,
    \end{equation}
    where in this case $\ndat$ is the number of discrete values of $t$ where the measurement of the 
    correlators yields a good signal -- a more quantitative definition of {\em good}\ signal will be 
    discussed below. 
\end{example}

\begin{example}
    In the case of scattering amplitudes, the observables are the discrete energy levels of the theory in a finite volume, 
    which are measured in lattice simulations from the large-time behaviour of Euclidean space 
    correlators. The theoretical predictions for these energy 
    levels are computed from the forward map as the solutions of the L\"uscher formula 
    in Eq.~\eqref{eq:Luscher}.
\end{example}

It is important to appreciate that in all these examples we only have access 
to a finite number of observables, $\ndat$,
and therefore the information that can be extracted from the data is limited. 
When working with 
a finite number of measurements, often subject to noise, we cannot aim to fully reconstruct the input 
functions. Looking at the examples that we introduced above, we can anticipate that 
\begin{enumerate}
    \item in the case of PDFs, the measurements of the structure functions only cover a finite kinematic
        range, leading to large uncertainties in the so-called extrapolation regions; 
    \item in the case of spectral densities, measuring the correlators for a finite number of discrete 
        values of the Euclidean time only allows the reconstruction of a limited number the states 
        contributing to $\rho(\omega)$.
    \item in the case of scattering amplitudes, the quantization condition only allows the 
        computation of the amplitude at a finite number of discrete energy values. The form 
        of the amplitude in between these values is not constrained by the data and the 
        reconstruction is regulated by the choice of prior~\cite{Salg:2025now}.                 
\end{enumerate}

Experimental data are subject to {\em observational noise}, which is modelled as 
\begin{equation}
    \label{eq:ObsNoise}
    Y_I = T_I[f] + \eta_I\, , \quad I = 1, \ldots, \ndat\, ,
\end{equation}
where $\eta$ is a random variable in $\mathbb{R}^{\ndat}$ with probability distribution 
$P_\eta$. Unless otherwise stated, we assume that $P_\eta$ is a multi-variate Gaussian distribution, 
centred at the origin and with a covariance matrix $C_Y$, given by the experiments, 
\begin{equation}
    \label{eq:ObsNoiseDist}
    P_\eta(\eta) = \frac{1}{\sqrt{|2 \pi C_Y|}}\, e^{-\frac{1}{2} \eta^T C_Y^{-1} \eta}\, .
\end{equation}
The central values quoted by the experiments are one particular draw from the
probability distribution $P_\eta$.

\subsection{A Solution on a Grid}
\label{sec:SimpleSolution}
For the case of observables that depend linearly on the input function, a simple solution to the inverse problem 
is obtained by minimizing the $\chi^2$, defined as
\begin{equation}
    \label{eq:Chi2Def}
    \chi^2[f] = \sum_{I,J=1}^{\ndat} \left(Y_I - T_I[f]\right) \left(C_Y^{-1}\right)_{IJ} 
        \left(Y_J - T_J[f]\right)\, ,
\end{equation}
or equivalently, suppressing the indices, 
\begin{equation}
    \label{eq:Chi2DefCompact}
    \chi^2[f] = (Y - T[f])^T C_Y^{-1} (Y - T[f])\, .
\end{equation}
The solution is then given by the function $\hat{f}$ that minimizes $\chi^2[f]$,
\begin{equation}
    \label{eq:Chi2Min}
    \hat{f} = \arg\min_f \chi^2[f]\, .
\end{equation}

We present a solution to this problem, obtained by discretizing the 
input function on a grid of $\ngrid$ points, and then minimizing $\chi^2$ 
with respect to the values of the function at these points. The unknown function is therefore 
represented as a vector of $\ngrid$ real numbers, $f \in \mathbb{R}^{\ngrid}$, while 
the forward map is represented as a $\ndat \times \ngrid$ matrix, 
\begin{equation}
    \label{eq:FKDef}
    T[f] = \FKtab f\, ,
\end{equation} 
where $\FKtab$ is the matrix representation of the forward map. Note that in this framework, 
we have reduced the infinite-dimensional space of input functions to a $\ngrid$ dimensional real 
space. 

The covariance matrix $C_Y$ is positive-definite, its SVD decomposition yields
\begin{equation}
    \label{eq:CovSVD}
    C_Y = R\, \Sigma^{1/2}\, \Sigma^{1/2} R^T\, ,
\end{equation}
where $R$ is orthogonal and $\Sigma$ is the diagonal matrix made of the eigenvalues of $C_Y$. 
We can then rewrite $\chi^2$ as
\begin{align}
    \chi^2[f] 
        &= (Y - \FKtab f)^T R\, \Sigma^{-1/2}\, \Sigma^{-1/2} R^T (Y - \FKtab f) \nonumber \\
    \label{eq:Chi2Cholesky}
        &= \| \Upsilon - A f \|^2\, ,
\end{align}
where we have introduced the uncorrelated, whitened data $\Upsilon$ and the re-scaled forward map $A$,
\begin{equation}
    \label{eq:RescaledData}
    \Upsilon = \Sigma^{-1/2} R^T Y\, , \quad A = \Sigma^{-1/2} R^T \FKtab\, .
\end{equation}
The solution to the minimization problem in Eq.~\eqref{eq:Chi2Min} is then given by acting with the 
pseudo-inverse of $A$, denoted as $A^+$, on the rescaled data~\cite{Penrose_1955},
\begin{equation}
    \label{eq:PinvSolution}
    \hat{f} = A^+ \Upsilon\, .
\end{equation}
Using the following identity, 
\begin{equation}
    \label{eq:PinvIdentity}
    A^+ = (A^T A)^{+} A^T\, ,
\end{equation}
allows us to rewrite the solution as
\begin{equation}
    \label{eq:PinvSolution2}
    \hat{f} = M^{+} \FKtabT C_Y^{-1} Y\, ,
\end{equation}
where we have introduced the matrix $M$,
\begin{equation}
    \label{eq:MDef}
    M = \FKtabT C_Y^{-1} \FKtab\, .
\end{equation}
In the physical applications of interest in these lectures, we will encounter two situations
\begin{itemize}
    \item the matrix $M$ is not invertible, and then the solution in 
        Eq.~\eqref{eq:PinvSolution2} is not unique;

    \item or $M$ has a large condition number, and the solution is unstable with respect to small
        variations of the data.
\end{itemize}
Note that a non-invertible matrix $M$ has a non-trivial kernel, 
\begin{equation}
    \label{eq:KernelM}
    \mathrm{Ker}(M) = \{ f \in \mathbb{R}^{\ngrid} : M f = 0 \}\, ,
\end{equation}
which coincides with the kernel of the forward map, $\mathrm{Ker}(\FKtab)$. In this case 
the solution is not unique, since any function of the form 
\begin{equation}
    \label{eq:NonUniqueSolution}
    \hat{f} + f_{\parallel}\, , \quad f_{\parallel} \in \mathrm{Ker}(M)\, 
\end{equation}
is also a solution.
In both examples, in order to get a well-defined, stable solution, some form of regularization is needed.

\subsection{Probabilistic Framework}
\label{sec:ProbFramework}
As we will discuss later in the lectures, it is useful to work in a probabilistic framework, 
where the unknown function $f$ is promoted to a stochastic process and the answer of the inverse 
problem is given in the form of a 
posterior probability for the solution $f$. 
In general, this is a probability measure in an infinite-dimensional space of 
functions and needs to be handled with (mathematical) care. In practice, following the discussion in
the above section, we aim to compute the joint posterior distributions~\footnote{
    We use a tilde to identify the posterior distributions in these lectures. As an alternative, 
    we also use the notation $P(f | D_{\ndat})$ to denote the posterior distribution of $f$ given the data $D_{\ndat}$, 
    where $D_{\ndat}$ is the set of $\ndat$ measurements. The two notations are equivalent, 
    and we will use them interchangeably.
    }
\begin{equation}
    \label{eq:postDistribution}
    \tilde{P}(f_1, \ldots, f_{\ngrid}) =
        P(f_1, \ldots, f_{\ngrid} | D_{\ndat})\, , 
\end{equation}
for any generic set of values of $f$, 
\begin{equation}
    \label{eq:fgrid}
    f_\alpha = f(x_\alpha)\, , 
        \quad \alpha = 1, \ldots, \ngrid\, , \quad x_\alpha \in \mathcal{I}\, , 
\end{equation}  
and $\mathcal{I}$ is the domain of the function $f$. Looking at the two examples we discussed,
we have 
\begin{itemize}
    \item $\mathcal{I} = [0,1]$ for the PDFs, since the Bjorken $x$ is defined in this range;
    \item $\mathcal{I} = [0,\infty)$ for the spectral density, since the energy of the states 
        contributing to $\rho(\omega)$ is positive.
\end{itemize} 
We can marginalize with respect to some of the values of $f$ and get the posterior distribution
for a subset of the values of $f$:
\begin{equation}
    \label{eq:postMarginalized}
    P_{\alpha_1 \ldots \alpha_n}(f_{\alpha_1}, \ldots, f_{\alpha_n} | D_{\ndat}) = 
    \int \prod_{\alpha \neq \alpha_1, \ldots, \alpha_n} df_\alpha\, 
        P(f_1, \ldots, f_{\ngrid} | D_{\ndat})\, .
\end{equation}
The central value and the variance of $f(x_\alpha)$ at a single point on the grid $x_{\alpha}$ 
are computed from the posterior $P_\alpha(f | D_{\ndat})$ as
\begin{align}
    \label{eq:PostMean}
    \bar{f}_\alpha &= \mathbb{E}[f(x_\alpha)] = \int df P_\alpha(f | D_{\ndat})\, f\, , \\
    \label{eq:PostVar}
    \sigma_\alpha^2 &= \mathbb{E}[\left(f_\alpha - \bar{f}_\alpha\right)^2] 
        = \int df P_\alpha(f | D_{\ndat})\, \left(f - \bar{f}_\alpha\right)^2 \, .
\end{align}
Correlations between the values of the function $f$ at multiple points can be computed using 
the joint distributions in Eq.~\ref{eq:postMarginalized}.

\clearpage

\section{Backus-Gilbert Method}
\label{sec:bg}

\subsection{Backus-Gilbert Method}
\label{sec:bg-original}

We begin this section by reviewing the general principles behind the 
Backus-Gilbert method, which is a widely used approach to solve inverse 
problems in a non-parametric way. The method was originally developed in 
the context of geophysics~\cite{Backus1967}, but has found applications in 
many other fields, including particle physics and lattice QCD. We follow here 
the presentation in Ref.~\cite{Valentine2019}, which leads naturally to a 
connection with Gaussian Processes, as we will see in the next section.

We start by considering the continuous version of the inverse problem, 
where we want to determine a function $f(x)$ defined on a real interval 
$\mathcal{I}$, from a finite set of data points 
$D = \{Y_I\, , I=1, \ldots, \ndat\}$. 
The Backus-Gilbert method provides a way to construct an estimator 
$\hat{f}^{A}(x)$ for the function $f(x)$, which is a linear combination of the 
data points,
\begin{equation}
    \label{eq:BGEstimator}
    \hat{f}^{A}(x) = \sum_{I=1}^{\ndat} A_I(x) Y_I \, ,
\end{equation}
defined by choosing a set of coefficients $A_I(x)$. For data that depend linearly
on the function $f(x)$, in the absence of statistical noise, 
\begin{equation}
    \label{eq:LinearDepIntegral}
    Y_I = \int_{\mathcal{I}} dx' \, C_I(x') f(x') \, ,
\end{equation}
where $C_I(x)$ are known functions that define the linear forward map for the 
problem under study.~\footnote{
    The reader can refer to the examples of the previous section for concrete instances 
    of the functions $C_I(x)$ in the case of PDFs and lattice QCD.
} 
The Backus-Gilbert estimator can be rewritten as
\begin{equation}
    \label{eq:BGEstimatorLinear}
    \hat{f}^{A}(x) = \int_{\mathcal{I}} dx' \, \mathcal{R}^{A}(x,x') f(x') \, ,
\end{equation}
where we introduced the {\em resolution function}
\begin{equation}
    \label{eq:BGResolutionFunction}
    \mathcal{R}^{A}(x,x') = \sum_{I=1}^{\ndat} A_I(x) C_I(x') \, .
\end{equation}
Eq.~\eqref{eq:BGEstimatorLinear} shows explicitly that the BG estimator 
$\hat{f}^{A}(x)$ is a smeared version of the true solution $f(x)$, 
where the smearing is dictated by the resolution function $\mathcal{R}^{A}(x,x')$. 

In general, the coefficients $A_I(x)$ can be chosen by requiring that the 
resolution function $\mathcal{R}^{A}(x,x')$ is as {\it close}\ as possible to some target 
smearing function
of our choice, which depends on the specific problem that is being solved. In 
order to quantify the closeness of $\mathcal{R}^{A}(x,x')$ to the target smearing function, 
we introduce a metric in the space of functions, 
\begin{equation}
    \label{eq:FuncDistanceDef}
    d[u,v] = \| u-v \|_k\, ,
\end{equation}
where the norm is defined with respect to a symmetric, positive-definite kernel 
$k(x,x')$ as
\begin{equation}
    \label{eq:FuncNormDef}
    \| u \|_k^2 = \int_{\mathcal{I}} dx\, dx'\, u(x) k(x,x') u(x') \, .
\end{equation}
Note that the arbitrariness in the choice of the kernel
$k(x,x')$ can be used to encode some prior information about the function 
$f(x)$. 

\paragraph{Point Estimation.} 
In order to get a point estimator for the function $f(x)$, we can choose the target 
smearing function to be as close as possible to
a delta function $\delta(x-x')$, which would correspond to an ideal estimator 
that recovers the true function $f(x)$ without any smearing. 
In practice, this is done by minimizing the distance between the resolution 
function $\mathcal{R}^{A}(x,x')$ and the delta function $\delta(x-x')$,
\begin{equation}
    \label{eq:BGPointEst}
    d\left[\mathcal{R}^{A}(x,x'), \delta(x-x')\right]^2\, ,
\end{equation}
as a function of the coefficients $A_I(x)$. Imposing 
\begin{equation}
    \label{eq:GradDistanceBG}
    \frac{\partial}{\partial A_I(x)} d\left[\mathcal{R}^{A}(x,x'), \delta(x-x')\right]^2 = 0 \, ,
\end{equation}
yields
\begin{equation}
    \label{eq:ACondition}
    W_{IJ} A_J(x) = R_I(x) \, ,
\end{equation}
where
\begin{align}
    \label{eq:WMatDef}
    W_{IJ} &= \int_{\mathcal{I}} dx\, dx'\, C_I(x) k(x,x') C_J(x') \, , \\
    \label{eq:RVecDef}
    R_I(x) &= \int_{\mathcal{I}} dx'\, k(x,x') C_I(x') \, .
\end{align}
Having introduced the matrix $W$ and the vector $R$ with explicit observable indices, we are going to
suppress the indices and use a compact notation, where $W$ is a matrix in the space of observables and 
$R$ is a vector in the same space. 
Assuming that the matrix $W$ is invertible, we can solve for the 
coefficients $A(x)$ as
\begin{equation}
    \label{eq:ASolution}
    A(x) = R(x)^T W^{-1}\, , 
\end{equation} 
where we also suppressed the index for the coefficients $A_I(x)$, which is now implicit in the matrix-vector 
multiplication.
Plugging this solution into Eq.~\eqref{eq:BGEstimator} yields the point estimator 
for the function $f(x)$,
\begin{equation}
    \label{eq:BGPointEstimator}
    \hat{f}^{\mathrm{BG}}(x) = R(x)^T W^{-1} Y \, .
\end{equation}
The resolution function for this estimator is given by
\begin{equation}
    \label{eq:BGPointResolutionFunction}
    \mathcal{R}^{\mathrm{BG}}(x,x') = R(x)^T W^{-1} C(x') \, ,
\end{equation}
where we introduced the vector $C(x)$ with components $C_I(x)$.

\paragraph{Noisy Data.} 
Statistical noise in the data is modelled by adding a stochastic term $\eta$ to the 
theoretical prediction,
\begin{equation}
    \label{eq:BGNoisyData}
    Y_I = \int_{\mathcal{I}} dx'\ C_I(x') f(x') + \eta_I \, ,
\end{equation}
where $\eta$ is a vector of random variables $\eta_I$ with zero mean and 
covariance matrix $C_Y$. The point estimator is obtained by minimizing
the regulated distance
\begin{equation}
    \label{eq:BGNoisyPointEst}
    d\left[\mathcal{R}^{A}(x,x'), \delta(x-x')\right]^2 + \lambda A^T(x) C_Y A(x) \, ,
\end{equation}
where $\lambda$ is a regularization parameter that controls the trade-off between
the closeness of the resolution function to the delta function and the stability 
of the solution with respect to noise in the data. The directions in data space
that correspond to the larger eigenvalues of the covariance matrix $C_Y$ are more 
affected by noise, and the regularization term in Eq.~\eqref{eq:BGNoisyPointEst} 
suppresses the contribution of these directions to the solution. 
The solution for the coefficients $A_I(x)$ is given by
\begin{equation}
    \label{eq:BGNoisyASolution}
    A(x)^T = R(x)^T \left(W + \lambda C_Y\right)^{-1}  \, ,
\end{equation}
and the point estimator is given by
\begin{equation}
    \label{eq:BGNoisyPointEstimator}
    \hat{f}^{\mathrm{BG}}(x) = R^T(x) \left(W + \lambda C_Y\right)^{-1} Y \, . 
\end{equation}
The resolution function for this estimator is given by
\begin{equation}
    \label{eq:BGNoisyPointResolutionFunction}
    \mathcal{R}^{\mathrm{BG}}(x,x') = R^T(x) \left(W + \lambda C_Y\right)^{-1} C(x') \, . 
\end{equation}
Note that, because the covariance matrix $C_Y$ is positive-definite, the 
regularized matrix $W + \lambda C_Y$ is also positive-definite and therefore 
invertible, so that the solution in Eq.~\eqref{eq:BGNoisyASolution} is 
well-defined for any value of the regularization parameter $\lambda > 0$.

\medskip
\noindent
\fbox{
    \begin{minipage}{0.9\textwidth}
        \begin{exercise}
            \label{ex:BGNoisyData}
            Derive the solution for the case of noisy data, starting from the minimization
            of the regulated distance in Eq.~\eqref{eq:BGNoisyPointEst}. If needed, write all 
            the indices explicitly to get familiar with the notation.
        \end{exercise}
    \end{minipage}
}
\medskip

\paragraph{A Comment on the Probabilistic Interpretation.}
Note that the BG solution does not start from a prior distribution for the solution. However, 
we can construct a posterior distribution for the solution by bootstrapping the point estimator
discussed above over the noise in the data. In this way, the statistical fluctuations of the data
translate into fluctuations of the solution. 

\subsection{BG on a Grid}
\label{sec:DiscretizedBG}

It is instructive to pause for a second and rewrite the BG derivation for 
the case discussed in Sect.~\ref{sec:SimpleSolution}, where the solution 
is represented as the finite-dimensional vector of values $f_\alpha$, 
and the theoretical prediction is computed as a matrix-vector multiplication,
\begin{equation}
    \label{eq:FKtabAgain}
    T[f] = \FKtab f \, .
\end{equation}
The estimator for the solution is a linear combination of the data, 
\begin{equation}
    \label{eq:DiscreteBGEstimator}
    \hat{f}^A(x_\alpha) = \hat{f}^A_\alpha 
        = \sum_{I=1}^{\ndat} A_{\alpha I} Y_I\, ,
\end{equation}
where the functions $A_I(x)$ that we encountered in the continuous case are now replaced 
by the coefficients $A_{\alpha I}$, \ie, by a matrix of coefficients $A$ with dimensions 
$\ngrid \times \ndat$. The norm in the space of functions is replaced by a norm in the space 
of vectors, defined in terms of a positive-definite matrix $K_{\alpha_1\alpha_2}$ as
\begin{equation}
    \label{eq:DiscreteFuncNorm}
    \| u \|_K^2 = u^T K u \, .
\end{equation}
Similarly, we introduce the matrices $W$ and $R$ as
\begin{align}
    \label{eq:DiscreteWMatDef}
    W_{IJ} &= \FKtab_{I\alpha_1} K_{\alpha_1\alpha_2} \FKtabT_{\alpha_2 J} \, , \\
    \label{eq:DiscreteRVecDef}
    R_{I\alpha}&= \FKtab_{I\alpha_1} K_{\alpha_1\alpha} \, .
\end{align}     
The interested reader can verify that the solution for the coefficients $A_{\alpha I}$ 
for the case of noisy data is given by
\begin{equation}
    \label{eq:DiscreteASolution}
    A = R^T \left(W + \lambda C_Y\right)^{-1} \, .      
\end{equation}

\medskip
\noindent
\fbox{
    \begin{minipage}{0.9\textwidth}
        \begin{exercise}
            \label{ex:DiscreteBGNoisyData}
            Derive the solution for the case of noisy data, starting from the minimization
            of the regulated distance in Eq.~\eqref{eq:BGNoisyPointEst}, with the 
            appropriate replacements for the discrete case.
        \end{exercise}
        % \begin{exercise}
        %     \label{ex:DiscreteBGPointEst}
        %     Derive the solution for the case of noiseless data, starting from the minimization
        %     of the distance in Eq.~\eqref{eq:BGPointEst}, with the appropriate replacements for 
        %     the discrete case.
        % \end{exercise}
        
    \end{minipage}
}
\medskip

The point estimator for the solution is given by
\begin{equation}
    \label{eq:DiscreteBGPointEstimator}
    \hat{f}^{\mathrm{BG}} = R^T \left(W + \lambda C_Y\right)^{-1} Y \, ,         
\end{equation}
and the resolution function is given by
\begin{align}
    \label{eq:DiscreteBGResolutionFunction}
    \mathcal{R}^{\mathrm{BG}} &= R^T \left(W + \lambda C_Y\right)^{-1} \FKtab \\
    \label{eq:DiscreteBGResolutionFunctionTwo}
    &= K \FKtabT \left(\FKtab K \FKtabT + \lambda C_Y\right)^{-1} \FKtab \, .
\end{align}
We have suppressed all the indices in the equations above. It may be useful to rewrite the 
resolution function with all the indices explicitly written as a way to get familiar with the 
notation. These expressions will be useful in connecting the BG result with other solutions 
that we will discuss in the next sections.

\subsection{Closure Tests}
\label{sec:ClosTest}

Closure tests are a powerful tool to validate the solutions of the inverse problem, and to
assess the quality of the reconstructed function. The idea is to generate synthetic data from 
a known solution, and then to apply, in this particular case, the BG method to these data to see 
if we can recover the original solution. 
This allows us to check that the implementation of the BG method is correct, and to understand how 
well the method performs in practice. The synthetic data are generated by applying the forward operator 
to a known solution, $f_0$, and adding noise to mimic the statistical uncertainty of the data. Using the 
grid notation introduced in Sect.~\ref{sec:DiscretizedBG}, the synthetic data are generated as 
\begin{align}
    \label{eq:SynthData}
    Y = \FKtab f_0 + \epsilon \, ,
\end{align}
where $f_0$ is the known solution, $\FKtab$ is the forward operator, and $\epsilon$ is a vector of random 
variables with zero mean and covariance matrix $C_Y$. The BG method is then applied to the synthetic 
data $Y$ to obtain an estimator $\hat{f}^{\mathrm{BG}}$ for the solution and compare it with the 
known solution $f_0$.

Writing explicitly all the matrices that appear in the BG solution in Eq.~\eqref{eq:DiscreteBGPointEstimator}, 
yields
\begin{align}
    \hat{f}^{\mathrm{BG}} &= K \FKtabT \left(\FKtab K \FKtabT + \lambda C_Y\right)^{-1} 
        \FKtab f_0 \nonumber \\
        \label{eq:BGClosureTestExplicit}
        & \qquad + K \FKtabT \left(\FKtab K \FKtabT + \lambda C_Y\right)^{-1} \epsilon \, .
\end{align}
The first term on the right-hand side of the equation above is the contribution to the BG solution 
that comes from the known solution $f_0$, while the second term is the contribution that comes from 
the noise in the data. The second contribution is linear in the noise $\epsilon$ and therefore averages
to zero over an ensemble of instances of the noise. The first contribution, as already 
discussed in Sect.~\ref{sec:bg-original}, is a smeared version of the known solution $f_0$. A few 
algebraic manipulations allow us to rewrite the BG solution as
\begin{align}
    \label{eq:BGSmearing}
    \hat{f}^{\mathrm{BG}} &= \mathcal{R}^{\mathrm{BG}} f_0 = K M \frac{1}{K M + \lambda} f_0\, ,
\end{align}
where 
\begin{equation}
    \label{eq:MDefAgain}
    M = \FKtabT C_Y^{-1} \FKtab \, ,
\end{equation}
is the matrix we already introduced in Eq.~\eqref{eq:MDef}. If $M$ were invertible, then we 
would recover the exact solution $f_0$ in the limit $\lambda \to 0$. As discussed in 
Sect.~\ref{sec:prelim}, the matrix $M$ is not invertible, and the BG solution is {\it one} prescription to 
regularize the problem. In the case where $M$ is invertible but ill-conditioned, the BG solution 
for non-vanishing $\lambda$ is a Tikhonov regularization of the problem. Eq.~\eqref{eq:BGSmearing} 
shows explicitly that the BG prescription yields a smeared version of the solution $f_0$, 
where the smearing is dictated by the matrix $K M \frac{1}{K M + \lambda}$.

\subsection{Lattice Spectral Densities and the HLT Method}
\label{sec:hlt}

A contemporary application of the BG method is in the determination of spectral densities from 
Euclidean correlators computed in lattice field theories. Here we focus on the HLT method as 
introduced in Ref.~\cite{Hansen:2019idp}, which is a particular implementation of the BG method, 
tailored to the specific problem of spectral density reconstruction. We begin by formulating
the problem and introducing the notation that is specific to this application. We follow the
notation used in the original publication, which is different from the one we used in 
the general formulation of the problem in Sects.~\ref{sec:notation}, \ref{sec:bg-original} 
and~\ref{sec:DiscretizedBG} . 

\paragraph{Observables.}
The observables used to extract the spectral densities are Euclidean correlators of field 
operators, which are defined as
\begin{equation}
    \label{eq:EuclideanCorrelatorDef}
    C_L(t) = \frac{1}{L^3} \sum_{\vec{x}} \langle O(t,\vec{x}) O(0) \rangle_L \, ,
\end{equation}
where $L$ is the spatial extent of the lattice, and $t>0$ is the Euclidean time. The subscript 
$L$ is a reminder of the fact that these correlators are computed in a finite volume. They are
evaluated as the mean over an ensemble of gauge configurations generate by Monte Carlo sampling. 
Their covariance matrix is also computed from the available sample and scales like the 
inverse of the number of statistically independent configurations. The correlators
are computed for a {\em finite}\ set of discrete time points $t = a, 2a, \ldots, T$, where $a$ is the lattice 
spacing and $T$ is the temporal extent of the lattice in physical units.\footnote{
We choose here to measure Euclidean time in physical units. The (dimensionless) 
number of points in the time direction is $N_T=T/a$.
} The index $I$ used to label the data points 
in the previous sections can be identified with the time index $t_I$ of the correlators, so that we 
can write $Y_I = C_L(t_I)$.

\paragraph{The Unknown Function.}
The function that we want to determine as the solution of the inverse problem is the spectral 
density,
\begin{equation}
    \label{eq:SpectralDensityDef}
    \rho_L(E) = \frac{1}{L^3} \sum_{\vec{x}}
    \langle O(0,\vec{x})\, \delta(E - H_L)\, O(0)\rangle_L\, ,
\end{equation}
where $H_L$ is the Hamiltonian of the theory in a finite volume. The spectral density is a 
positive-definite function of the energy $E$, and it contains information about the spectrum 
of the theory, as well as about the matrix elements of the operator $O$ between the vacuum and 
the energy eigenstates. Comparing again with the notation used in the previous sections for the 
general formulation of the inverse problem, we can identify the unknown function $f(x)$ with the 
spectral density $\rho_L(E)$. The argument of the unknown function in this case is the energy $E$; 
when we discretize the problem on a gird of points, the index $\alpha$ is used to label the 
grid points with the energy variable $E_\alpha$.

\paragraph{Formulation of the Problem.}
The Euclidean correlators, computed on a lattice with infinite time extent, and the spectral 
density are related by a Laplace transform,
\begin{equation}
    \label{eq:LaplaceTransform}
    C_L(t) = \int_0^\infty dE\, e^{-tE} \rho_L(E) \, .
\end{equation}
Note that in a finite volume the spectrum of the Hamiltonian is discrete, which implies that
the spectral density is a sum of delta functions, 
\begin{equation}
    \label{eq:SpectralDensityDiscrete}
    \rho_L(E) = \sum_{n} \rho_n(L) \delta(E - E_n) \, ,
\end{equation}
and therefore it is a distribution rather than a regular function. As usual when dealing 
with distributions, we need to introduce a smearing function $\Delta_\sigma(E,E')$ in order to 
define a regular, smeared version of the spectral density,
\begin{equation}
    \label{eq:SmearedSpectralDensity}
    \hat{\rho}_L(\sigma,E_*) = \int_0^\infty dE \, \Delta_\sigma(E,E_*) \rho_L(E) \, .
\end{equation}
The smearing function $\Delta_\sigma(E,E')$ is a regular function, which is peaked around $E = E'$ 
and has a width $\sigma$. The infinite-volume, unsmeared spectral density can be obtained 
by taking the limit $\sigma \to 0$ and $L \to \infty$ in the smeared spectral density,
\begin{equation}
    \label{eq:InfiniteVolumeLimit}
    \rho(E) = \lim_{\sigma \to 0} \lim_{L \to \infty} \hat{\rho}_L(\sigma,E) \, ,
\end{equation}
where the order of the limits matters. 

Note that the limit $\sigma \to 0$ may not be needed in practice, since the smeared spectral 
density $\hat{\rho}_L(\sigma,E)$ is a regular function of $E$ that can be directly compared 
with experimental data, which can also be smeared. 

\paragraph{HLT Formalism on a Finite Lattice.} 
For correlators computed on a lattice with finite time extent $T$, the relation between the 
correlators and the spectral density is given by
\begin{equation}
    \label{eq:HLTFiniteT}
    Y_I = C_L(t_I) = \int_0^\infty dE\, b_T(t_I,E) \rho_L(E) \, ,
\end{equation}
where the basis function $b_T(t,E)$ is given by
\begin{equation}
    \label{eq:HLTKernelFiniteT}
    b_T(t,E) = e^{-tE} + e^{-(T-t)E} \, .
\end{equation}
The basis function $b_T(t,E)$ is a generalization of the Laplace kernel $e^{-tE}$ that takes into account 
the finite time extent of the lattice. In the limit $T \to \infty$, the basis function $b_T(t,E)$ reduces 
to the Laplace kernel,
\begin{equation}
    \label{eq:HLTKernelInfiniteT}
    \lim_{T \to \infty} b_T(t,E) = e^{-tE} \, .
\end{equation}
The function $b_T(t_I,E)$ is plays the role of the function $C_I(x)$ that we introduced in the 
general formulation of the inverse problem -- remember that $I$ identifies the data point and $E$ is
the argument of the unknown function. 

\paragraph{HLT Solution.}
In the HLT formulation, as in all BG formulations, the solution is expanded as a linear 
combination of the data, 
\begin{equation}
    \label{eq:HLTExpansion}
    \hat{\rho}_{L}(E^*) = \sum_{t=0}^{t_{\mathrm{max}}} g_t(E^*)\, C(t+a)\, ,
\end{equation}
so that the inverse problem is reduced to determining the coefficients $g_t(E^*)$. As discussed
in Sect.~\ref{sec:bg-original}, any BG estimator yields a smeared version of the unknown 
function. In the notation of HLT, the smearing kernel is 
\begin{equation}
    \label{eq:HLTSmearKern}
    \bar{\Delta}(E^*,E) = 
        \sum_{t=0}^{t_{\mathrm{max}}} g_t(E^*)\, b_T(t+a, E)\, .
\end{equation}
The coefficients $g_t(E^*)$ are determined by requiring that the smearing kernel approximates
a specific function, \eg, 
\begin{equation}
    \label{eq:HLTGaussSmear}
    \Delta_{\sigma}(E^*,E) = \frac{1}{\mathcal{N}} \, 
        \exp\left[
            - \frac{\left(E - E^*\right)^2}{2\sigma^2}
        \right]\, ,
\end{equation}
with the {\em unusual}\ normalization
\begin{equation}
    \label{eq:SemiGaussSmearNorm}
    \mathcal{N} = \int_{0}^{\infty} dE\, 
        \exp\left[
            - \frac{\left(E - E^*\right)^2}{2\sigma^2}
        \right]\, .
\end{equation}
Following the derivation in Sect.~\ref{sec:bg-original}, for each value of $E^*$, 
we minimize a cost function $W[\lambda,g]$ with respect to the parameters $g_t(E^*)$. For 
each value of $E^*$, we thereby determine $(t_{\mathrm{max}} + 1)$ coefficients, which are 
necessary to construct $\hat{\rho}_{L}(E^*)$ according to Eq.~\ref{eq:HLTExpansion}.
which is the sum of the distance between $\Delta_{\sigma}$ and $\bar{\Delta}$ and a term that takes into account the covariance of the data,
\begin{equation}
    \label{eq:HLTDistance}
    W\left[\lambda, g\right] = \left(1-\lambda\right) A[g] + \lambda \frac{B[g]}{C(0)^2}\, ,
\end{equation}
where 
\begin{align}
    \label{eq:HLTAdef}
    A[g] &= 
        \int_{E_0}^{\infty} dE\, 
            \left| \Delta_{\sigma}(E^*,E) - \bar{\Delta}(E^*,E) \right|^2\ , \\
    \label{eq:HLTBdef}
    B[g] &=
        \sum_{t,t'} g_t(E^*) \left(C_Y^{-1}\right)_{tt'} g_{t'}(E^*)\, , 
\end{align}
and $C_Y$ is the covariance of the Euclidean correlators
\begin{equation}
    \label{eq:CorrCov}
    \left(C_Y\right)_{tt'} = \cov [C(t), C(t')]\, .
\end{equation}
Note that the integral in Eq.~\eqref{eq:HLTAdef} has a lower integration limit $E_0$, which
cuts off the small values of $E$. 

The two terms, $A[g]$ and $B[g]$, are the equivalent of the two terms that appear in 
Eq.~\ref{eq:BGNoisyPointEst}; in the case of HLT the coefficients are chosen to approximate a
well-behaved smearing kernel rather than the singular delta distribution of 
Eq.~\eqref{eq:BGNoisyPointEst}. The main idea is that the HLT solution aims directly for 
a smeared solution, with the emphasis being on approximating the smearing kernel as best as
possible, rather than trying to approximate a Dirac delta. 

\medskip
\noindent
\fbox{
    \begin{minipage}{0.9\textwidth}
    \begin{exercise}
        Show that the solution of the minimization yields 
        \begin{align}
            \label{eq:HLTSolution}
            g^{\mathrm{HLT}}_t(E^*; \lambda, \sigma) 
                &= W^{-1}(\lambda)_{tt'} \left[ f_{t'}(E^*; \lambda, \sigma) +
                R_{t'} \frac{1 - R^T W^{-1}(\lambda) f(E^*; \lambda,\sigma)}{R^T W^{-1}(\lambda) R} \right]\, ,
        \end{align}
    where we introduced
    \begin{align}
        \label{eq:HLTRtDef}
        & R_t = \int_0^\infty dE\, b_T(t+a, E)\, , \\
        \label{eq:HLTftDef}
        & f_t(E^*; \lambda, \sigma) = \left(1 - \lambda\right)\, 
            \int_{E_0}^\infty dE\, b_t(t+a, E) \Delta_{\sigma}(E^*, E)\, ,
    \end{align}
    and 
    \begin{equation}
        \label{eq:HLTWDef}
        W(\lambda)_{tt'} = (1-\lambda) M_{tt'} + \lambda \frac{(C_Y)_{tt'}}{C(0)^2}\, , 
    \end{equation}
    where 
    \begin{equation}
        \label{eq:HLTMDef}
        M_{tt'} = \int_{E_0}^\infty dE\, b_T(t+a, E)\, b_T(t'+a, E)\, .
    \end{equation}
    The matrix $M$ is an ill-conditioned matrix; as discussed in Sect.~\ref{sec:bg-original} adding
    to the distance the $B$ term, which takes into account the covariance of the data, acts as a regulator. 
    \end{exercise}
\end{minipage}
}\medskip

\paragraph{A Comment on $A$.}   
For $T\to\infty$, the function $b_T$ becomes an exponential, and therefore the $A$ term in 
the distance is 
\begin{equation}
    \label{eq:AatTInfty}
    A[g] = \int_{E_0}^\infty dE\, \left|
        \sum_{t=0}^{t_{\mathrm{max}}} g_t(E^*;\lambda, \sigma) e^{-(t+a) E} - 
        \Delta_{\sigma}(E^*,E)
    \right|^2 \, .
\end{equation}

\paragraph{Some Insight into the HLT Solution.}
The HLT approach focuses on approximating the smearing kernel $\Delta_{\sigma}(E^*,E)$ as best as possible, 
rather than trying to find the best point estimator for the spectral density. It is possible, 
and desirable, to monitor the quality of the approximation of the smearing kernel using 
as an estimator
\begin{equation}
    \label{eq:HLTQualityEstimator}
    \delta_{\sigma}(E^*, E) = 1 - 
        \frac{\bar{\Delta}(E^*,E)}{\Delta_\sigma(E^*,E)}\, . 
\end{equation}
The quality of the approximation of the smearing kernel can be improved by increasing 
the number of data points, \ie, by increasing $t_{\mathrm{max}}$. It is worthwhile to 
emphasize the quantitative character of this estimate; for a given quality of the 
kernel approximation, \eg, fixing
\begin{equation}
    \label{eq:HLTEstimatorBound}
    \delta_{\sigma}(E^*, E^*) = 0.05 \, ,
\end{equation}
allows to explore regions in the $\sigma$ vs. $t_{\mathrm{max}}$ plane where the 
approximation of the smearing kernel is good enough, and therefore to determine the optimal 
choice of $\sigma$ for a given number of data points.

However, the quality of the approximation is also affected by the choice of the 
regularization parameter $\lambda$. 
Increasing $\lambda$ improves the stability of the solution with respect to noise in 
the data, but it also worsens the approximation of the smearing kernel. The optimal 
choice of $\lambda$ is therefore a trade-off between the quality of the approximation 
of the smearing kernel and the stability of the solution with respect to noise in the data.

\clearpage

\section{Gaussian Processes}
\label{sec:gp}

\subsection{A Brief Introduction to Gaussian Processes}
\label{sec:gp-intro}

A {\em stochastic process} is a collection of random variables 
$\{f(x) : x \in \mathcal{I}\}$, 
where $\mathcal{I}$ is the set on which $f$ is defined. For any discrete set of points 
$\{x_1, \ldots, x_n\}$, the random variables $\{f(x_1), \ldots, f(x_n)\}$ have a joint 
probability distribution. A {\em Gaussian process} is a stochastic process such that any 
finite collection of random variables has a multivariate Gaussian distribution. As such, 
a Gaussian process is completely specified by its mean function $m(x)$ and its covariance 
function $k(x,x')$, defined as
\begin{equation}
    \label{eq:GPMeanCov}
    m(x) = \mathbb{E}[f(x)]\, , \qquad k(x,x') = \mathbb{E}[(f(x) - m(x))(f(x') - m(x'))]\, .
\end{equation}
In the context of the inverse problems discussed in these lectures, we use Gaussian 
processes to define a prior distribution over functions and then use the observed data to 
obtain a posterior distribution, as suggested in Sect.~\ref{sec:ProbFramework}.

\paragraph{Setup of the Problem.} Following the conventions in Sect.~\ref{sec:SimpleSolution}, 
we represent the function $f$ as an $\ngrid$-dimensional real vector, made of the values
of the function $f$ at selected points on a grid, 
\begin{equation}
    \label{eq:fGridReminder}
    f_{\alpha} = f(x_{\alpha})\, \quad \alpha = 1, \ldots, \ngrid\, .
\end{equation}
These values are promoted to stochastic variables with a joint prior probability 
distribution that is a multi-dimensional Gaussian with respective mean and 
covariance given by
\begin{align}
    \label{eq:GPPriorMean}
    m_{\alpha} &= m(x_{\alpha}) = \mathbb{E}_{P}[f_{\alpha}]\, , \\
    \label{eq:GPCovPrior}
    K_{\alpha_1\alpha_2} &= k(x_{\alpha_1},x_{\alpha_2}) = \cov_{P}[f_{\alpha_1},f_{\alpha_2}]\, ,
\end{align}
where we introduced the subscript $P$ to remind the reader that here we consider averages 
over the prior distribution $P$,
\begin{align}
    \label{eq:GPExplicitP}
    P(f | h) = 
    \frac{1}{\sqrt{\det\left(2\pi K\right)}}
    \exp\left\{
        - \frac12 \left(f-m\right)^T
        K^{-1} \left(f - m\right)
    \right\}\, .
\end{align}
The covariance matrix of the Gaussian prior in this section is denoted by $K$, a symbol we used in 
Sect.~\ref{sec:bg} to denote the metric in the space of functions. The reason for this 
choice is that the covariance of the prior distribution plays a similar role to the metric in the 
Backus-Gilbert method.

Any parameter that appears in the definitions of $m$ and $k$ is called 
an {\em hyperparameter}\ 
and is collectively denoted by $h$. For the PDF determinations that we are going 
to analyse later, we choose $m(x) = 0$ and a Gibbs kernel,
\begin{equation}
    \label{eq:DefGibbsKernel}
    k(x,x') = \sigma^2 \sqrt{\frac{2 \ell(x) \ell(x')}{\ell(x)^2 + \ell(x')^2}}\, 
        \exp\left[
            - \frac{(x-x')^2}{\ell(x)^2 + \ell(x')^2}
        \right]\, ,
\end{equation}
where 
\begin{equation}
    \label{eq:EllFunDef}
    \ell(x) = \ell_0 (x + \delta)\, .
\end{equation}
There are two hyperparameters in this kernel, $\sigma$ and $\ell_0$, while 
$\delta$ is a small regulator, which is introduced in order to avoid the 
singularity at $x=0$. By inspecting Eq.~\eqref{eq:DefGibbsKernel}, we see that $\sigma^2$ is the 
variance of $f(x)$, which 
is the same for all values of $x$. This is a {\em choice}\ that is encoded in the 
prior. Other choices are possible, including the possibility of an $x$-dependent
prefactor. The correlation between two values $f(x)$ and $f(x')$ decreases 
exponentially with the square of the distance between $x$ and $x'$. The rate
of the exponential decay is dictated by $\ell(x)$ and $\ell(x')$. For this particular
choice of the prior, $\ell(x)$ grows linearly with $x$, which implies that the 
correlation between points decreases as $x\to 0$. The values of $f$ for nearby 
points at small $x$ are almost uncorrelated, which implies that the function is
free to fluctuate more than at higher values of $x$. We will discuss later 
how to constrain the hyperparameters $\sigma$ and $\ell_0$.

Remembering that the theoretical predictions are computed by multiplying the 
vector $f$ by the forward map $\FKtab$, 
\begin{equation}
    T = \FKtab f\, ,
\end{equation}
it is clear that these are also Gaussian variables, and one can readily show 
that the average and the covariance of the theoretical 
prediction $T$, induced by the prior probability distribution of $f$, 
are given by 
\begin{align}
  \label{eq:MeanThDiscr}
  E[T_I] &= \FKtab_{I\alpha} m_\alpha\, ,\\
  \label{eq:CovThDiscr}
  \cov[T_I,T_J] &= \FKtab_{I\alpha_1} K_{\alpha_1\alpha_2} \FKtabT_{\alpha_1J}\, .
\end{align}
Clearly, $T$ is a vector of dimensions $\ndat$, with components $T_I$, 
$I = 1, \ldots, \ndat$ -- there is one theoretical prediction
for every data point. 

Finally, we have $\ndat$ data points. Their central values, as given by the 
experiments, are denoted $Y_I$, while their fluctuations are assumed to be 
Gaussian, with a covariance matrix $C_Y$, which also taken from experimental
analyses. The fluctuations of the data around the central values are described by a stochastic 
variable, 
\begin{equation}
    \label{eq:EpsilonDistr}
    \epsilon \sim \mathcal{N}\left(0, C_Y\right)\, .
\end{equation}
Remember once again that $\epsilon$ is an $\ndat$-dimensional vector, with components
$\epsilon_I$. The set of central values and the covariance matrix of the data 
together constitute what we call the {\em dataset}\ $D_{\ndat}$.

\subsection{The Gaussian Process Solution}
\label{sec:gp-solution}

The solution of the inverse problem in this formalism is encoded in the {\em joint}\ 
posterior distribution, 
\begin{equation}
    \label{eq:GPPosterior}
    \tilde{P}(f, h) 
        = P\left(f, h | D_{\ndat}\right)
\end{equation}
Using simple probabilities, we can factorize
\begin{equation}
    \label{eq:BGPostFactorized}
    P\left(f, h | D_{\ndat}\right) = 
        P(f | h, D_{\ndat})\, P(h | D_{\ndat})\, ;
\end{equation}
the joint posterior is the product of two factors. The first one is the 
posterior probability induced by the data at fixed values of the 
hyperparameters $h$, while the second factor is the probability of the 
hyperparameters given the data. Let us analyse each factor on the right-hand 
side in turn. 

The {\bf first factor} is computed using Bayesian inference, 
\begin{equation}
    \label{eq:GPBayesFixedHyper}
    P(f | h, D_{\ndat}) = 
        \frac{P(D_{\ndat} | f, h)\, P(f | h)}{P(D_{\ndat} | h)}\, .
\end{equation}
The denominator in Eq.~\eqref{eq:GPBayesFixedHyper} is just a normalization 
factor, independent of $f$, which can be computed a posteriori. We focus here on 
on the expression in the numerator, which is the product of two factors, the likelihood and 
the prior. The likelihood is the probability of the data given the function $f$ and the 
hyperparameters $h$. Assuming that the data are normally distributed around the theoretical 
predictions, we have 
\begin{equation}
    \label{eq:GPLikelihood}
    P(D_{\ndat} | f, h) = 
        \frac{1}{\sqrt{\det\left(2\pi C_Y\right)}}
        \exp\left\{
            - \frac12 \left(Y - \FKtab f\right)^T
            C_Y^{-1} \left(Y - \FKtab f\right)
        \right\}\, .
\end{equation}
For linear data, \ie\ for theoretical predictions that depend linearly on the function $f$, 
the likelihood is a Gaussian distribution in $f$. The prior is also a Gaussian distribution in $f$, 
as given by Eq.~\eqref{eq:GPExplicitP}. The product of two Gaussian distributions is another
Gaussian distribution, so that the posterior distribution at fixed hyperparameters is also a 
Gaussian distribution in $f$. The mean and the covariance of this posterior distribution can be 
computed analytically, and are given by
\begin{align}
    \label{eq:GPPostMean}
    \tilde{m} &= m + K \FKtabT \left(C_{YT}\right)^{-1} (Y - \FKtab m)\, , \\
    \label{eq:GPPostCov}
    \tilde{K} &= K - K \FKtabT \left(C_{YT}\right)^{-1} \FKtab K\, , 
\end{align}
where we introduced the matrix $C_{YT}$
\begin{equation}
    \label{eq:CYTDef}
    C_{YT} = C_Y + \FKtab K \FKtabT\, .
\end{equation}
Note that, because $C_{YT}$ is positive definite, the diagonal elements of $\tilde{K}$ are 
smaller than the diagonal elements of $K$. Recalling that the diagonal elements of $K$ and 
$\tilde{K}$ are respectively the variances of the prior and posterior distributions, we see that the
that Bayesian inference reduced the uncertainty on the function $f$ at fixed hyperparameters. 
The information contained in the data leads to a reduction of the uncertainty on the function $f$.

It is useful to prove that 
\begin{align}
    \label{eq:CYTIdentity}
    \FKtabT \left[ C_Y + \FKtab K \FKtabT \right]^{-1} = 
        \frac{1}{1 + M K} \FKtabT C_Y^{-1}\, ,
\end{align}
with the usual definition of the matrix $M$ in Eq.~\eqref{eq:MDef}, and therefore
note that the covariance of the posterior can be rewritten in a form that is more 
similar to the expression for the covariance of the BG solution,
\begin{equation}
    \label{eq:GPPostCovAlternative}
    \tilde{K}^{-1} = K^{-1} + M \, .
\end{equation}            
This expression makes it clear that 
the posterior covariance is the result of the combination of the prior covariance and the information 
contained in the data, as encoded in $M$. The posterior covariance is smaller than both the prior 
covariance and the covariance of the BG solution, which is given by $M^{-1}$, since 
$\tilde{K}^{-1} > K^{-1}$ and $\tilde{K}^{-1} > M$.

\medskip
\noindent
\fbox{
    \begin{minipage}{0.9\textwidth}
        \begin{exercise}
            Prove Eqs.~\eqref{eq:GPPostMean}, \eqref{eq:GPPostCov}, 
            \eqref{eq:CYTIdentity} and \eqref{eq:GPPostCovAlternative}.
        \end{exercise}
    \end{minipage}
}
\medskip

\noindent 
Using Eq.~\eqref{eq:CYTIdentity}, the mean of the posterior distribution can be rewritten as
\begin{equation}
    \label{eq:GPPostMeanAlternative}
    \tilde{m} = m + K \frac{1}{1 + M K} \FKtabT C_Y^{-1} (Y - \FKtab m)\, .
\end{equation}
This expression makes it clear that the posterior mean is the result of the combination of the prior mean 
and the information contained in the data, with a corrections that is proportional to the difference between 
the data and the theoretical predictions computed using the prior mean. For $m=0$, we can rewrite 
the posterior mean as
\begin{equation}
    \label{eq:GPPostMeanAlternative2}
    \tilde{m} = \left[K \frac{1}{1 + M K}\right]\, \FKtabT C_Y^{-1} Y\, .
\end{equation}

\paragraph{A Note on Tikhonov Regularization.}
By choosing a prior covariance of the form $K = \sigma^2 I$, where $\lambda$ is a positive parameter, 
we can rewrite the posterior mean as
\begin{equation}
    \label{eq:TikhonovRegulated}
    \tilde{m} = \frac{1}{M + \lambda}\, \FKtabT C_Y^{-1} Y\, ,
\end{equation}
which is the Tikhonov regularized solution of the inverse problem, with regularization parameter 
$\lambda = 1/\sigma^2$. The Tikhonov regularized solution is a particular case of the GP solution, 
characterized by a diagonal prior covariance, \ie\ a prior that does not encode any 
correlation between the values of the function at different points. 

The {\bf second factor} in Eq.~\eqref{eq:BGPostFactorized} is the probability of the hyperparameters given the data. This is computed using Bayes' theorem as well,
\begin{equation}
    \label{eq:GPBayesHyper}
    P(h | D_{\ndat}) = 
        \frac{P(D_{\ndat} | h)\, P(h)}{P(D_{\ndat})}\, .
\end{equation}
Once again, we only need to focus on the numerator, since the denominator is just a normalization factor. 
At fixed values of the hyperparameters, the data can be modeled as 
\begin{equation}
    \label{eq:YDistrAtFixedHyper}
    Y = \FKtab f + \epsilon\, ,
\end{equation}
where $f$ is a stochastic variable with distribution given by the prior, and $\epsilon$ is a stochastic variable 
with distribution given by Eq.~\eqref{eq:EpsilonDistr}. 
Therefore the data is the sum of two Gaussian variables, and is itself a Gaussian variable,
\begin{equation}
    \label{eq:GPHyperLikelihood}    
    P(D_{\ndat} | h) = 
        \frac{1}{\sqrt{\det\left(2\pi C_{YT}\right)}}
        \exp\left\{
            - \frac12 \left(Y - \FKtab m\right)^T
            C_{YT}^{-1} \left(Y - \FKtab m\right)
        \right\}\, .   
\end{equation}
Choosing a uniform prior for the hyperparameters, the posterior distribution of the hyperparameters is proportional to 
the likelihood in Eq.~\eqref{eq:GPHyperLikelihood}. The most likely values of the hyperparameters are those that maximize 
this likelihood. Note that $P(D_{\ndat} | h)$ is a complicated function of the hyperparameters, since $C_{YT}$ depends on $h$ 
in a non-trivial way. However, the distribution can be sampled using standard Markov Chain Monte Carlo methods, which 
allow us to obtain a characterization of the posterior distribution of the hyperparameters. If the 
posterior distribution of the hyperparameters is sharply peaked around its maximum, we can approximate 
it with a delta function, and use the most likely values of the hyperparameters in Eqs.~\eqref{eq:GPPostMean} 
and \eqref{eq:GPPostCov} to obtain an approximation of the posterior distribution of $f$.

\subsection{GP determination of one PDF}
\label{sec:gp-pdf}

A simple application of the GP methodology is the determination of the PDF
\begin{align}
    \label{eq:T3Def}
    T_3(x) = u(x) + \bar{u}(x) - d(x) - \bar{d}(x)\, ,
\end{align}
using DIS data from the BCDMS experiment. The difference between the proton and the deuteron structure 
functions, $F_2^p - F_2^d$, is proportional to $T_3$ only, so that we can write
\begin{equation}
    \label{eq:F2pmd}
    F_2^p(x,Q^2) - F_2^d(x,Q2) =  C_{T_3} \otimes T_3(x)\, .
\end{equation}
This is a simple example of an inverse problem in the realm of PDFs determinations, where only one 
PDF is involved, and the forward map is given by a convolution with a kernel that is known in perturbation 
theory. The solution of this problem can be obtained using the GP methodology described in the 
previous section.

All results presented here are obtained using synthetic data generated from a known solution. The 
approach follows very closely the results presented in Ref.~\cite{Candido:2024hjt}. After applying 
standard kinematic cuts, we have 248 data points for the structure functions with their corresponding
covariance matrix, $C_Y$, as described by the experimental analyses~\cite{BCDMS:1989qop}. In the case
of $T_3$, we want to implement the fact that the function has to be integrable at $x=0$. This can be 
achieved by introducing and additional hyperparameter $\alpha$, used to rescale the kernel as 
\begin{equation}
    \label{eq:ScaledIntegrabKernel}
    k(x,x') \to \phi(x) \, k(x,x') \, \phi(x')\, , \quad \phi(x) = x^\alpha\, ,
\end{equation}
and imposing $\alpha \in (-1,0]$.

\begin{figure}[ht!]
    \center
    \includegraphics[scale=0.4]{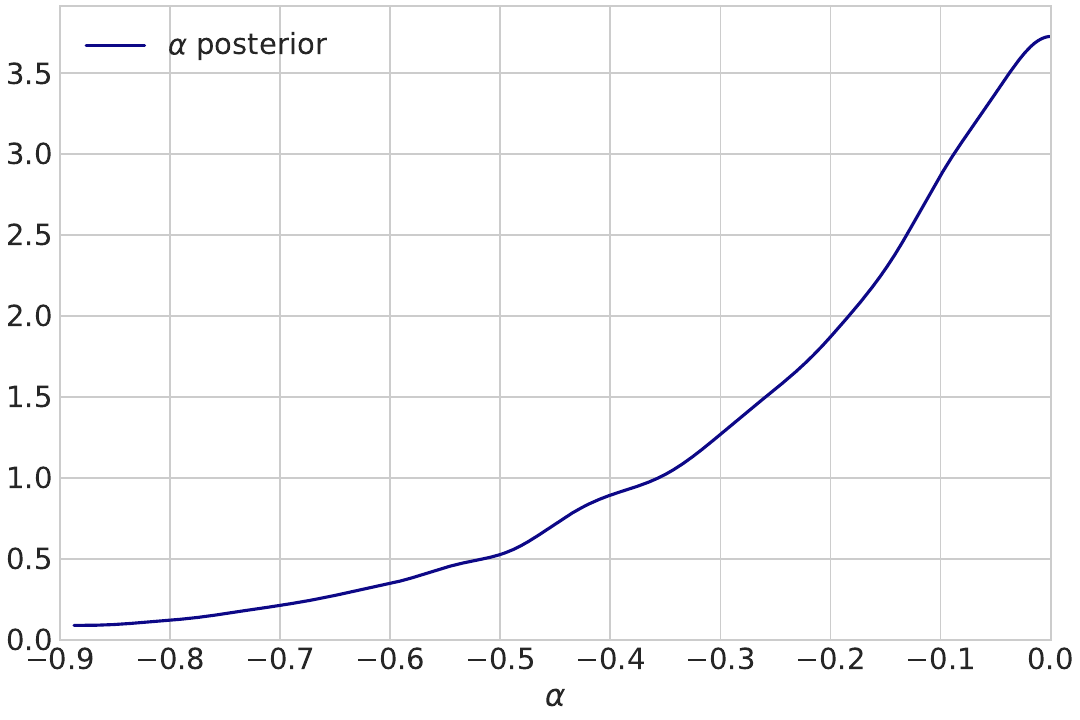} 
    \includegraphics[scale=0.4]{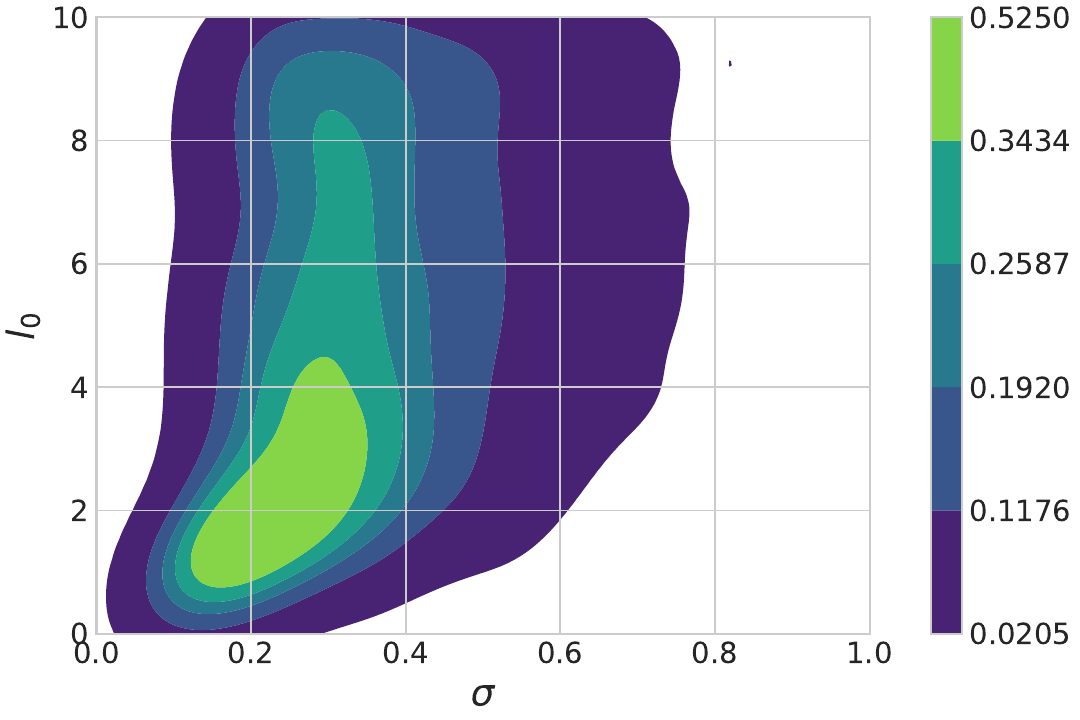}
    \includegraphics[scale=0.4]{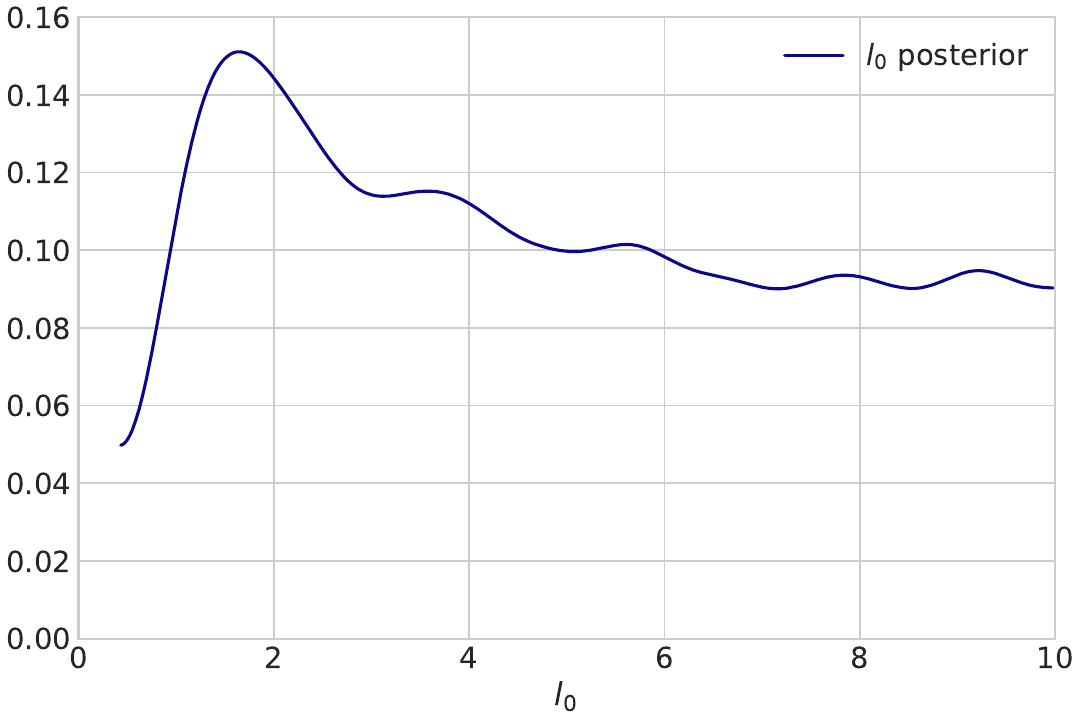} 
    \includegraphics[scale=0.4]{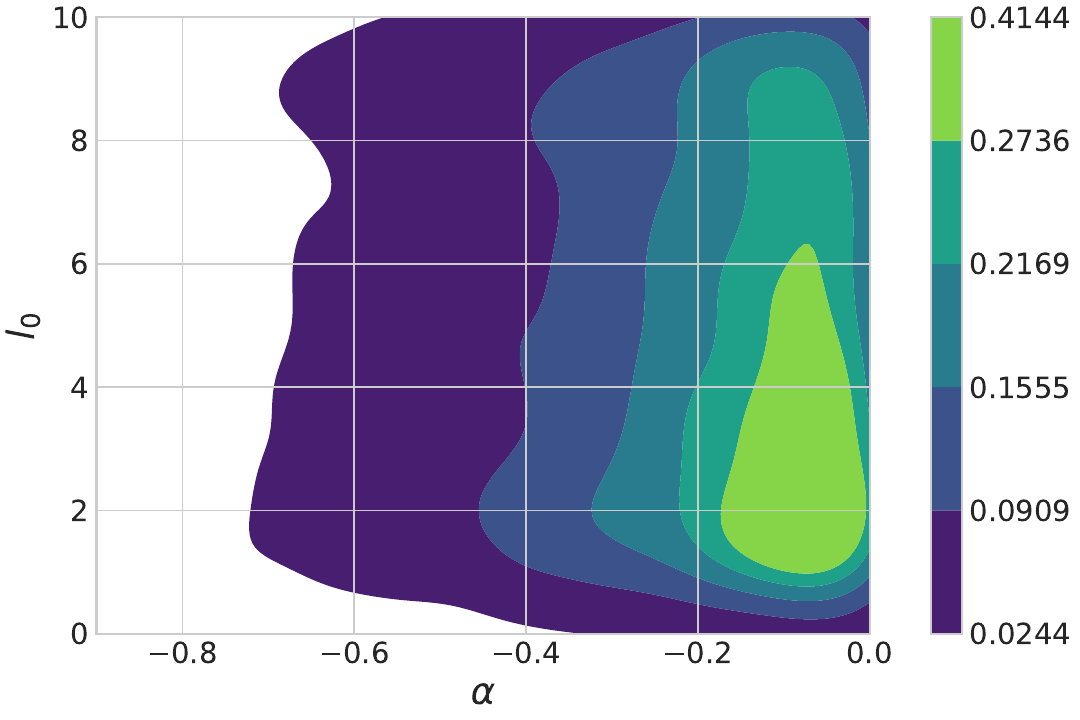}
    \includegraphics[scale=0.4]{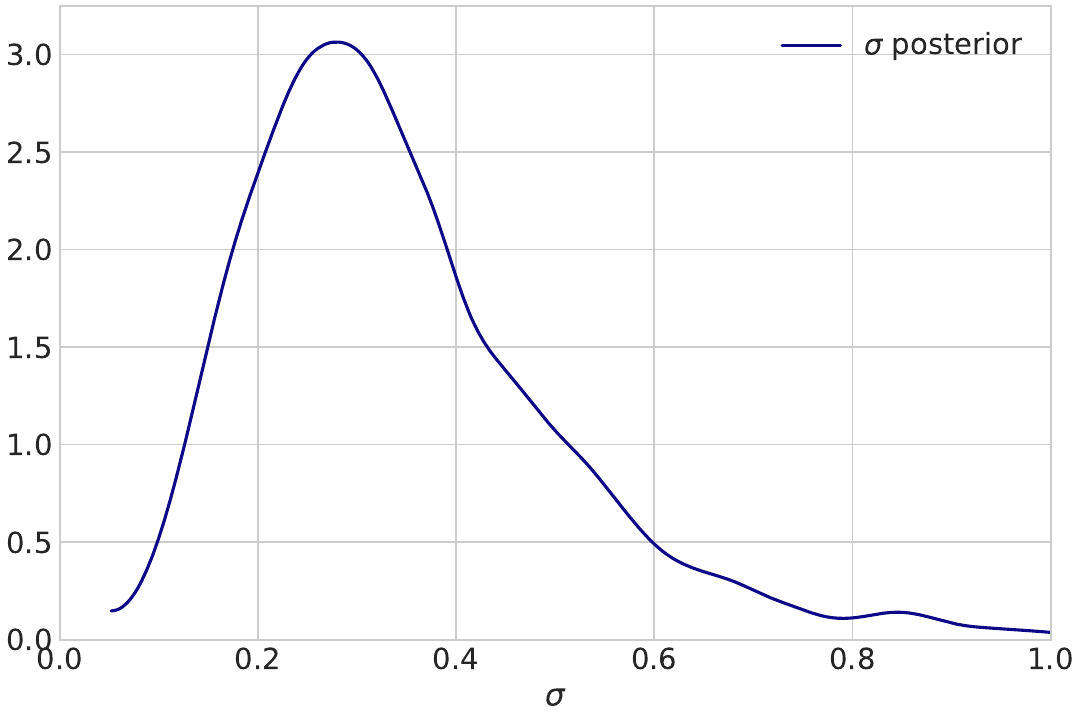} 
    \includegraphics[scale=0.4]{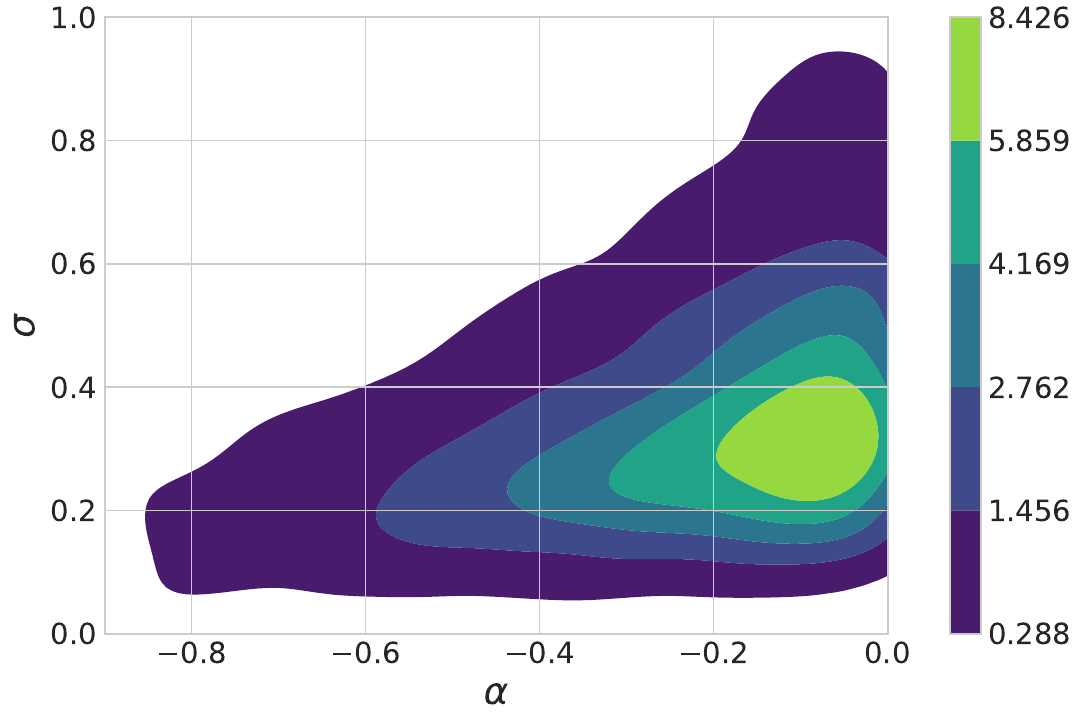}
    \caption{\small 1-dimensional (left panel) and 2-dimensional (right panel) 
    posteriors of the hyperparameters $\alpha$, $l_0$ and $\sigma$.
    The hyperparameter $\sigma$ is characterized by a sharply peaked posterior
    located around $\sigma\sim 0.25$ which quickly decays to zero 
    (for this reason in the posterior plot only the region $\left(0,1\right)$ is shown,
    even if the support of the prior is $\left(0,10\right)$); 
    $\alpha$ tends to sit closer to $0$, with a slow decay for smaller values towards $-1$;
    the $l_0$ posterior discards the smaller values of the correlation length, 
    it shows a peak for $l_0\sim 1.7$ and then remains fairly constant.  
    (Plot taken from Ref.~\cite{Candido:2024hjt}).
    \label{fig:GPHyperPosterior}
    }
\end{figure}

\begin{figure}[h!]
    \center
    \includegraphics[scale=0.4]{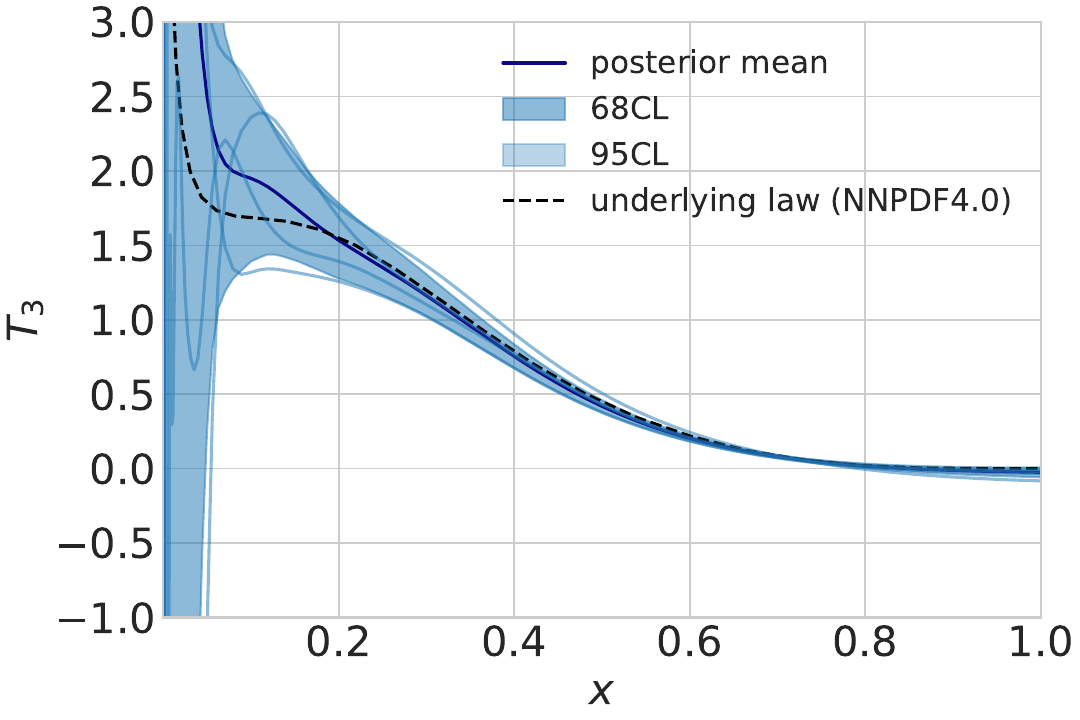}
    \includegraphics[scale=0.4]{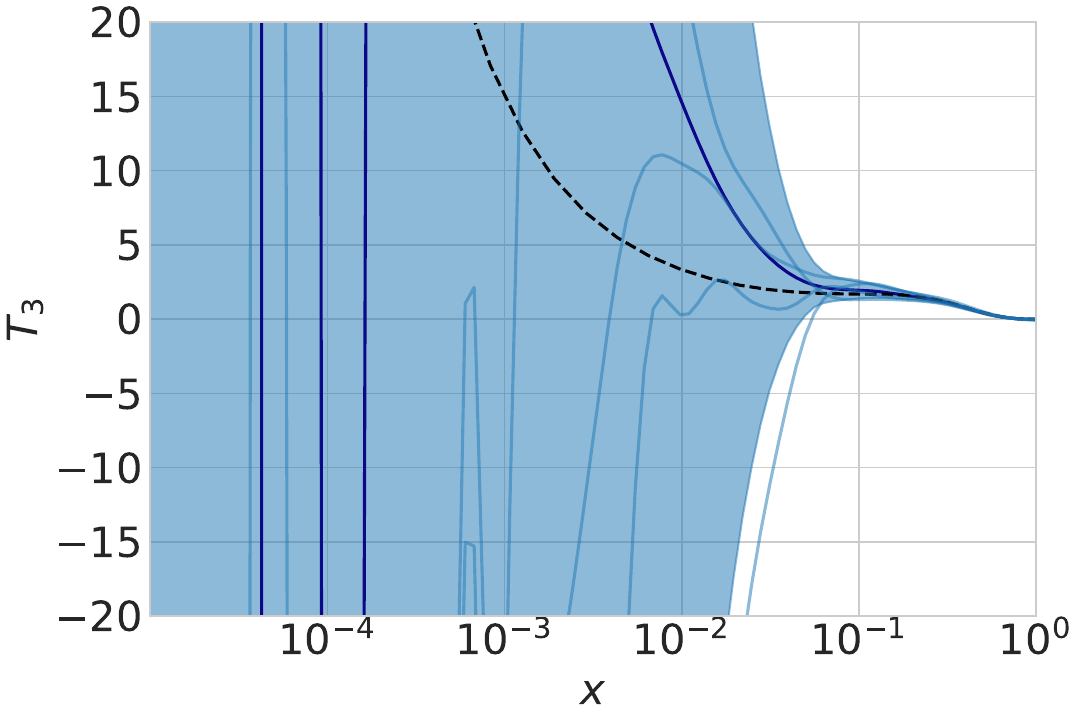}
    \caption{\small Samples from the posterior distribution of $T_3$,
    plotted in linear (left panel) and log (right panel) scale.
    The dark blue line represent the mean of the distribution, while the black dotted line the 
    input PDF, $f_0$, used to generate pseudo-data. 
    The shaded regions represent the 68CL and 95CL intervals, and in light blue we plot a few representative samples
    from the distribution. 
    The posterior displays a smaller variance in the regions sensitive to experimental data, 
    and an increasing spread in the small and large-$x$ extrapolation regions, 
    where the results are mostly determined by the chosen prior. 
    (Plot taken from Ref.~\cite{Candido:2024hjt}).
    \label{fig:GPPosteriorT3}
    }
\end{figure}

The Bayesian inference is performed in two steps. First, we sample the posterior of the hyperparameters, 
$P(h | D_{\ndat})$, using Markov Chain Monte Carlo methods. Then, for each set of hyperparameters drawn from 
this posterior, we use the analytical expressions in Eqs.~\eqref{eq:GPPostMean} and \eqref{eq:GPPostCov} 
and generate a replica of $T_3$. The ensemble of replicas generated in this 
way provides a characterization of the posterior distribution of $T_3$, which can be used to compute the 
mean and the uncertainty of $T_3$ at any value of $x$. Results for the posterior distribution of the 
hyperparameters are shown in Fig.~\ref{fig:GPHyperPosterior}. 
The mean value and the uncertainty of $T_3$ obtained from the posterior distribution are shown in 
Fig.~\ref{fig:GPPosteriorT3}. The mean value is taken by averaging over the ensemble of replicas
at each value of $x$, while the uncertainties are obtained from the 68\% and 95\% confidence intervals of 
the distribution of replicas. The uncertainty due to the fluctuations of the hyperparameters is included
in the final uncertainty of $T_3$, since the replicas are generated by sampling the posterior distribution 
of the hyperparameters and then using the normal distribution of $T_3$ at fixed hyperparameters. The Bayesian
inference is able to reconstruct the functional form of the PDF, as can be seen by comparing the dark 
blue and the dotted lines in the figure, which represent the mean of the posterior distribution and the input PDF, respectively.
The uncertainty on $T_3$ is smaller in the regions where the data are more constraining, and increases 
in the extrapolation regions at small and large $x$, where the results are mostly determined by the 
choice of the prior.

\FloatBarrier

\subsection{Closure Tests}
\label{sec:gp-clos}
\begin{figure}[ht!]
    \center
    \includegraphics[scale=0.4]{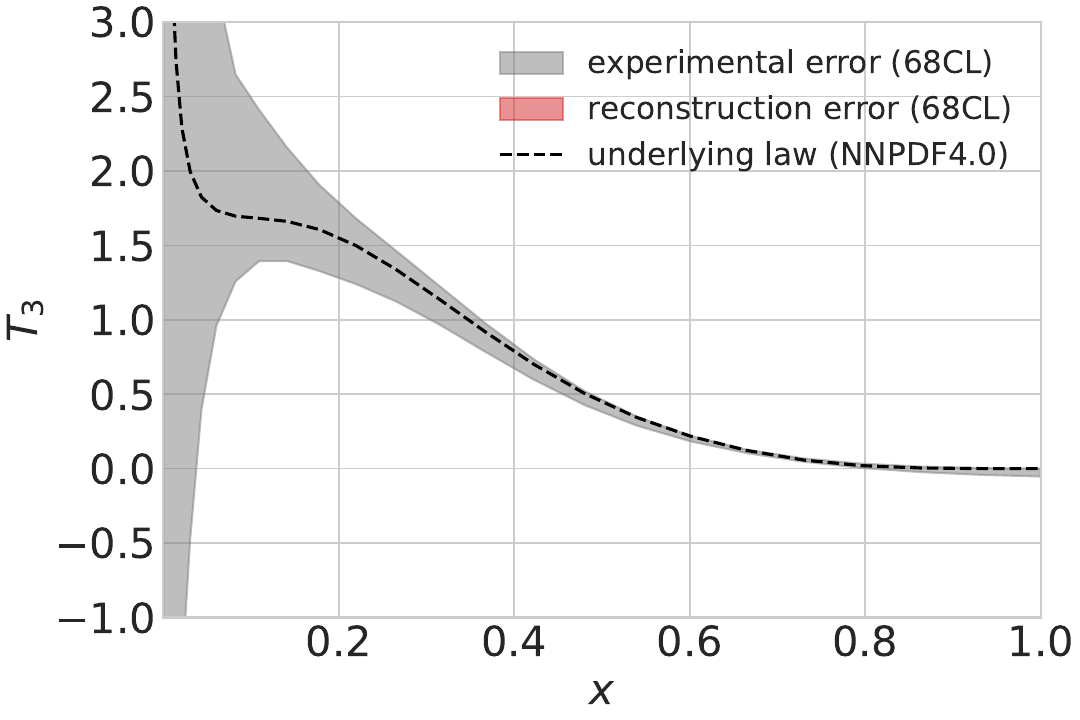}
    \includegraphics[scale=0.4]{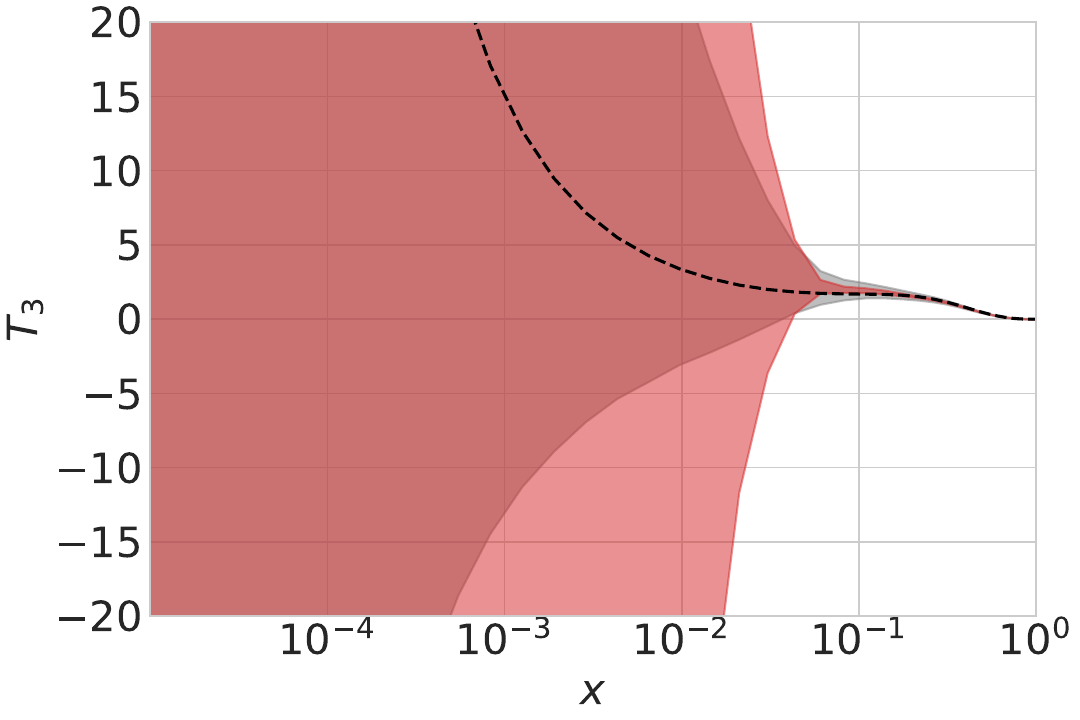}
    \caption{\small GP reconstruction of the input function $f_0$, with the 68\% CL band 
    for the experimental and reconstruction errors plotted separately in grey and red respectively,
    according to Eq.~\eqref{eq:cov_decomposition_cy}. The error bands are overlapped 
    rather than stacked. Plot taken from Ref.~\cite{Candido:2024hjt}.
    \label{fig:exp_meth_cov}
    }
\end{figure}

Following the discussion in Sect.~\ref{sec:ClosTest}, we can perform closure tests to validate the 
implementation of the GP method, and to assess the quality of the solution. Ignoring statistical 
fluctuations, \ie\ setting $C_Y=0$, synthetic data are generated 
from a known solution, $f_0$, by applying the forward operator,
\begin{equation}
    \label{eq:GPClosData}
    Y = \FKtab f_0\, .
\end{equation}
In this case the posterior mean is given by
\begin{equation}
    \label{eq:GPClosMean}
    \tilde{m} = K \FKtabT \left[\FKtab K \FKtabT \right]^{+} \FKtab\, f_0\, ,
\end{equation}
where the superscript $+$ denotes the pseudo-inverse, which is needed since $\FKtab K \FKtabT$ is 
not invertible when $C_Y=0$. By comparing with Eq.~\eqref{eq:BGClosureTestExplicit}, we see that 
this solution corresponds to the solution obtained using the BG method 
with the kernel $K$ as the metric. The two solutions coincide in the limit of zero noise and 
when the hyperparameters are fixed to the values used to generate the data. It is interesting to note that, 
even when there is no statistical noise, the solution of the inverse problem does not coincide with the input 
function $f_0$, but it is a smeared version of it, with the same smearing kernel given by the resolution function of 
the BG method. The solution of the inverse problem is not exact, even in the absence of noise, because of the 
smoothing effect of the forward map and the regularization effect of the prior. The resolution function
in this case is given by
\begin{equation}
    \label{eq:GPClosResolutionFunction}
    \mathcal{R}^{\mathrm{GP}}(x,x') = K(x,x') \FKtabT \left[\FKtab K \FKtabT\right]^{+} \FKtab
        \, .
\end{equation}
A non vanishing experimental error can be readily added to the closure test by adding a noise term 
to the data, which is normally distributed around zero, with a covariance given by $C_Y$. In this case, 
the mean of the posterior distribution is given by Eq.~\eqref{eq:GPPostMean} and the reconstruction 
kernel is
\begin{align}
    \label{eq:ReconstructionKernelWithExpErrors}
    \mathcal{R} = K\, 
        \FKtabT \left[\FKtab K \FKtabT + C_Y\right]^{-1} \FKtab\,,
\end{align}
so that the posterior mean and covariance can be rewritten as
\begin{align}
    \label{eq:bias_function_space_cy}
    &\tilde{m} - f_0 = \left[\mathcal{R} - \mathds{1}\right] f_0 + 
        a^T \epsilon\, , \\
    \label{eq:cov_decomposition_cy}
    &\tilde{K} = \left(\mathds{1} - \mathcal{R}\right) K
        \left(\mathds{1} - \mathcal{R}\right)^T 
        + a^T C_Y a\, ,
\end{align} 
where we have introduced
\begin{align}
    \label{eq:aOpDef}
    a^T = K\,\FKtabT \left[\FKtab \, K \,
         \FKtabT + C_Y\right]^{-1}\, ,
\end{align}
so that 
\begin{align}
    \label{eq:aRRelation}
    \mathcal{R} = a^T \FKtab\,.
\end{align}
The decomposition in Eqs.~\eqref{eq:bias_function_space_cy},~\eqref{eq:cov_decomposition_cy} 
highlights the fact that
there are two types of contributions to the bias and to the posterior covariance matrix. 
The first term in both Eqs.~\eqref{eq:bias_function_space_cy} and~\eqref{eq:cov_decomposition_cy} 
comes from the limited reconstruction of the central value and indeed would vanish 
when $\mathcal{R}=\mathds{1}$. If $\mathcal{R} \neq \mathds{1}$,
this term survives in the limit where $C_Y \to 0$, \ie\ in the limit of no 
experimental errors on the data.
The second term is the propagation of the covariance of the data into the
covariance of the model. In the case $\mathcal{R} = \mathds{1}$, the only error
fluctuations in the posterior distribution come from this term.
In Fig.~\ref{fig:exp_meth_cov}, Monte Carlo samples generated 
according to the reconstruction and experimental components are plotted separately for the BCDMS results,
in red and grey respectively.
In the medium-$x$ region, where more experimental data are available, 
the PDF uncertainty is dominated by the experimental error, yet a smaller reconstruction error is still present;
when moving to the small and large-$x$ extrapolation regions the reconstruction error becomes the dominant one, 
pointing out the lack of experimental information. 
We stress once more how these qualitative considerations are precisely quantified 
in Eqs.~\eqref{eq:bias_function_space_cy},~\eqref{eq:cov_decomposition_cy}, giving the analytical expression 
for the posterior covariance matrices associated to the experimental and reconstruction error, 
making it possible to quote different component of the PDF error in the context of a 
phenomenology analysis.

Note that in this formalism it is straightforward to assess the impact of a reduced uncertainty 
on the data, simply by reducing the covariance matrix $C_Y$ and recomputing the posterior 
distribution.

\subsection{GP for Spectral Densities}
\label{sec:gp-bg}
Let us now discuss how to apply the GP formalism to the problem of reconstructing
spectral densities from Euclidean correlators computed on the lattice. The notation 
in this section follows very closely the one used in Ref.~\cite{DelDebbio:2024lwm}. 

\subsubsection{GP Solution and Connection with BG}
\label{sec:GPSolOne}

Following the formalism introduced in this section, we represent the spectral density 
as a stochastic field $\rho(\omega)$, which is described by a GP centred around
$\rho^{\rm prior}(\omega)$ and with covariance $\mathcal{K}^{\rm
  prior}(\omega,E)$. The prior probability measure is therefore
\begin{equation}
    \label{eq:RhoDistr}
    \Pi[\rho] = \frac{1}{\mathcal{N}} \, \exp \biggr(
    -\frac{1}{2} \left| \rho - \rho^{\rm prior}
    \right|^2_{\mathcal{K}^{\rm prior}} \biggr) \; ,
\end{equation}
where
\begin{equation}
    \left| \rho- \right.
    \left.\rho^{\rm prior} \right|^2_{\mathcal{K}^{\rm prior}} =    \int dE_1 \int dE_2 
    \left[ \rho(E_1) - \rho^{\rm prior}(E_1) \right] 
    \mathcal{K}_{\rm prior}^{-1}(E_1,E_2)
    \left[\rho(E_2) - \rho^{\rm prior}(E_2) \right]\, ,
\end{equation}
and the normalization
\begin{equation}
    \mathcal{N} = \int \mathcal{D} \rho \; \Pi[\rho] \; .
\end{equation}
In the previous expressions, $\mathcal{D} \rho$ represents the functional
integration measure over the field variable $\rho$.
%We have so defined a field theory for the stochastic variable $\mathcal{R}(E)$. 
By definition, the expectation value for $\rho(E)$ from the prior distribution is
\begin{equation}
    \rho^{\rm prior}(\omega) = \int \mathcal{D}\rho\, \Pi[\rho] \; 
        \rho(\omega)  \; .
\end{equation}
We then introduce the noise $\vec{\eta} \in \mathbb{R}^{\tau_{\rm
    max}}$ which takes into account the uncertainty in the lattice
data and is represented as a real-valued stochastic variable, for
which we assume a multivariate Gaussian distribution with zero mean,
\begin{equation}
    \label{eq:eta_gauss_noise}
    \mathbb{G}[\vec{\eta}; C_Y] = \frac{1}{\sqrt{\text{det}
        (2\pi C_Y)}} \exp \left( -\frac{1}{2} \vec{\eta}\;
        C_Y^{-1} \; \vec{\eta} \right) \, ,
\end{equation}
where $C_Y$ is the covariance matrix of the correlators. 
We also introduce a stochastic variable associated to the lattice data, 
$\vec{{\mathcal{C}}} \in \mathbb{R}^{\tau_{\rm max}}$ with entries:
\begin{equation}
    \label{eq:laplace_transform_with_noise}
    \mathcal{C}(t) = \int dE \, b_T(t,E) \, \rho(E) + \eta(t) \; .
\end{equation} 
Given the distributions of Eq.~\eqref{eq:RhoDistr} and 
Eq.~\eqref{eq:laplace_transform_with_noise}, we can evaluate the covariance
associated to the variable $\vec{\mathcal{C}}$. It is straightforward
to show that
\begin{equation}\label{eq:CCSigmaHatPlusB}
    \begin{split}
        \braket{\mathcal{C}(t_1)\mathcal{C}(t_2)} 
        & = \int d\vec{\eta}\;\mathcal{D}\rho\; 
            \mathcal{C}(t_1) \, \mathcal{C}(t_2) \; 
            \mathbb{G}[\vec{\eta}, \text{Cov}_d] \,
            \Pi[\rho] \, \\[8pt] 
        & = \Sigma_{t_1t_2} + \left(C_Y\right)_{t_1t_2} \; ,
    \end{split}
\end{equation}
where we defined
\begin{equation}\label{eq:SigmaDef}
    \Sigma_{t_1t_2} = \int dE_1 \int \; dE_2 \; b_T(t_1,E_1) \,
    \mathcal{K}^{\rm prior}(E_1,E_2) \, b_T(t_2,E_2)\, .
\end{equation}

In order to predict the value of the spectral density at the energy
$\omega$, we need to extend the dimensionality of the covariance of
Eq.~\eqref{eq:CCSigmaHatPlusB} to include the indirect observation of
$\rho(\omega)$. To this end, we introduce the vector $\vec{F} \in
\mathbb{R}^{\tau_{\rm max}}$ of components
\begin{equation}\label{eq:Fvector}
    F_t(\omega) = 
    \langle \mathcal{C}(t) \rho(\omega)\rangle
    =
    \int d\vec{\eta}\, \mathcal{D}\rho \; 
    \mathcal{C}(t) \, \rho(\omega)\,  
    \mathbb{G}[\vec{\eta}, C_Y] \,\Pi[\rho]\;   , 
\end{equation}
together with the scalar $F_*$,
\begin{equation}\label{eq:f_star}
    F_*(\omega) = \int \mathcal{D}\rho\; 
    \rho(\omega)^2\; \Pi[\rho]  = 
    \mathcal{K}^{\rm prior}(\omega,\omega) \; .
\end{equation}
The total covariance can be written as a block matrix,
\begin{equation}
    \label{eq:total_covariance}
    \Sigma^{\rm tot} = \begin{pmatrix} F_*(\omega) & \vec{F}(\omega)^T
      \\ \vec{F}(\omega) & \Sigma+C_Y 
    \end{pmatrix}\; .
\end{equation}
Let $C^{\rm obs}(t)$ be the correlator measured on the lattice by
averaging over a gauge ensemble, and $C^{\, \rm prior}(t)$ the vector 
$\vec{\mathcal{C}}$ evaluated at $\rho= \rho^{\rm prior}$, whose components are:
\begin{equation}
     C^{\rm prior}(t) = \mathcal{C}(t) \, |_{\rho^{\rm prior}} 
        = \int dE \, b_T(t,E) \, \rho^{\rm prior}(E) \; .
\end{equation}
The joint probability density for $\rho(\omega)$ and
$\mathcal{C}(t)$, is also a Gaussian distribution,
\begin{equation}
    \mathbb{G} \left[\rho - \rho^{\rm prior} , \,
      \vec{\mathcal{C}} - \vec{C}^{\; \rm prior} ; \, \Sigma^{\rm tot}
      \right] \; ,
\end{equation}
and can be factorized as the product of the posterior probability density 
for $\rho(\omega)$ and the 
likelihood of the data. Both distributions are Gaussian, and therefore
\begin{equation}
    \label{eq:conditional_probability_factorised}
    \mathbb{G} \left[\rho - \rho^{\rm prior} , \,
        \vec{\mathcal{C}} - \vec{C}^{\; \rm prior} ; \, \Sigma^{\rm tot}
        \right] 
      = \mathbb{G} \left[ \rho- \rho^{\rm post}; \,
        \mathcal{K}^{\rm post}\right] \; \mathbb{G} \left[
        \vec{\mathcal{C}} - \vec{C}^{\; \rm prior} ; \, \Sigma +
        C_Y \right] \; .
\end{equation}

\medskip
\noindent
\fbox{
    \begin{minipage}{0.9\textwidth}
        \begin{exercise}
            Prove Eq.~\eqref{eq:conditional_probability_factorised}.
        \end{exercise}
    \end{minipage}    
}
\medskip

\noindent
The posterior Gaussian distribution for the spectral density is centred around:
\begin{equation}
    \label{eq:naive_gp_mean}
    \left. \rho^{\rm post}(\omega) \right|_{\mathcal{C}=C^{\rm obs}} 
        = \rho^{\rm prior}(\omega) 
            + \vec{F}^T(\omega) \frac{1}{\Sigma + C_Y} 
            \left( \vec{C}^{\;\rm obs} - \vec{C}^{\; \rm prior} \right) \; ,
\end{equation}
and has variance
\begin{equation}
    \label{eq:naive_gp_variance}
    \left. \mathcal{K}^{\rm post}(\omega,\omega) \right|_{\mathcal{C}=C^{\rm obs}} 
        = \mathcal{K}^{\rm prior}(\omega,\omega) 
            - \vec{F}^T(\omega) \frac{1}{\Sigma + C_Y} \vec{F}(\omega) \; .
\end{equation}
Note that the posterior mean obtained in Eq.~\ref{eq:naive_gp_variance} is a linear combination of the data, 
as in the Backus-Gilbert method discussed in the previous section. In order to explicitly make contact with the 
BG result that we obtained in the previous section, we introduce the coefficients
\begin{equation}
    \label{eq:gt_naive_GP}
    \vec{g}^{\rm \, GP} (\omega) =  \vec{F}^T  \frac{1}{\Sigma + C_Y} \; , 
\end{equation}
where the superscript GP reminds us that these are the coefficients obtained using the GP method. 

Equation~\eqref{eq:gt_naive_GP} shows that the problem is regularized by formulating it in terms of
probability distributions. The numerical instability of the BG solution, which
manifests itself in the very large coefficients $\vec{g}^{\rm \, GP} (\omega)$, 
is the consequence of the large condition
number of the model covariance $\Sigma$ of Eq.~\eqref{eq:SigmaDef},
which grows exponentially for acceptable choices of the model
covariance $\mathcal{K}$. In the absence of errors on the data, the
magnitude of the coefficients is determined by the inverse
of the matrix $\Sigma$. For noisy data, the
covariance $C_Y$ is added to the matrix $\Sigma$, thereby providing a
cut-off for its lower modes. The magnitude of the coefficients is now
determined by the condition number of $\Sigma + C_Y$. We will later show 
how the choice of $\mathcal{K}^{\rm prior}$ can be used to tune the cut-off 
on the low modes of $\Sigma$.

Another important remark concerns the smearing kernel that one has
implicitly introduced in Eq.~\eqref{eq:naive_gp_mean},
\begin{equation}
    \label{eq:GP_smearing_kernel}
    \mathcal{S}^{\rm GP}(\omega,E) = \sum_{\tau=1}^{\tau_{\rm max}} \vec{g}^{\rm \, GP} (\omega) \,b_T(a\tau ,E) \; .
\end{equation}
The centre of the posterior distribution, $\rho^{\rm post}(\omega)$ is an estimator
for a spectral density that is smeared
with $\mathcal{S}^{\rm GP}$. The unsmeared spectral density can be
obtained if the prior is engineered such that $\mathcal{S}^{\rm
  GP}(\omega-E)$ is a Dirac delta. Such a distribution cannot be
obtained by a finite linear combination of the regular functions
$b_T(t,E)$, and one would need to extrapolate the result at vanishing
smearing radius.

We conclude this section with a crucial remark on the nature of the
prior. The functional behaviour of the prediction is related to the
prior through Eqs.~\eqref{eq:naive_gp_mean}
and~\eqref{eq:naive_gp_variance}. In particular, the covariance of the
prior is known to affect the typical correlation length of the
posterior. On the lattice finite-volume spectral densities are
targeted, therefore this length should be smaller than the typical
spacing between energy levels in order to capture the features of the
underlying physics.

\subsubsection{Choice of the priors}
\label{sec:ChoiceOfGPPrior}

The prior model covariance used in this lectures is a variation of a Gaussian covariance, 
\begin{equation}
    \label{eq:Kprior_gaussian}
    \mathcal{K}(\omega,E) 
        = \frac{e^{\alpha E}}{\lambda} 
            \frac{e^{-\frac{(\omega-E)^2}{2\epsilon^2}}}{\sqrt{2\pi} \epsilon} 
        \equiv \frac{e^{\alpha E}}{\lambda} G_\epsilon(\omega-E) \; ,
\end{equation}
where $\epsilon$, $\alpha$ and $\lambda$ are the hyperparameters that fully 
specify the prior. The motivations behind the choice made in
Eq.~\eqref{eq:Kprior_gaussian} are the following. First, a Gaussian is
a common choice in the
literature~\cite{Horak:2021syv,10.1093/gji/ggz520}, with width
$\epsilon$ and amplitude $\lambda^{-1}$ as parameters to be chosen. In
addition, the Gaussian distribution has a simple limit in which the
covariance becomes diagonal by changing the parameter $\epsilon$. The
term $e^{\alpha E}$ allows us to control the deviations from the Gaussian
case, and provides a link with the results obtained using the method of
Ref.~\cite{Hansen:2019idp}. In order to set the notation, we
specialize the previous equations to this choice of the prior,
\begin{equation}
    \label{eq:SigmaMat_GP_specialised}
    \frac{\Sigma^\epsilon_{tr}}{\lambda} 
        = \int dE_1\, dE_2\;
            b_T(E_1,t) \, b_T(E_2,r)\, e^{\alpha E_1} \,
            \frac{G_\epsilon(E_1-E_2)}{\lambda} \; ,
\end{equation}
\begin{equation}
    \label{eq:Fvector_GP_specialised}
    \frac{F^{\epsilon}_t(\omega)}{\lambda} 
        = \int dE \, b_T(t,E) \,
            e^{\alpha E} \; \frac{G_\epsilon(E-\omega)}{\lambda} \; .
\end{equation}
The resulting expression for the coefficients of
Eq.~\eqref{eq:gt_naive_GP} is
\begin{equation}
    \label{eq:gt_naiveGP}
    \vec{g}^{\rm GP}(\epsilon; \omega) 
        = \vec{F}^{\, T}(\omega) \,
            \frac{1}{\Sigma^\epsilon + \lambda \, C_Y} \; .
\end{equation}
It is clear that $\lambda$ is parametrizing the cut-off on the low
modes of $\Sigma^\epsilon$, thus introducing a bias. When $\lambda$
becomes larger the role of the regularizing term, $C_Y$, is
enhanced, and the coefficients $g_\tau$ become increasingly
smaller. At smaller values of $\lambda$, the bias decreases, but it
cannot be eliminated. Its dependence must be therefore addressed, its
effect quantified and controlled. The role of such parameter is
extensively discussed in the context of Backus-Gilbert methods, and in
particular in its formulation of Ref.~\cite{Hansen:2019idp}, where it
is usually prescribed to choose $\lambda$ such that the prediction for
the spectral density is stable within statistical
noise~\cite{Bulava:2021fre,DelDebbio:2022qgu,ExtendedTwistedMassCollaborationETMC:2022sta}
upon variations of $\lambda$. From this ``stability analysis'', the
bias is assumed to be absorbed into the statistical error, a procedure
that has been validated numerically, see for instance
Refs.~\cite{Bulava:2021fre,
  ExtendedTwistedMassCollaborationETMC:2022sta, Bennett:2024cqv}. In
the context of GPs, on the other hand, the hyperparameters (including
$\lambda$) are selected so that the resulting probability of observing
the data,
\begin{equation}
\mathbb{G} \left[ \vec{C}^{\, \rm obs} - \vec{C}^{\; \rm prior} ; 
    \, \Sigma + C_Y \right] \; , 
\end{equation}
is
maximized~\cite{Horak:2021syv,10.1111/j.1365-246X.1968.tb00216.x}.
\footnote{We discussed at the beginning of this section the possibility of 
sampling the posterior distribution of the hyperparameters, which would be 
the most rigorous way to proceed. This has not been done in Ref.~\cite{DelDebbio:2024lwm}, 
where the hyperparameters were kept fixed at the mode of the posterior distribution.}
Equivalently,
one minimizes the "negative logarithmic likelihood" (NLL)
\begin{multline}\label{eq:NLL}
\frac{\tau_{\rm max}}{2}\, \text{Log} (2\pi) +\frac{1}{2} \,
\text{Log}\, \text{det} \left( \Sigma + \text{Cov}_d \right)
+\frac{1}{2} \, (\vec{C}^{\, \rm obs}- \vec{C}^{\; \rm prior})
\frac{1}{\Sigma+\text{Cov}_d} (\vec{C}^{\, \rm obs}- \vec{C}^{\; \rm
  prior}) \; .
\end{multline}

\begin{figure}[tbh]
    \centering
    \includegraphics[width=0.49\textwidth]{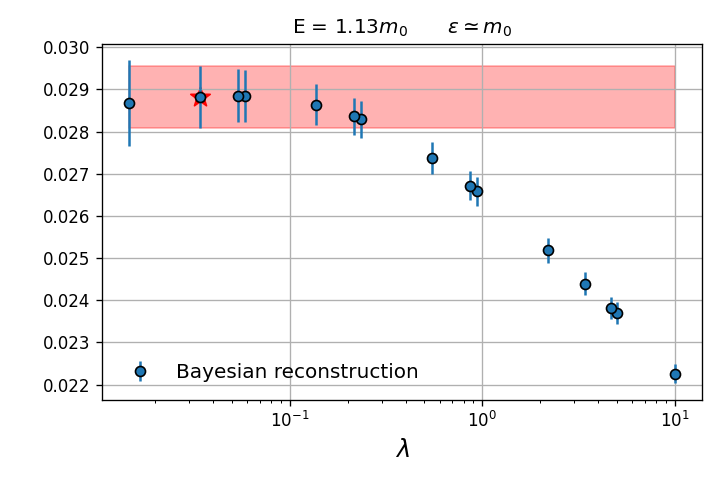}
    \includegraphics[width=0.49\textwidth]{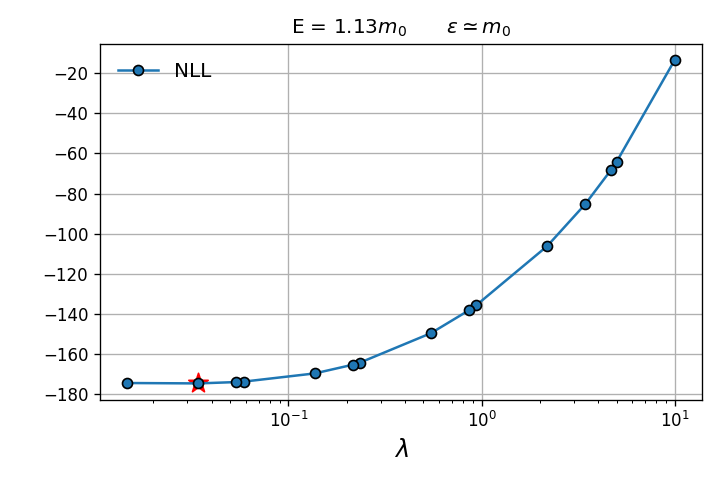}
    \caption{Combination of the stability analysis used (left) and a
      scan of the NLL (right). The smeared density shown in the left
      panel is obtained at a specific energy, and is evaluated from
      GPs, i.e. from Eqs.~\eqref{eq:naive_gp_mean}
      and~\eqref{eq:naive_gp_variance} for the central value and the
      error respectively. The prior is the modified Gaussian of
      Eq.~\eqref{eq:Kprior_gaussian}, with $\epsilon \simeq m_0$, the
      latter being the ground state of the channel. We only display
      value corresponding to $\alpha=0$ because these correspond to a
      systematically smaller NLL. The purpose of this figure is to
      show that the treatment of the parameter $\lambda$ from
      Ref.~\cite{Hansen:2019idp} and GPs can lead to compatible
      results. Details about the lattice data used for this example
      are found in the main text.}
    \label{fig:naiveGPscan}
\end{figure}

In this setting, the fate of the bias introduced by $\lambda$ could be
considered opaque. We therefore assess whether porting the ``stability
analysis'' that is carried in Ref.~\cite{Hansen:2019idp} into the
Bayesian setup, can add insights in this regard. In
Fig.~\ref{fig:naiveGPscan}, we show how the choice of $\lambda$
affects the spectral reconstruction (left panel)\footnote{The result
shown in Fig.~\ref{fig:naiveGPscan} is obtained from a pseudoscalar
correlator of fermions in a higher representation, computed within the
ensemble B3 generated by the authors in
Ref.~\cite{DelDebbio:2022qgu}.} at a specific energy. The
corresponding values of the NLL are shown in the right panel of the
same figure, with the minimum value highlighted. For large values of
$\lambda$, the smeared spectral density changes considerably, showing
a large dependence on the prior. The corresponding values of the NLL
are large. Remarkably, as the NLL approaches its minimum, the
dependence on the prior softens. The horizontal band in the left panel
of Fig.~\ref{fig:naiveGPscan} is obtained at the value of $\lambda$
that minimizes the NLL, which is flagged by a star in the right
panel. This result suggests that the treatment of the bias might not
be drastically different in these two cases, despite the very
different approaches.

\begin{figure}[tbh!]
    \centering
    \includegraphics[width=0.49\textwidth]{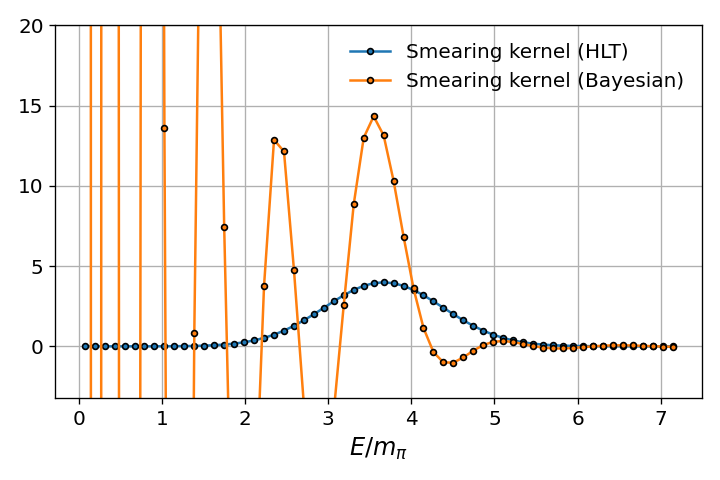}
    \includegraphics[width=0.49\textwidth]{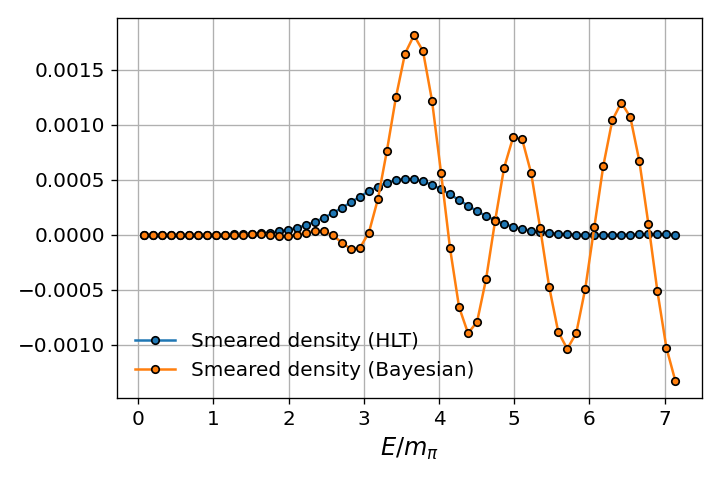}
    \caption{Left panel: examples of the function smearing the
      spectral density at the energy $E = 3.7 m_\pi$ in the Bayesian
      setup (orange) and from the HLT procedure (blue) using exact
      data. The latter targets a Gaussian kernel with a width of
      approximately $0.75$ in units of $m_\pi$, which is reconstructed
      with great precision given the lack of uncertainties on the
      input data. For the Bayesian calculation, we use the same
      Gaussian function as a prior, but we have less control on the
      output function, which in this case features oscillations with a
      length scale determined by the prior. The right panel displays
      the reconstructed smeared spectral densities from the same
      data. This example uses $t_{\rm max}=32$ data points.}
    \label{fig:naive_smearing_kernel}
\end{figure}

Another important remark, already stated in the previous Section, is
that the value obtained at the minimum of the NLL (horizontal band in
Fig.~\ref{fig:naiveGPscan}) has to be interpreted and understood as a
spectral density that is smeared with the appropriate smearing kernel,
which is not known a priori in the Bayesian formulation given in this
section. In this respect, this method is closer to the original
proposal of Backus and Gilbert rather than the modification of
Ref.~\cite{Hansen:2019idp}. For this reason we find it instructive to
show, in the left panel of Figure~\ref{fig:naive_smearing_kernel}, an
example of the smearing kernel at a specific energy ($\omega= 3.7
m_\pi$) obtained with the Bayesian setup (orange), as opposed to the
smearing kernel obtained from the method (in blue) of
Ref.~\cite{Hansen:2019idp} (HLT in short) that will be described in
the next Section. The Bayesian kernel is not known a priori: its
behaviour is constrained by the covariance of the prior, which can
produce a smoother or more rapidly changing function. In the example
of Fig.~\ref{fig:naive_smearing_kernel} the prior is defined in
Eq.~\ref{eq:Kprior_gaussian}, with $\epsilon= 0.75 m_\pi$. The output
thus features, around $\omega= 3.7 m_\pi$, oscillation that have
roughly a wavelength of $\epsilon$. In the right panel of the same
figure, the smeared spectral density obtained with the two methods is
also displayed. In this example, we have used synthetic data without
any statistical noise, in order to showcase what each method does in
the ideal limit of exact data. The input correlator contains a single
state with $E/m_\pi = 3.7$.

\newpage

\section{Neural Networks}
\label{sec:nn}

\subsection{Neural Networks Basics}
\label{sec:nn-basics}

In the context of PDF fitting, the NNPDF collaboration has pioneered the 
use of neural networks as a flexible parametrization of PDFs,
which allows to reduce the bias introduced by the choice of a specific 
functional form. The NNPDF approach is based on the idea of using a 
large number of parameters (the weights and biases of the neural network) 
to fit the data, while monitoring the fitting process in order to 
prevent overfitting, see \eg\ Ref.~\cite{NNPDF:2021njg} for details. In this 
language, the ensemble of NN replicas at initialization defines the prior 
probability distribution of the model, and the training process is used to
update the probability distribution in light of the data, thus defining the
posterior distribution. Note that, while we have prior and posterior distributions, 
the inference is not performed according to Bayes' theorem. 

In these section we briefly recall some basic notions about neural networks, 
and develop a statistical field theory description of the behaviour 
of neural networks at initialization, which will be the foundation for 
the discussion of the training process in the next section. 

A neural network is one way of parametrizing a function, $f(x)$, using 
interconnected layers of computing units, which are called {\em neurons}\ 
by analogy with the biological neurons. We are going to focus here on the 
simplest type of neural network, which is the feed-forward neural network, 
where the information flows in one direction from the input layer, 
identified by $\ell=0$ to the output layer, identified by $\ell=L$. The
neuron in each layer are labelled by an index $i$, and the number of neurons 
in layer $\ell$ is denoted by $n_\ell$. The input of the network is $x$, 
which is a vector of dimension $n_0$, and the output is $f(x)$, which is 
a vector of dimension $n_L$. If we are parametrizing a function that returns
a scalar, then $n_L=1$. At each neuron we associated a so-called 
{\em pre-activation}\ function, $\phi^{(\ell)}_i(x)$. The indices $\ell$ and 
$i$ identify the neuron, and we explicitly wrote the dependence on the 
input $x$. The output of the neuron is obtained by applying the 
{\em activation function}\ $\rho$ to the pre-activation, 
\begin{equation}
    \label{eq:ActivFunction}
    \rho^{(\ell)}_i(x) = \rho\left( \phi^{(\ell)}_i(x) \right)\, .
\end{equation}
The pre-activation function are recursively defined, 
\begin{equation}
    \label{eq:PreActivFunction}
    \phi^{(\ell)}_i(x) = \sum_{j=1}^{n_{\ell-1}} W^{(\ell)}_{ij} \, 
        \rho^{(\ell-1)}_j(x) + b^{(\ell)}_i\, , 
\end{equation}
with $\phi^{(0)}_i(x) = x_i$.
The $W^{(\ell)}$ are the {\em weight}\ matrices, $b^{(l)}$ are the {\em bias}\ 
vectors. For each input $x$, the recursion relations above allow us to 
compute the pre-activation and the output of each neuron, and ultimately 
the output of the network, $f_i(x)=\phi^{(L)}_i(x)$.

The weights and biases are the parameters of the neural network, 
they are initialized according to some prescription and they are 
then determined by the training process. Neural networks provide a 
flexible parametrization of functions and are known to be universal 
approximators for sufficiently large number of parameters. They were originally
introduced for PDF fitting precisely in order to avoid any bias introduced 
by a fixed functional form with insufficient degrees of freedom. 

\begin{figure}[!ht]
\begin{center}
  \begin{tikzpicture}[x=3.2cm,y=0.2cm]
    \message{^^JNeural network large}
    \readlist\Nnod{2,25,20,8} % array of number of nodes per layer
    
    \message{^^J  Layer}
    \foreachitem \N \in \Nnod{ % loop over layers
      \def\lay{\Ncnt} % alias of index of current layer
      \pgfmathsetmacro\prev{int(\Ncnt-1)} % number of previous layer
      \message{\lay,}
      \foreach \i [evaluate={\y=\N/2-\i; \x=\lay; \n=\nstyle;
                             \nprev=int(\prev<\Nnodlen?min(2,\prev):3);}] in {1,...,\N}{ % loop over nodes
        
        % NODES
        %\node[node \n,outer sep=0.6,minimum size=18] (N\lay-\i) at (\x,\y) {};
        \coordinate (N\lay-\i) at (\x,\y);
        
        % CONNECTIONS
        \ifnum\lay>1 % connect to previous layer
          \foreach \j in {1,...,\Nnod[\prev]}{ % loop over nodes in previous layer
            \draw[connect,white,line width=1.2] (N\prev-\j) -- (N\lay-\i);
            \draw[connect] (N\prev-\j) -- (N\lay-\i);
            %\draw[connect] (N\prev-\j.0) -- (N\lay-\i.180); % connect to left
            \node[node \nprev,minimum size=6] at (N\prev-\j) {}; % draw node over lines
          }
          \ifnum \lay=\Nnodlen % draw last node over lines
            \node[node \n,minimum size=6] at (N\lay-\i) {};
          \fi
        \fi % else: nothing to connect first layer
        
      }
    }  
  \end{tikzpicture}   
  \end{center}
  \caption{A feed-forward neural network with $L=3$ layers and $n_0=2$, 
  $n_1=25$, $n_2=20$, $n_3=8$. The input layer is in green on the left, the
  output layer is in red on the right, and the hidden layers are in blue. 
  This is the actual architecture of the neural network used by the 
  NNPDF collaboration for PDF fitting, see Ref.~\cite{NNPDF:2021njg}.}  
  \label{fig:ExampleOfANN}
\end{figure}
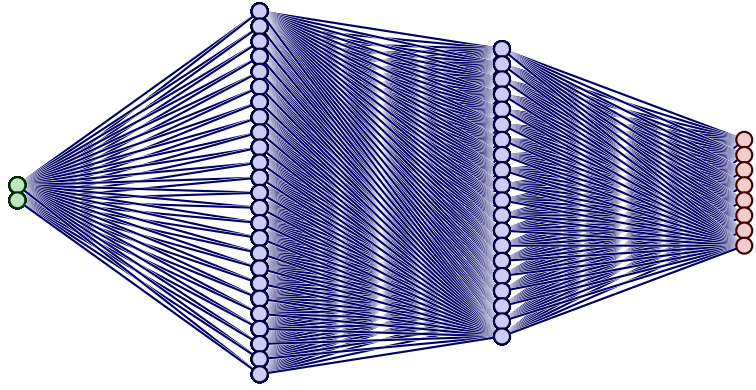

Typically, we consider a discrete set of input points, labelled by 
an index $\alpha$ as discussed in the previous sections. The input points 
are therefore $x_\alpha$. In order to simplify the notation, we introduce
\begin{equation}
    \label{eq:AlphaIndexNotation}
    \phi^{(\ell)}_{i\alpha} = \phi^{(\ell)}_i(x_\alpha)\, , \qquad
    \rho^{(\ell)}_{i\alpha} = \rho^{(\ell)}_i(x_\alpha)\, .
\end{equation}
It is important to distinguish between the index $\alpha$, which labels 
the input points, and the index $i$, which labels the neurons. For each 
$\alpha$, the input $x$ could be a vector, whose components are labelled by 
the index $i=1, \ldots, n_0$. A typical example of a NN is shown in 
Fig.~\ref{fig:ExampleOfANN}. Each dot in the figure represents a 
neuron, so that for each dot we have a pre-activation 
$\phi^{(\ell)}_{i\alpha}$ and an output $\rho^{(\ell)}_{i\alpha}$. The actual 
values of the pre-activation and the output depend on the input 
$x_\alpha$, which is not explicitly shown in the figure.
The connections between the dots represent the weights $W^{(\ell)}_{ij}$, 
and the bias $b^{(\ell)}_i$ is associated to each neuron. 
The input layer is in green on the left, the output layer is in red on the right, 
and the hidden layers are in blue. Note that the NN shown in the figure has two 
neurons in the input layer. Having two neurons in the input layer allows 
us to use both $x$ and $\log x$ as inputs, which enhances the sensitivity of the 
NN to the small-$x$ region. 

\subsection{Neural Networks at Initialization}
\label{sec:nn-init}

The weights and biases of the NN are initialized as independent Gaussian 
variables, 
\begin{equation}
    \label{eq:NNInitialization}
    W^{(\ell)}_{ij} \sim \mathcal{N}\left(0, \frac{C_W}{n_{\ell-1 }}\right) \, , 
    \qquad
    b^{(\ell)}_i \sim \mathcal{N}(0, C_b) \, .
\end{equation}
Both distributions have zero mean. The variance of the weights in layer $\ell$ 
is scaled by the number of neurons in the previous layer, $n_{\ell-1}$. 
This is a common choice, sometimes referred to as the Glorot 
initialization~\cite{pmlr-v9-glorot10a}.
By inspecting the recursion relation in Eq.~\ref{eq:PreActivFunction}, it is
clear that the scaling of the variance guarantees that the 
pre-activations have a finite limit when the number of neurons in the 
previous layer becomes large. The average over the probability 
distributions in Eq.~\ref{eq:NNInitialization} yields 
\begin{align}
    \label{eq:NNInitWMean}
    & \mathbb{E}[W^{(\ell)}_{ij}] = 0 \, ,  \\
    \label{eq:NNInitBMean}
    & \mathbb{E}[b^{(\ell)}_i] = 0 \, , \\
    \label{eq:NNInitWVar}
    & \mathbb{E}[W^{(\ell)}_{ij} W^{(\ell)}_{i'j'}] 
      = \delta_{ii'}\delta_{jj'} \frac{C_W}{n_{\ell-1}} \, , \\
    \label{eq:NNInitBCov}
    & \mathbb{E}[b^{(\ell)}_i b^{(\ell)}_{i'}] 
      = \delta_{ii'} \, C_b \, , 
\end{align}
and all other correlations -- in particular correlations between different
layers -- vanish. The $\phi^{(0)}$ are given by the values of the input, and 
therefore are deterministic quantities. All other pre-activations are 
stochastic variables and we are interested in their joint probability 
distribution, 
\begin{equation}
  \label{eq:JointPriorProbability}
  P\left(\phi^{(L)} \ldots \phi^{(1)}\right) = 
    P\left(\phi^{(L)}|\phi^{(L-1)}\right) \ldots 
    P\left(\phi^{(1)}|\phi^{(0)}\right) \, .
\end{equation}
The following derivations are based on the recent review in Ref.~\cite{Roberts_2022}.

\paragraph{A Forward Recursion Relation.}
The probability distribution of the pre-activations in layer $\ell$ obeys 
a forward recursion relation, which allows to compute it from the 
probability distribution of the pre-activations in layer $\ell-1$, 
\begin{equation}
  \label{eq:RecursionRelationForPhi}
  P\left(\phi^{(\ell+1)} | \phi^{(0)}\right) = \int d\phi^{(\ell)} \, 
    P\left(\phi^{(\ell+1)} | \phi^{(\ell)}\right) 
    P\left(\phi^{(\ell)} | \phi^{(0)}\right) \, .
\end{equation}
The conditional probability distribution $P\left(\phi^{(\ell+1)} | \phi^{(\ell)}\right)$ 
can be computed by integrating over the weights and biases,
\begin{align}
    P\left(\phi^{(\ell+1)}|\phi^{(\ell)}\right) 
      &= \int \prod_{i,j} dW^{(\ell+1)}_{ij}
        \prod_i db^{(\ell+1)}_i \, p\left(W^{(\ell+1)}\right) p\left(b^{(\ell+1)}\right) 
        \, \nonumber \\
    \label{eq:ForwardRecursion} 
      & \qquad \times \
        \prod_i \delta\left( \phi^{(\ell+1)}_i - \sum_j W^{(\ell+1)}_{ij} 
        \rho^{(\ell)}_j - b^{(\ell+1)}_i \right) \, ,
\end{align}
where $p\left(W^{(\ell+1)}\right)$ and $p\left(b^{(\ell+1)}\right)$ are the probability
distributions of the weights and biases in layer $\ell+1$ given in Eq.~\ref{eq:NNInitialization}. The delta functions in the last line 
enforce the definition of the pre-activations in terms of the weights, biases, and 
activations in the previous layer. For fixed values of the 
pre-activations in layer $\ell$, the pre-activations in layer $\ell+1$ are 
linear combinations of the weights and biases,
and therefore their probability distribution is Gaussian, 
\begin{equation}
  \label{eq:GaussianDistributionForPhi}
  P\left(\phi^{(\ell+1)}|\phi^{(\ell)}\right) = 
  \frac{1}{\left|2\pi \hat{G}^{(\ell+1)}\right|^{n_{\ell+1}/2}} \,
  \exp\left( -\frac{1}{2} \phi^{(\ell+1)}_{\alpha_1} \cdot 
    \phi^{(\ell+1)}_{\alpha_2}
    \left(\hat{G}^{(\ell+1)}\right)^{-1}_{\alpha_1 \alpha_2}
   \right) \, ,
\end{equation}
where the scalar product indicates the contraction of the neuron indices, 
and we introduced the {\em propagator}
\begin{equation}
  \label{eq:Propagator}
  \hat{G}^{(\ell+1)}_{\alpha_1 \alpha_2} = 
    C_b + C_W \, \frac{1}{n_\ell} 
    \rho^{(\ell)}_{\alpha_1} \cdot \rho^{(\ell)}_{\alpha_2} \, . 
\end{equation}
The factor of $1/n_\ell$ in Eq.~\ref{eq:Propagator} comes from the 
scaling of the variance of the weights in Eq.~\ref{eq:NNInitialization}. 
The scalar product between the outputs $\rho^{(\ell)}$ involves a sum 
over the neuron indices, and therefore it is of order $n_\ell$; 
the propagator has a finite limit when the number of neurons in 
layer $\ell$ becomes large. Clearly, $\hat{G}^{(\ell+1)}$ depends on 
the pre-activations $\phi^{(\ell)}$.

\medskip
\noindent
\fbox{
    \begin{minipage}{0.9\textwidth}
    \begin{exercise}
        Show that the correlators of the pre-activations in any layer $\ell$ are 
        symmetric under rotations in the space of neurons, \ie\ 
        \begin{align}
          \label{eq:NeurRotInv}
          \mathbb{E}\left[
            R_{i_1j_1} \phi^{(\ell)}_{j_1 \alpha_1} \ldots
            R_{i_nj_n} \phi^{(\ell)}_{j_n \alpha_n}
          \right] =
          \mathbb{E}\left[
            \phi^{(\ell)}_{i_1 \alpha_1} \ldots
            \phi^{(\ell)}_{i_n \alpha_n}
          \right]\, ,
        \end{align}
        where $R$ is an orthogonal matrix in $\text{SO}(n_{\ell})$. Eq.\eqref{eq:NeurRotInv} implies
        that the probability distribution is also invariant under rotations,
        and therefore it can only be a function of $\text{SO}(n_{\ell})$ invariants, \ie\ it 
        can only depend on 
        the scalar product of the pre-activations.
    \end{exercise}
\end{minipage}
}
\medskip

In the first hidden layer, $\ell = 1$, the pre-activations are 
linear combinations of the weights and biases, 
\begin{equation}
  \label{eq:PhiFirstLayer}
  \phi^{(1)}_{i\alpha} = \sum_j W^{(1)}_{ij} \rho\left(x_{\alpha j}\right) + b^{(1)}_i \, ,
\end{equation}
and therefore they are Gaussian variables, with zero mean and covariance
\begin{equation}
  \label{eq:CovPhiFirstLayer}
  \mathbb{E}\left[ \phi^{(1)}_{i\alpha} \phi^{(1)}_{i'\alpha'} \right] = 
    \delta_{ii'} \left( C_b + C_W \, \frac{1}{n_0} 
    \sum_j \rho\left(x_{\alpha j}\right) \rho\left(x_{\alpha' j}\right) \right) \, .
\end{equation}
One can readily check, by writing down the explicit expression, that the distributions are 
no longer Gaussian as soon as we move to $\ell > 1$. 

\paragraph{Effective Theory Approach.}
It is useful to rewrite the recursion relation in~\eqref{eq:RecursionRelationForPhi} by 
introducing an effective action, $S^{(\ell)}$, defined by
\begin{equation}
  \label{eq:EffectiveAction}
  P\left(\phi^{(\ell)} | \phi^{(0)}\right) = \frac{1}{Z^{(\ell)}} \, 
    e^{-S^{(\ell)}(\phi^{(\ell)})} \, , 
\end{equation}
where $Z^{(\ell)}$ is a normalization factor. The recursion relation in Eq.~\ref{eq:RecursionRelationForPhi} can be rewritten as
\begin{equation}
  \label{eq:RecursionRelationForEffectiveAction}
  \frac{e^{-S^{(\ell+1)}(\phi^{(\ell+1)})}}{Z^{(\ell+1)}} 
    = \int d\phi^{(\ell)} \, \frac{e^{-S^{(\ell)}(\phi^{(\ell)})}}{Z^{(\ell)}} \,
      \frac{\exp\left( -\frac{1}{2} \phi^{(\ell+1)}_{\alpha_1} \cdot 
    \phi^{(\ell+1)}_{\alpha_2}
    \left(\hat{G}^{(\ell+1)}\right)^{-1}_{\alpha_1 \alpha_2}
   \right)}{\left|2\pi \hat{G}^{(\ell+1)}\right|^{n_{\ell+1}/2}} \, .
\end{equation}

In order to extract further information from the recursion relation, we need to 
parametrize the effective action. The most general form of the effective action is
\begin{equation}
  \label{eq:GeneralEffectiveAction}
  S^{(\ell)}(\phi^{(\ell)}) = \frac{1}{2}\, \gamma^{(\ell)}_{\alpha_1 \alpha_2} \, 
    \left(\phi^{(\ell)}_{\alpha_1} \cdot \phi^{(\ell)}_{\alpha_2}\right) 
     + \frac{1}{8 n_{\ell-1}} \, \gamma^{(\ell)}_{\alpha_1 \alpha_2; \alpha_3 \alpha_4} \,
    \left( \phi^{(\ell)}_{\alpha_1} \cdot \phi^{(\ell)}_{\alpha_2} \right)
    \left( \phi^{(\ell)}_{\alpha_3} \cdot \phi^{(\ell)}_{\alpha_4} \right) + \ldots \, ,      
\end{equation}
where we introduced the {\em couplings} $\gamma^{(\ell)}_{\alpha_1 \alpha_2}$ and
$\gamma^{(\ell)}_{\alpha_1 \alpha_2; \alpha_3 \alpha_4}$, which are functions of the 
layer $\ell$ and of the input points. Note that because of symmetry reasons, the effective 
action can only depend on the scalar product of the pre-activations. The factors in the 
action reflect the symmetries of each term, and the scaling with the number of neurons in 
the previous layer is chosen so that the couplings have a finite limit when $n_{\ell-1}$ 
becomes large. We have omitted higher order terms in the effective action; they involve
higher powers of the scalar product of the pre-activations, and they are suppressed by 
higher powers of $1/n_{\ell-1}$.

At leading order, the second and fourth cumulants are respectively
\begin{align}
    \label{eq:SecondOrderCumulantLeading}
    &\langle \phi^{(\ell)}_{i_1,\alpha_1} \phi^{(\ell)}_{i_2,\alpha_2}\rangle
      = \delta_{i_1 i_2} K^{(\ell)}_{\alpha_1\alpha_2} + O(1/n_{\ell-1})\, , \\
    \label{eq:FourthOrderCumulantLeading}
    &\langle \phi^{(\ell)}_{i_1,\alpha_1} \phi^{(\ell)}_{i_2,\alpha_2}
      \phi^{(\ell)}_{i_3,\alpha_3} \phi^{(\ell)}_{i_4,\alpha_4}\rangle_c
      = O(1/n_{\ell-1})\, ,
\end{align}
where\footnote{
    The notation here refers to the matrix element $(\alpha_1,\alpha_2)$ of the inverse matrix of $\gamma^{(\ell)}$, and {\em not}\ to the inverse
    of the matrix element $\gamma^{(\ell)}_{\alpha_1\alpha_2}$.
}
\begin{equation}
    \label{eq:DefineKmat}
    K^{(\ell)}_{\alpha_1\alpha_2} = \left(\gamma^{(\ell)}\right)^{-1}_{\alpha_1\alpha_2}\, .
\end{equation}

\medskip
\noindent
\fbox{
    \begin{minipage}{0.9\textwidth}
        \begin{exercise}
            Derive the Feynman rules for the effective theory defined  by 
            the action in Eq.~\eqref{eq:GeneralEffectiveAction}, and use them to 
            compute the second and fourth cumulant of the pre-activations at 
            leading order in $1/n_{\ell-1}$, thus showing that they are given by 
            the expressions in Eqs.~\eqref{eq:SecondOrderCumulantLeading} and 
            \eqref{eq:FourthOrderCumulantLeading}.
        \end{exercise}
    \end{minipage}    
}
\medskip

\begin{figure}[!htb]
    \centering
    \includegraphics[scale=0.58]{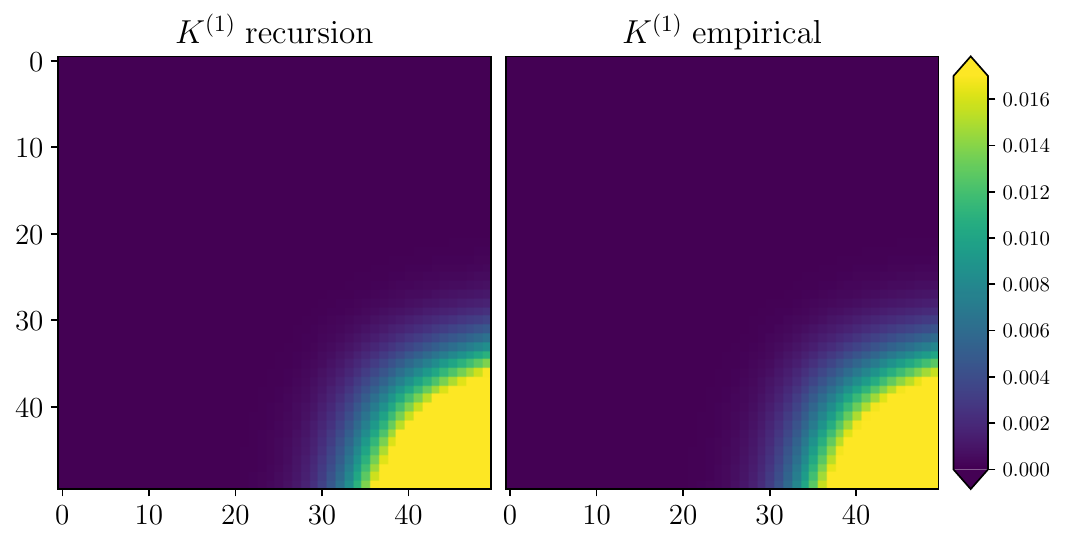}
    \includegraphics[scale=0.58]{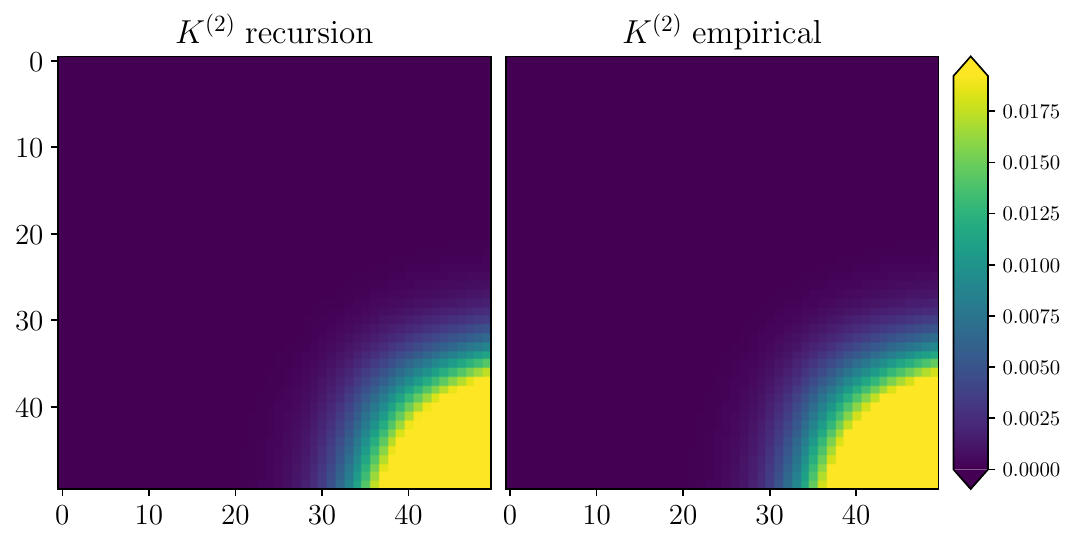}
    \includegraphics[scale=0.58]{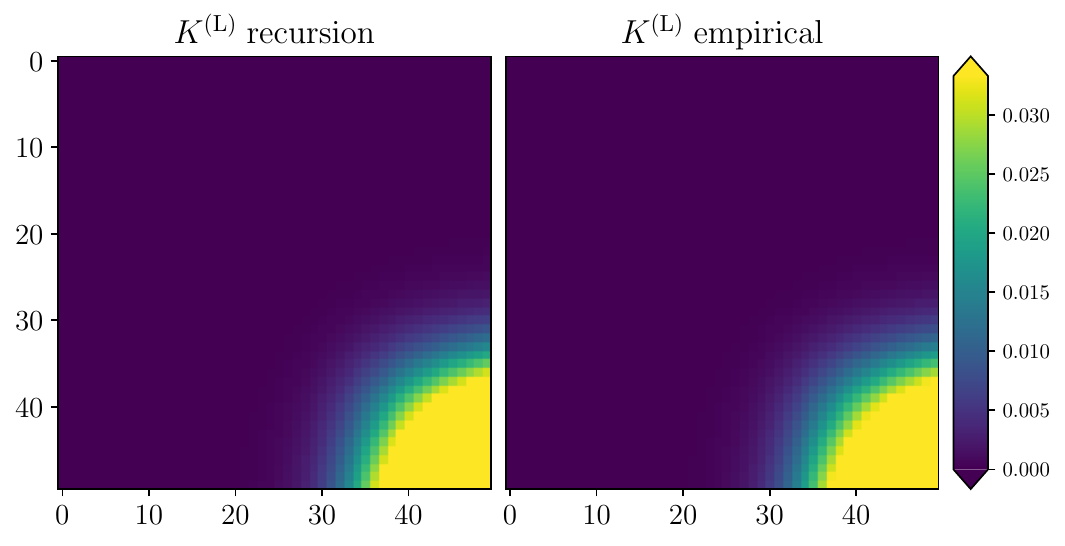}
    \caption{The empirical (left) and analytical (right) covariance matrices of
    the first, second and output layers of the NNPDF architecture (top to
    bottom). The covariance in the left panel is computed ``bootstrapping'' over
    an ensemble of replicas, initialized using the Glorot normal distribution.
    The covariance in the right panel is obtained by solving
    Eq.~\eqref{eq:RecursionForK} numerically. In order to reduce the bootstrap
    errors in the empirical covariance, an ensemble of 1000 replicas has been
    used for this figure.}
    \label{fig:KRecursion}
\end{figure}

\begin{figure}[!htb]
    \centering
    \includegraphics[width=0.86\textwidth]{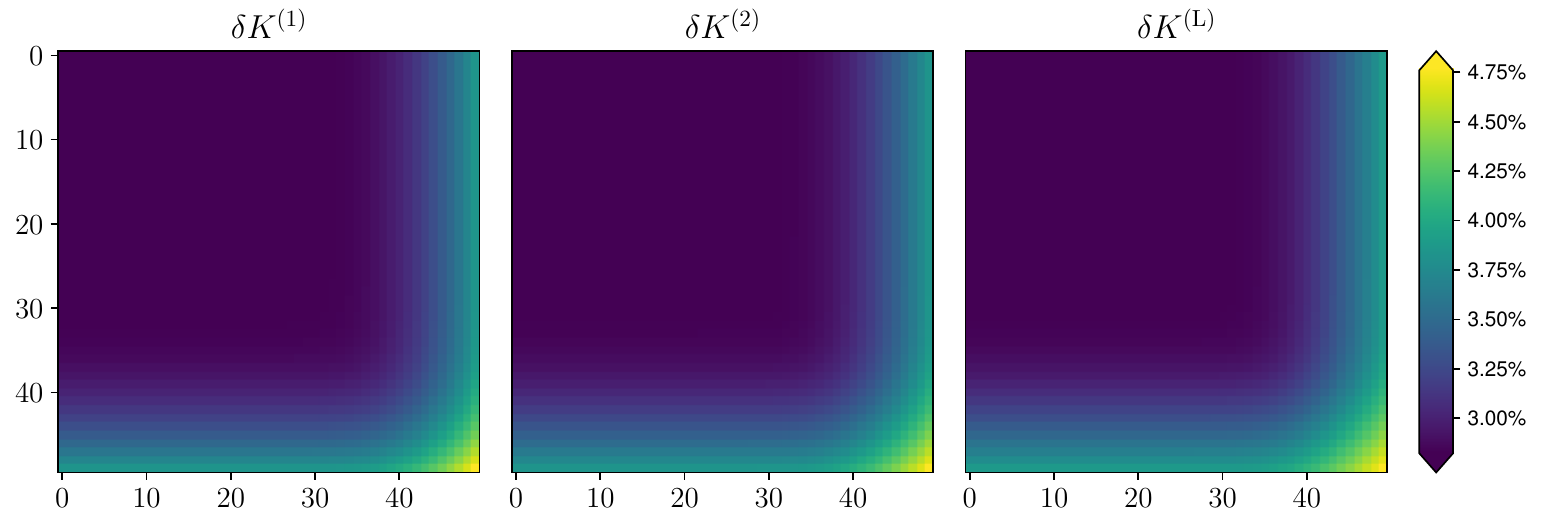}
    \caption{Relative difference between the empirical kernel,
    computed from an ensemble of networks at initialization, and the recursive
    kernel obtained by iterating Eq.~\eqref{eq:RecursionForK} for the three layers
    of the NNPDF architecture. An ensemble of 1000 replicas has been used to
    reduce the bootstrap errors in the empirical covariance.}
    \label{fig:delta_K}
\end{figure}

The ``evolution'' of the couplings as we go deep in the NN, \ie, the dependence of the couplings on
$\ell$, is governed by Renormalization Group (RG) equations, which preserve the power counting in
powers of $1/n_{\ell}$. At leading order,
\begin{align}
    K^{(\ell+1)}_{\alpha_1\alpha_2} &=
      \left.
      C_b^{(\ell+1)} + C_w^{(\ell+1)} \frac{n_\ell}{n_\ell+n_{\ell+1}}\frac{1}{n_\ell}
      \langle \vec{\rho}^{\,(\ell)}_{\alpha_1} \cdot
      \vec{\rho}^{\,(\ell)}_{\alpha_2} \rangle
      \right|_{O(1)} \\
      \label{eq:RecursionForK}
      &= C_b^{(\ell+1)} + C_w^{(\ell+1)} \frac{n_\ell}{n_\ell+n_{\ell+1}}\frac{1}{n_\ell}
      \langle \vec{\rho}^{\,(\ell)}_{\alpha_1} \cdot
      \vec{\rho}^{\,(\ell)}_{\alpha_2} \rangle_{K^{(\ell)}}\, ,
\end{align}
where
\begin{align*}
    \frac{1}{n_\ell}
      \langle \vec{\rho}^{\,(\ell)}_{\alpha_1} \cdot
      \vec{\rho}^{\,(\ell)}_{\alpha_2} \rangle_{K^{(\ell)}} =
    \int \mathcal{D}\phi\,
      \frac{e^{-\frac12 \left(K^{(\ell)}\right)^{-1}_{\beta_1\beta_2}
        \phi_{\beta_1} \phi_{\beta_2}}}
        {\left|2\pi K^{(\ell)}\right|^{1/2}}\,
        \rho(\phi_{\alpha_1}) \rho(\phi_{\alpha_2})\, , 
\end{align*}
and
\begin{align}
    \label{eq:FunctIntDef}
    \mathcal{D}\phi = \prod_{\alpha=1}^{\ngrid} d\phi_\alpha\, .
\end{align}
Note that the integration variables in Eq.~\eqref{eq:FunctIntDef} do not have a
neuron index and the integrals are $\ngrid$ dimensional integrals.
Eq.~\eqref{eq:RecursionForK} is iterated for the NNPDF architecture, yielding
$K^{(\ell)}$ for arbitrary $\ell$, \ie, the covariance at initialization for
various depths. These are compared with the empirical covariance computed from
an ensemble replicas in Fig.~\ref{fig:KRecursion} for the first two hidden layers
and the output layer. Furthermore, the relative difference between the empirical
covariance and the theoretical prediction is shown in Fig.~\ref{fig:delta_K}. In
order to reduce the bootstrap errors in the empirical covariance, an ensemble
with $\nreps=1000$ has been used for these figures. The agreement between the
theoretical prediction and the empirical computation is excellent, confirming
the validity of the large-network expansion even for networks of moderate size,
as those used in the NNPDF fits.


\begin{figure}
  \centering
  \includegraphics[width=0.95\textwidth]{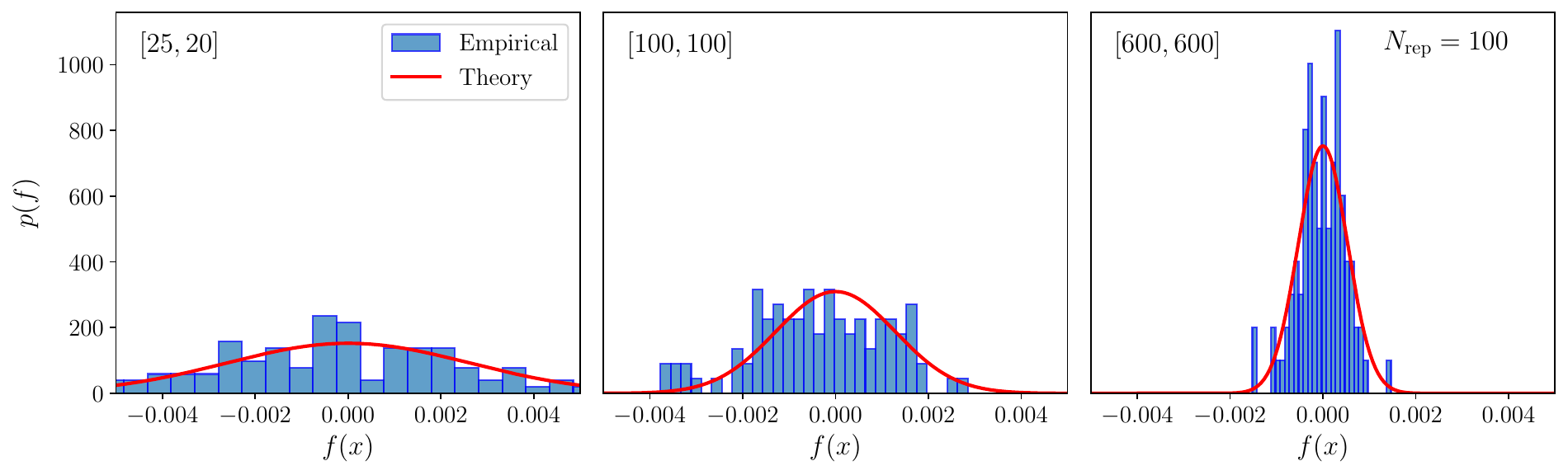}
  \includegraphics[width=0.95\textwidth]{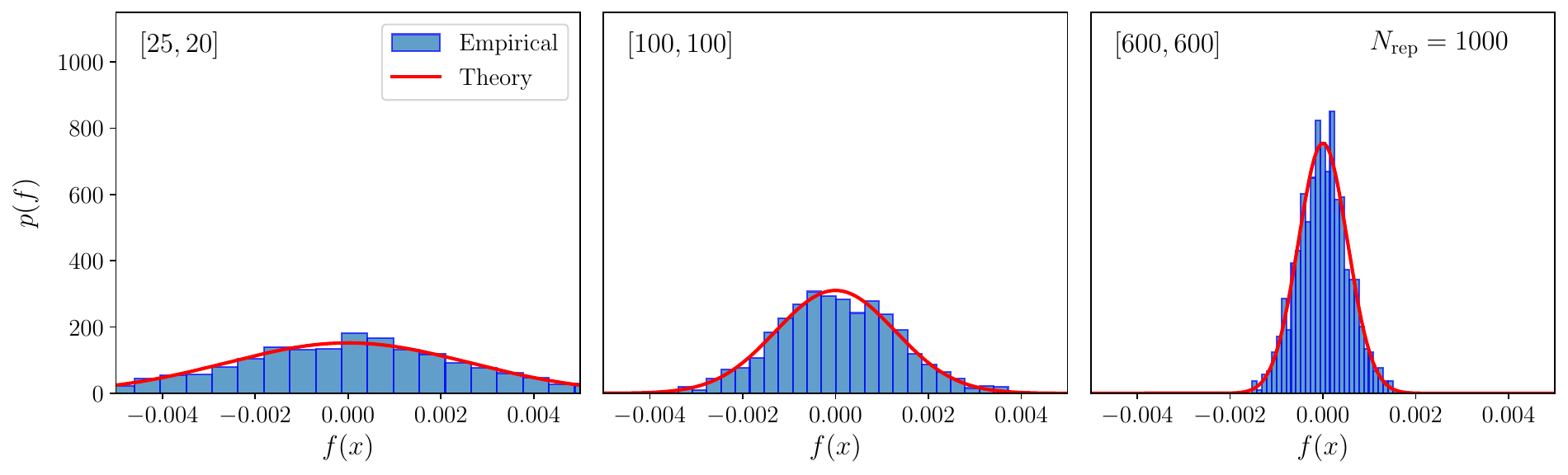}
  \caption{Sampled distribution of the output $xT_3$ at $x=0.0065$ for two
  different ensemble sizes, $\nreps=100$ (top) and $\nreps=1000$ (bottom). Each
  column shows the distribution for a different network architecture, the latter
  displayed in the top left corner of each panel. The red line the represents
  the predicted Gaussian distribution as dictated by the kernel recursion
  formula in Eq.~\eqref{eq:RecursionForK}.
  \label{fig:OutputDist}}
\end{figure}

\begin{figure}[!t]
    \centering
    \includegraphics[width=0.45\textwidth]{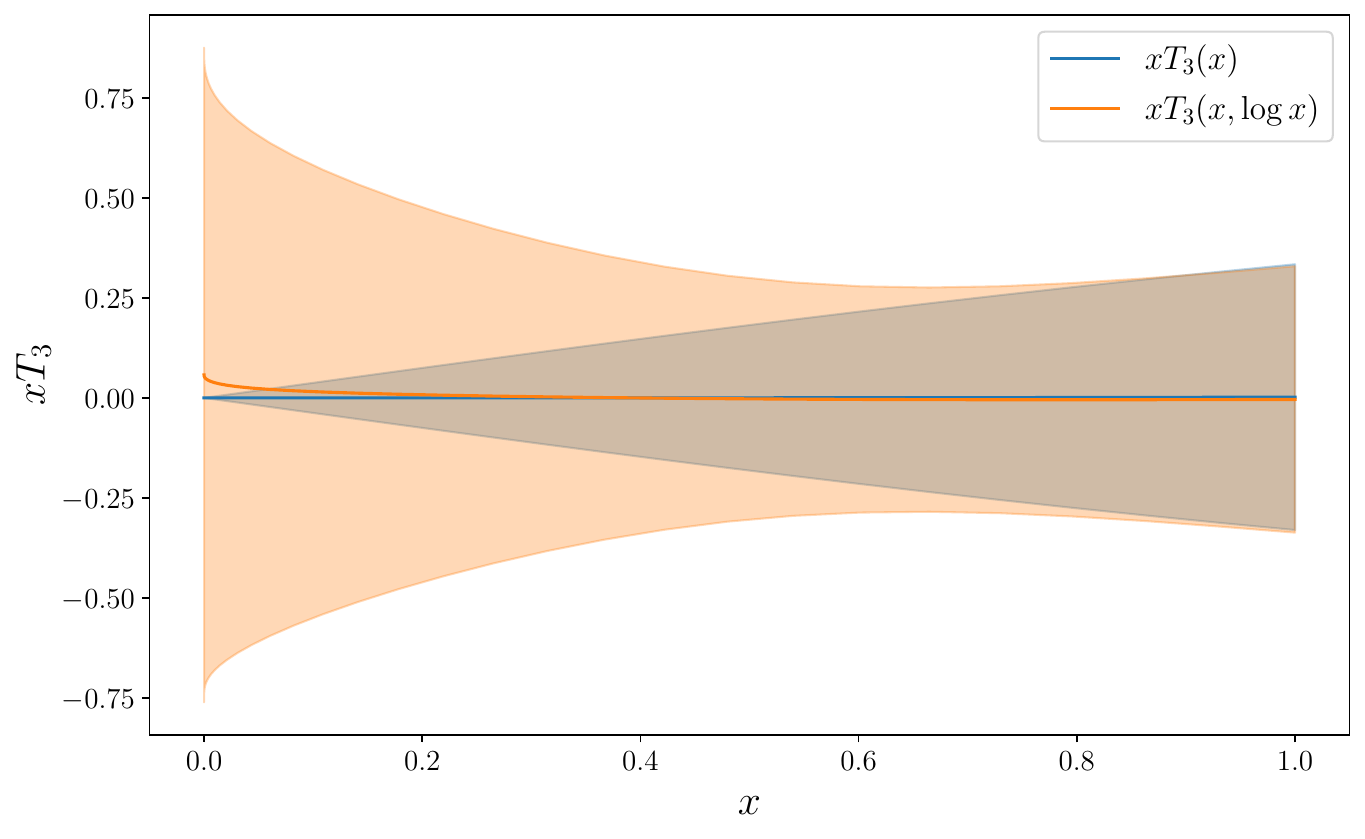}
    \includegraphics[width=0.45\textwidth]{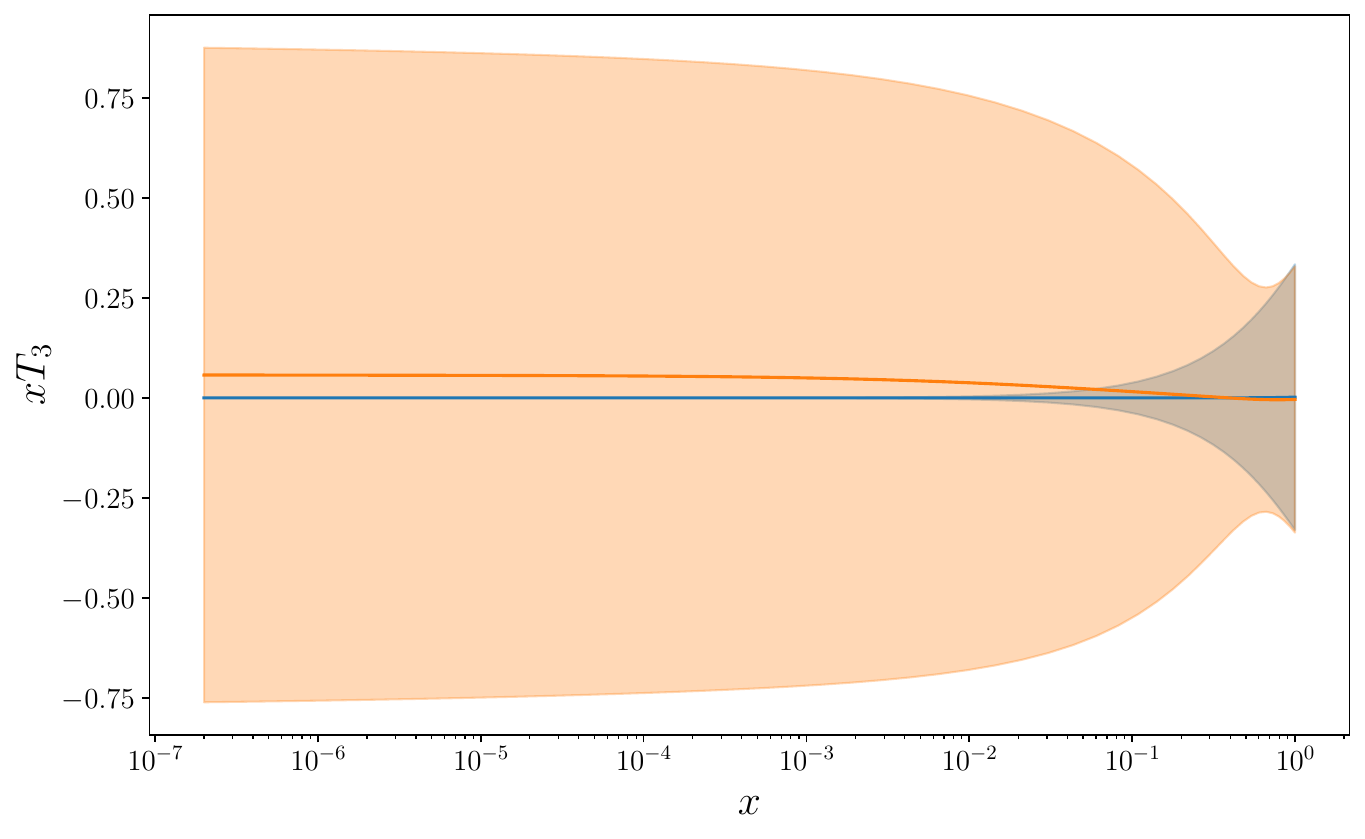}
    \caption{The output of the ensemble of neural networks at initialization
    using the NNPDF architecture in linear (left) and logarithm (right) scale.
    We compare the case of linear input $f(x)$ (blue) and the case of
    scaled input $f(x, \log x)$ (orange). The solid lines represent
    the mean value computed over an ensemble of 100 replicas, while the shaded
    bands represent the one-sigma uncertainty computed as the variance over the
    same ensemble. In the figure, we show $xT_3$ as used in the following
    sections.}
    \label{fig:prior}
\end{figure}

As a consequence of the symmetry of the probability distribution, the mean value
of the fields at initialization needs to vanish, while their variance at each
point $x_\alpha$ is given by the diagonal matrix elements of $K^{(\ell)}$. In
Fig.~\ref{fig:OutputDist}, the expected distribution is compared against the
empirical distribution of output fields for a selected value of $x$, using two
ensembles of replicas with $\nreps=100$ and $\nreps=1000$, respectively.
Inspecting the figures, we conclude that the recursion formula,
Eq.~\eqref{eq:RecursionForK}, accurately reproduces the output distribution of
the NNPDF networks at initialization, provided that a sufficiently large
ensemble of replicas is used to sample the distribution. Finally,
Fig.~\ref{fig:prior} shows the mean and variance of the output at initialization
across all values of $x$ for an ensemble of $\nreps=100$ neural networks
generated using the NNPDF architecture. We compare two cases: linear input
$f(x)$ and scaled input $f(x, \log x)$. 
The central value is computed as the average over replicas, 
\begin{align}
    \label{eq:MeanValAtInit}
    \bar{f}_{i\alpha} = \bar{f}_{i}(x_\alpha) 
    = \frac{1}{\nreps} \sum_{k=1}^{\nreps} f^{(k)}_i(x_\alpha)\, ,
\end{align}
and, likewise, the variance $\sigma^2_{i\alpha}$ is computed over the 
same ensemble of replicas, 
\begin{align}
    \label{eq:VarAtInitEmp}
    \sigma^2_{i\alpha} = \frac{1}{\nreps-1} \sum_{k=1}^{\nreps} 
    \left(f^{(k)}_i(x_\alpha) - \bar{f}_{i}(x_\alpha)\right)^2\, ,  
\end{align}
As is clear from the figure, the choice of input scaling has a significant impact
on the prior uncertainty, especially in the small-$x$ region. In the following,
we neglect this effect and focus on the case of linear input $f(x)$.

\FloatBarrier

\clearpage

\section{Training Neural Networks}
\label{sec:training}

We have summarized some properties of the neural network at initialization,
showing that the normal Glorot initialization leads to a prior distribution
that is a Gaussian processes with small corrections. In the final 
part of these lectures, we analyse the training dynamics of the neural network, 
with the aim of understanding how the process of training yields the 
posterior distribution of the solution to the inverse problem. 
We follow very closely the presentation of Ref.~\cite{Chiefa:2025cap}.

In the context of machine learning, specifically when dealing with
neural networks, optimization is an iterative algorithm that updates the
parameters of the network in order to minimize a figure of merit defined
appropriately. The direction towards the minimum is defined by the gradient
of the figure of merit. Due to the large number of parameters that characterize a neural
network, and the complex functional form induced by the recursive definition of the network, 
the figure of merit (also known as \textit{error function},
\textit{loss function}, or simply \textit{loss}) is a non-convex
high-dimensional function of the parameters, 
leading to numerical challenges in the minimization task. In
addition, in order to avoid \textit{over-} and \textit{under-learning}, 
training algorithms require a \textit{stopping criterion},
which specifies the optimal condition to end the training process.
While these algorithms have achieved remarkable empirical
success, a theoretical understanding of the optimization process remains
elusive. In these lectures we want to focus on understanding the 
dynamics, without being bogged down by unnecessary technical details; 
therefore,we work with the simplest gradient method, \ie, Gradient Descent (GD). 

We work out in detail one specific example, in the context of PDF 
determinations, where we
consider a reduced dataset for which predictions can be computed using 
just one flavor combination of the PDFs, so that the 
theoretical prediction for the data remains simply
\begin{equation}
    \label{eq:TheoryPredSingleFlav}
    T_I[f] = \int dx\, C_{I}(x) f(x)\,.
\end{equation}
The details of the dataset and the definition of the flavour combination
are not relevant for these lectures, the interested reader can find more
details in Ref.~\cite{Chiefa:2025cap}. 

\subsection{Training in Functional Space}
\label{sec:GradFlow}

For analytical tractability, GD is described as a continuous flow of the
parameters $\theta$ in training time $t$ along the negative gradient of the loss
function $\mathcal{L}$. For sufficiently small learning rates $\eta$, this
continuous flow approximates the discrete GD trajectory in parameter space, as
extensively discussed in Ref.~\cite{barrett2022igr}. The continuous Gradient
Flow (GF) is then given by
\begin{align}
    \label{eq:GradientFlowDef}
    \ddt &\theta_{t,\mu} = -\nabla_\mu \mathcal{L}_t\, ,
\end{align}
where $\theta_{t,\mu}$ and $\mathcal{L}_t$ identify respectively the parameters
and the loss function at training time $t$. We distinguish between the
continuous training time $t$ and the discrete epochs of GD, the latter denoted
using the capital letter $T$. For the purpose of this lecture, the two are related 
through a constant learning rate, $t = \eta T$ with $\eta = 10^{-5}$, 
and we will use them interchangeably.

We focus here on quadratic loss functions that are obtained as the negative
logarithm of Gaussian data distributions around their theoretical predictions;
using the notation introduced in Sect.~\ref{sec:prelim}, the loss function is given by
\begin{align}
    \label{eq:QuadLoss}
    \mathcal{L}_t = \frac12 \left(Y - T[f_t]\right)^T C_Y^{-1} \left(Y - T[f_t]\right)\, ,
\end{align}
where $f_t$ is the output of the network at training time $t$, obtained
from the time-dependence of the internal parameters, and $C_Y$ is the
covariance of the data. Note that the loss function
at training time $t$ is computed using the theoretical prediction $T[f_t]$, 
computed using a grid of points for $f_t$ at training
time $t$. For a quadratic loss, the gradient is
\begin{align}
    \nabla_\mu \mathcal{L}_t = - \left(\nabla_\mu f_t\right)^T \left(\frac{\partial T}{\partial f}\right)_t
      C_Y^{-1} \epsilon_t\, ,
\end{align}
where, writing explicitly the data index,
\begin{align}
    \label{eq:EpsDef}
    \epsilon_{t,I} = Y_I - T_I[f_t]\, , \quad I=1, \ldots, \ndat\, .
\end{align}
For the specific case of a quadratic loss function, the gradient is proportional
to $\epsilon_t$, which is the difference between the theoretical prediction and
the data at training time $t$. If at some point during the training the
theoretical predictions reproduce all the data, the training process ends. 

A further simplification is obtained in the case of data that depend linearly on
the unknown function $f$. In this case, the Jacobian of the theoretical 
prediction with respect to the fields is independent of the training time, 
\begin{align}
    \label{eq:dTbydf}
    \left(\frac{\partial T_I}{\partial f_{i\alpha}}\right)_t =
        \FKtab_{Ii\alpha}\, .
\end{align}
A few algebraic steps allow the flow of parameters $\theta$ to be translated 
into a flow for the fields,
\begin{align}
    \label{eq:NTKFlow}
    \ddt &f_{t,i_1\alpha_1} = (\nabla_\mu f_{t,i_1\alpha_1}) \ddt \theta_\mu =
      \Theta_{t,i_1\alpha_1i_2\alpha_2}
      \FKtabT_{i_2\alpha_2I} \left(C_Y^{-1}\right)_{IJ} \epsilon_{t,J}\, ,
\end{align}
where we have defined the Neural Tangent Kernel~\cite{jacot2018neural}
\begin{align}
    \label{eq:NTKDef}
    \Theta_{t,i_1\alpha_1i_2\alpha_2} = \sum_\mu
    \nabla_\mu f_{t,i_1\alpha_1} \nabla_\mu f_{t,i_2\alpha_2}\, .
\end{align}
If we introduce the matrix
\begin{equation}
  \label{eq:JacobMat}
  \left(J_t\right)_{i\alpha,\mu} = \nabla_\mu f_{t,i\alpha}\, ,
\end{equation}
the NTK can be written as $\Theta_t = J_t J_t^T$. 
The NTK is a positive semidefinite matrix that encodes the dependence of the 
training dynamics on the architecture of the network. The flow equation, 
Eq.~\eqref{eq:NTKFlow}, is a non-linear integro-differential equation that describes 
the evolution of the fields during training.

In order to facilitate the discussion in Sec.~\ref{sec:Lazy},
Eq.~\eqref{eq:NTKFlow} can be rewritten in a more compact form. We first omit
the indices and write, for instance,
\begin{align}
  \left(\frac{\partial T}{\partial f}\right)_t = \FKtab\, , \quad
  \Theta_t = \left(\nabla_\mu f_t\right) \left(\nabla_\mu f_t\right)^T\, .
  \label{eq:dTdfForLinearObs}
\end{align}
Then, using the definition of the error in Eq.~\eqref{eq:EpsDef}, we can rewrite
Eq.~\eqref{eq:NTKFlow} as 
\begin{align}
    \label{eq:FlowEquationNoIndices}
    \ddt f_t = -\Theta_t M f_t + b_t\, ,
\end{align}
where
\begin{align}
    M &= \FKtabT C_Y^{-1} \FKtab\, , \quad b_t = \Theta_t \FKtabT C_Y^{-1} Y\, .
    \label{eq:MandBDef}
\end{align}

\medskip
\noindent
\fbox{
    \begin{minipage}{0.9\textwidth}
        \begin{exercise}
            Derive the flow equation, Eq.~\eqref{eq:FlowEquationNoIndices}, 
            starting from the continuous Gradient Flow equation.
        \end{exercise}
    \end{minipage}
}
\medskip

\noindent 
Here $M$ is a positive-semidefinite matrix that depends only on the data covariance 
and the FK tables that enter the theoretical predictions, 
while $b$ is a vector that depends (amongst other quantities) on the central value 
of the data. Note that any vector $f$ that is in the kernel of $\FKtab$ is necessarily
in the kernel of $M$, $\ker M$. In turn, the vectors in $\ker M$ do not contribute 
to the flow evolution, as seen explicitly in Eq.~\eqref{eq:FlowEquationNoIndices}.

There are two important points that we want to emphasize. 
First, although derived in the context of neural networks, these equations do
not refer to a specific parametrization. Indeed, these remain valid even when
an explicit functional form is chosen to parametrize the unknown function $f$. 
Second, note that the flow equation,
Eq.~\eqref{eq:FlowEquationNoIndices}, depends on two matrices, $\Theta$ and $M$.
The former encodes the model dependence, while the latter contains the physical
information. The interplay between these two matrices is crucial for
understanding the training dynamics, as discussed in
Sec.~\ref{sec:NTKAlign}. Finally, the NTK derived in Eq.~\eqref{eq:NTKDef} is
inherently time-dependent in a complex way, which precludes any attempt at
integrating Eq.~\eqref{eq:FlowEquationNoIndices} analytically. We come back
to this point in Sec.~\ref{sec:Lazy}, after discussing the properties of the
NTK during training.

\FloatBarrier

\subsection{Inside the Training Dynamics: an NTK perspective}

From Eqs.~\eqref{eq:NTKDef} and~\eqref{eq:FlowEquationNoIndices}, we observe
that the NTK encodes the dependence on the architecture of the network and
governs its training dynamics. The analysis of the NTK properties is one way to 
unravel the behaviour of the network during training. 

\subsubsection{NTK at initialisation}
\label{sec:NTKAtInit}

Before training, the NTK is blind to data and depends on the $x$-grid of input
and on the architecture, as shown in Eq.~\eqref{eq:NTKDef}. The NTK
is a function of the fields $f$, which are stochastic variables described by
their joint probability distribution.
Therefore the NTK is also a stochastic variable, with its own probability
distribution, which we represent as usual as a set of replicas. 

\begin{figure}[ht!]
  \centering
  \includegraphics[width=0.90\textwidth]{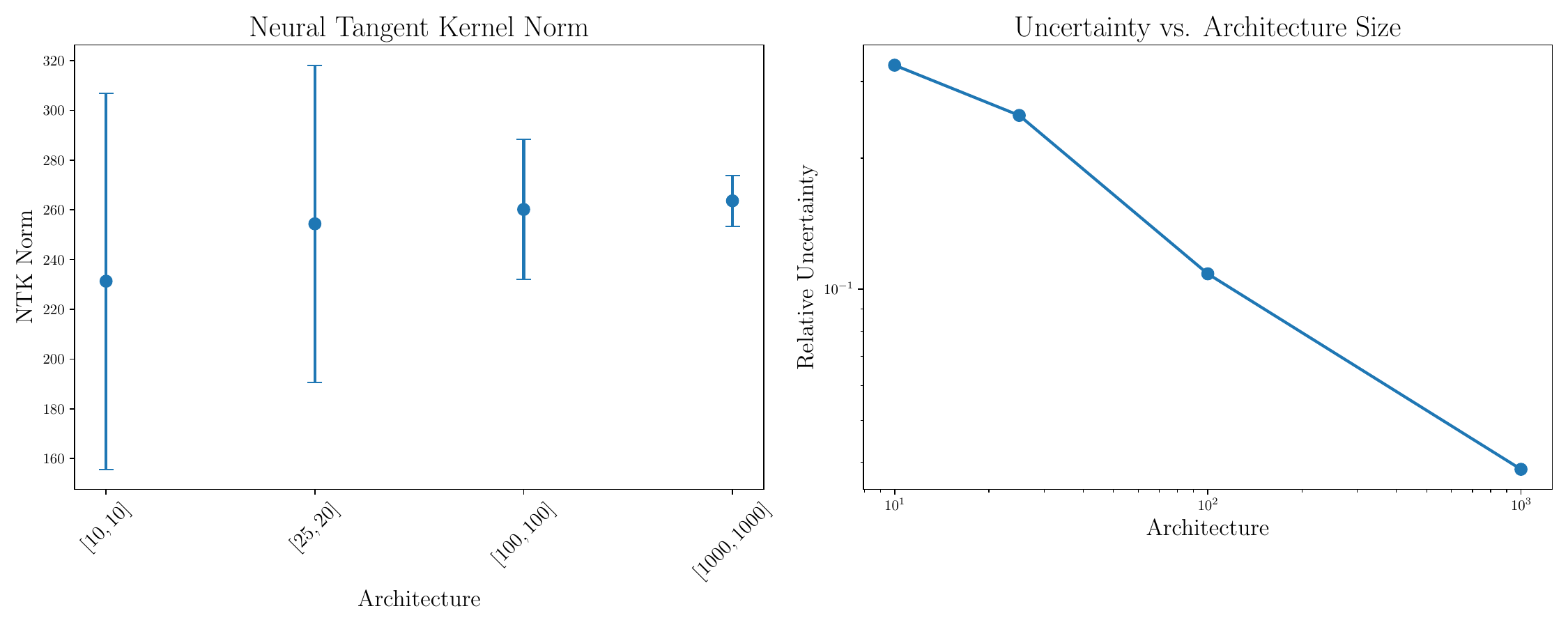}
  \caption{Frobenius norm of the NTK at initialization, $\lVert \Theta_0
  \rVert$, as a function of the width of the network. On the left, the central
  values and uncertainty bands are obtained as the mean and one-sigma deviation
  of the ensemble of networks. The labels on the horizontal axis indicate the respective 
  widths of the two hidden layers in the networks considered.The plot on the right shows the relative
  uncertainty. It is interesting to note the decrease of the relative uncertainty as the 
  architecture of the network is increased. For larger networks, the sensitivity to a change
  of the network parameters fluctuates less.}
  \label{fig:NTKInit}
\end{figure}

In the large-width limit, the variance of
the NTK over the set of replicas is expected to go to zero with the width of the hidden
layers. In order to quantify the
variation of the NTK, we start by computing the Frobenius norm of the NTK over
an ensemble of networks for different architectures. For each architecture, we
consider the standard deviation of the norm as a statistical
estimator of the variations of the NTK. The result is displayed in
Fig.~\ref{fig:NTKInit}. Even though the Frobenius norm is a coarse indicator of
the variations of the NTK, the figure shows clearly that the variance of the
norm becomes smaller with the size of the network, which is consistent with the
theoretical expectation that the NTK should not fluctuate for infinite-width
networks.\footnote{
  Note that, in addition to the scaling $\mathcal{O}(1/n)$
  theoretically predicted for large networks, the uncertainty bands include
  bootstrap errors due to the finite size of the ensemble. Using an ensemble of
  100 replicas, the bootstrap error on the standard deviation is $\sim 10\%$.
}

\begin{figure}[t]
  \centering
  \includegraphics[width=0.7\textwidth]{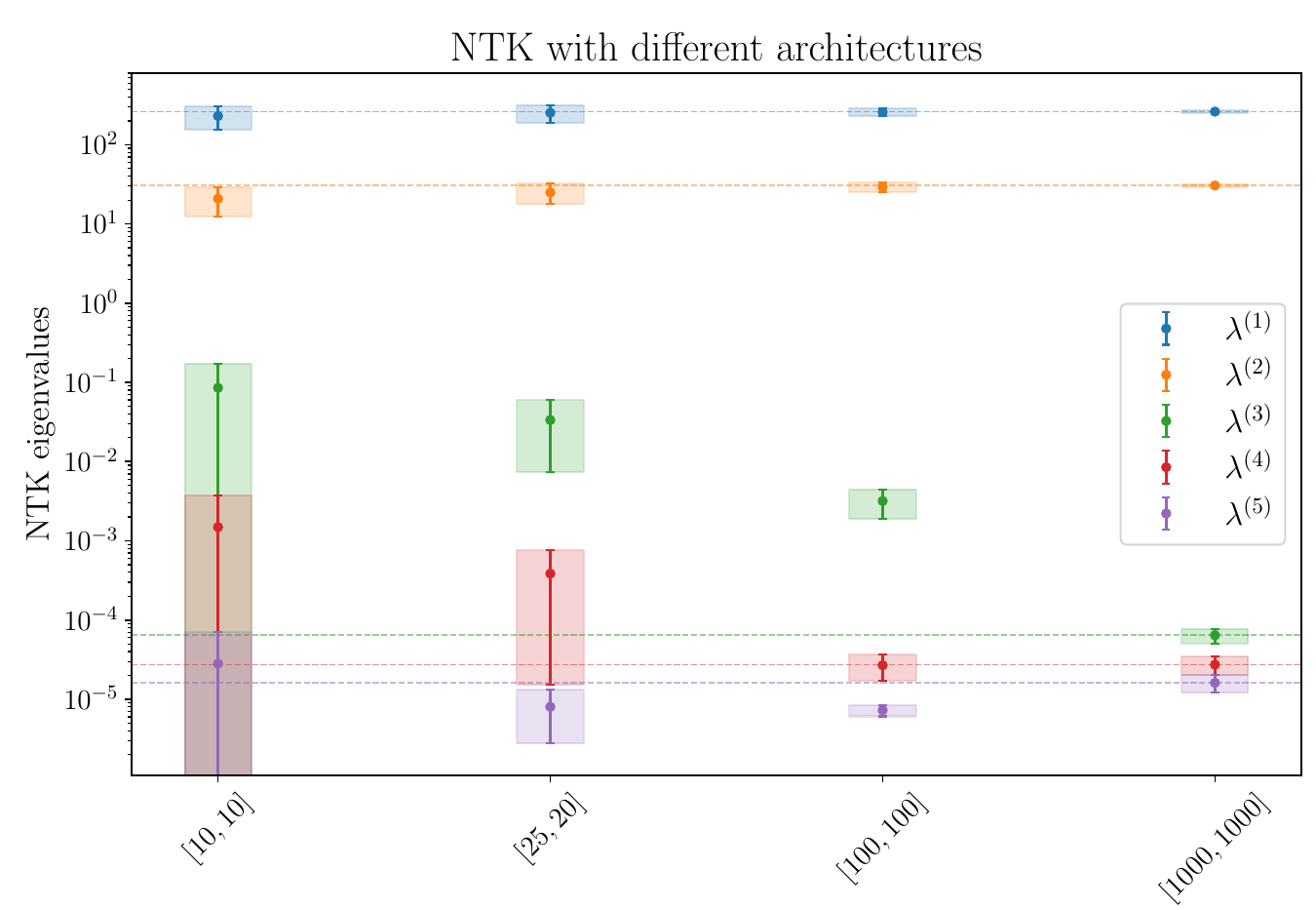}
  \caption{Spectrum of the NTK at initialization for the architectures shown in
  Fig.~\ref{fig:NTKInit}. The labels on the horizontal axis indicate the respective 
  widths of the two hidden layers in the networks considered. 
  Error bands correspond to one-sigma uncertainties over
  the ensemble of networks. The hierarchy of the eigenvalues is independent of the size
  of the network. In agreement with the data in Fig.~\ref{fig:NTKInit}, the fluctuations 
  of the eigenvalues decrease as the width of the layers is increased. 
  }
  \label{fig:NTKSpectrum}
\end{figure}

A more quantitative description of the NTK at initialization is provided by its
spectrum, which is shown in Fig.~\ref{fig:NTKSpectrum} for four different
architectures. Inspecting the figure, we see that the spectrum of the NTK is
heavily hierarchical, and only few eigenvalues are actually
non-zero. Such a hierarchy 
in the eigenvalues means that
only a small subset of active directions can inform the network during training,
as it will be discussed later. Note that, at least at initialization, these
observations do not depend on the architecture: the eigenvalues in
Fig.~\ref{fig:NTKSpectrum} are mostly independent of the size of the network.
Even though the logarithmic scale on the vertical axis may hide some small variations, it 
is clear that most eigenvalues remain constant within the error bars. On the other hand, 
the logarithmic scale emphasizes that there are several orders of magnitude between eigenvalues 
for a given architecture; that hierarchical structure does not depend on the architecture. 

\FloatBarrier

\subsubsection{NTK During Training}
\label{sec:NTKDuringTraining}

The behaviour of the NTK during training is best studied through {\em closure tests}, 
which we introduced in previous sections. The results of the training are then compared to
the known input PDFs; the performance of the training algorithm and the NN
architecture are assessed by quantifying the comparison between trained PDFs and
input PDFs. Following the original presentation in Ref.~\cite{NNPDF:2014otw}, we
distinguish three levels of closure tests, which are defined by the complexity
of the data used to train the NNs. We use the standard NNPDF nomenclature and
refer to these three levels as level-0 (L0), level-1 (L1), and level-2 (L2)
closure tests, and we denote the input PDFs used to generate the data as $f_0$.
We refer the reader to the original publication for the details of the data 
generation at each level. 

For each of the closure-test data given above, we perform a fit of the
triplet combination $T_3$ using the simplified version of the NNPDF methodology that we 
discussed above. We initialize an ensemble of $\nreps = 100$ replicas with identical 
architecture, training each replica independently using GD optimization. Throughout the
training process, we track the evolution of the NTK to understand how the
network's effective dynamics changes as it learns the target function.

\paragraph{Onset of Lazy Training} 

\begin{figure}[t]
  \centering
  \includegraphics[width=0.60\textwidth]{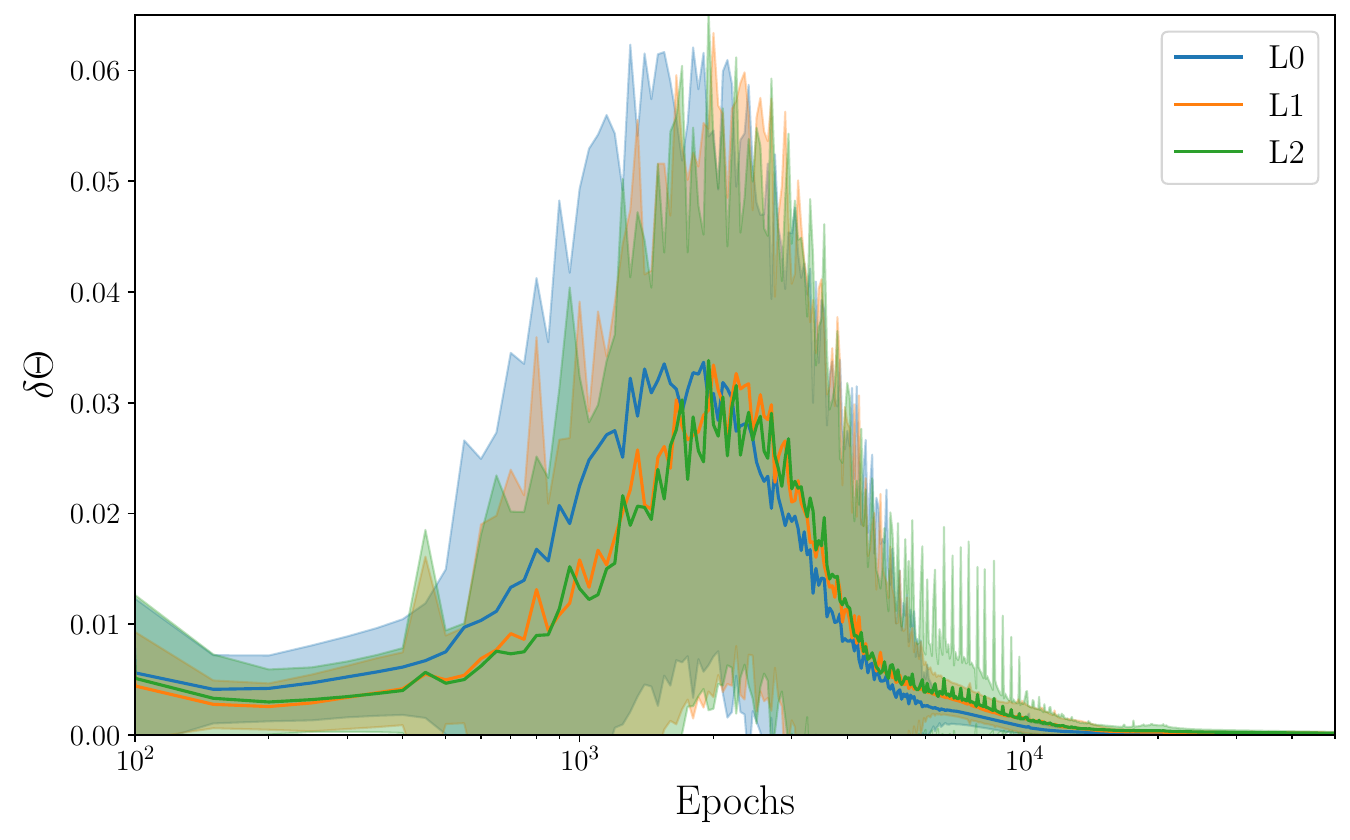}
  \caption{Relative variation of the NTK during training for L0, L1, and
  L2 data. Error bands correspond to one-sigma uncertainties over the ensemble
  of networks.}
  \label{fig:NTKTime}
\end{figure}

As a first estimator of the variation of the NTK, we show in
Fig.~\ref{fig:NTKTime} the Frobenius norm of the variation during training,
normalized by the Frobenius norm of the NTK itself, 
\begin{equation}
\delta \Theta_t = \frac{\lVert \Theta_{t+1} - \Theta_t \rVert}{\lVert \Theta_t \rVert} \;,
\label{eq:DeltaNTK}
\end{equation}
for the three different datasets, L0, L1, and L2. Inspecting the plot reveals that
the NTK undergoes significant changes during the initial phase of training, with
the relative variation $\delta \Theta_t$ reaching values as high as $6\%$. This
indicates that our settings differ from the standard picture of lazy training in
the context of very wide networks, as discussed, \eg, in
Refs.~\cite{jacot2018neural,Roberts_2022,lee2019wide}, where the NTK is expected to 
be independent of the flow time $t$. Remarkably, we do not
observe a dependence on how data have been generated, indicating that the NTK
dynamics is basically unaffected by the noise in the data. 

After this initial phase -- corresponding approximately to the first 20,000
epochs in our experiment -- the NTK tends to stabilize. These two regions will
be referred to as the \textit{rich} and \textit{lazy} training regimes,
respectively, in keeping with the standard terminology adopted in the literature
(see, \eg, Ref.~\cite{fort2020dlvk} where two similar regimes were also
identified). The lazy regime is discussed in detail in Sec.~\ref{sec:Lazy}.

\FloatBarrier

\paragraph{Eigenvalues During Training}

\begin{figure}[ht]
  \centering
  \includegraphics[width=0.3\textwidth]{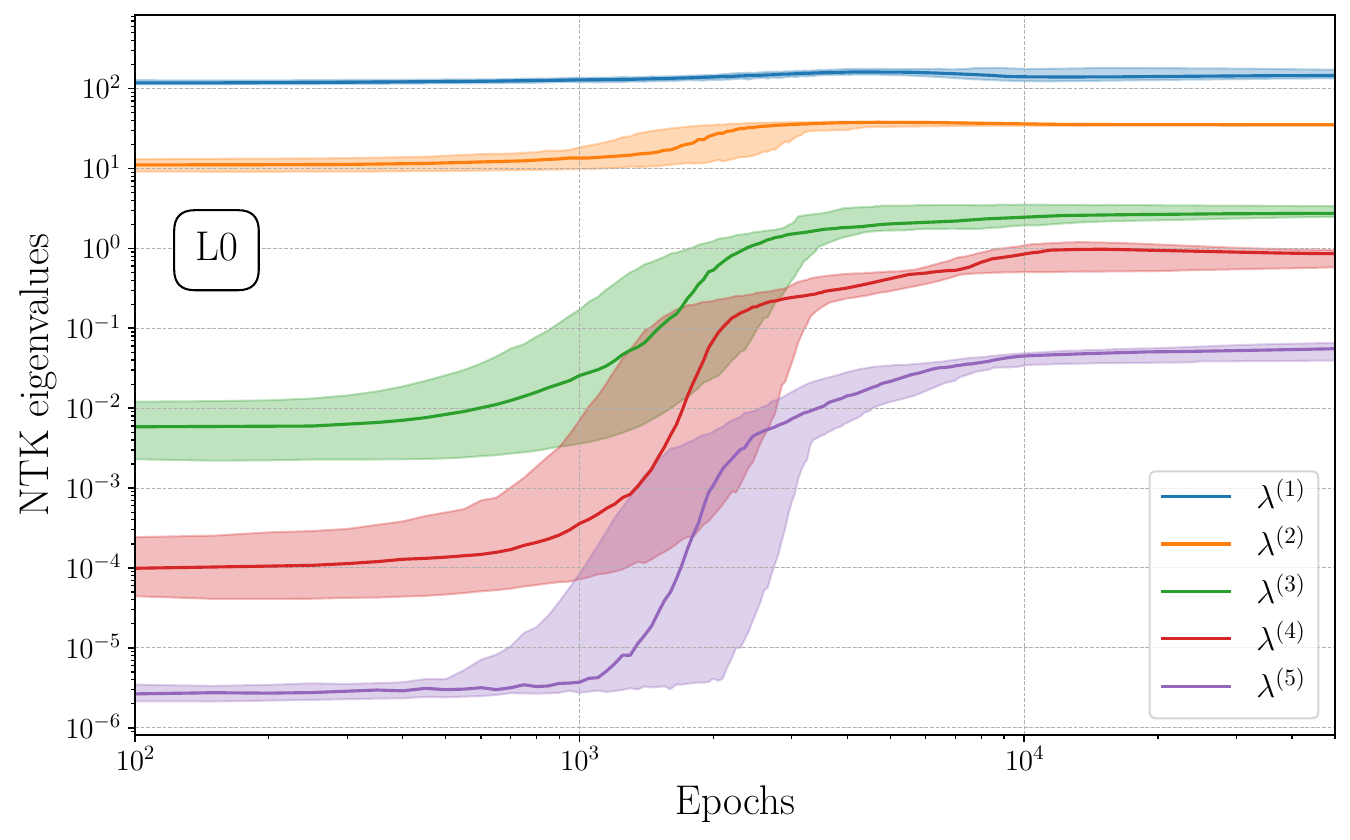}
  \includegraphics[width=0.3\textwidth]{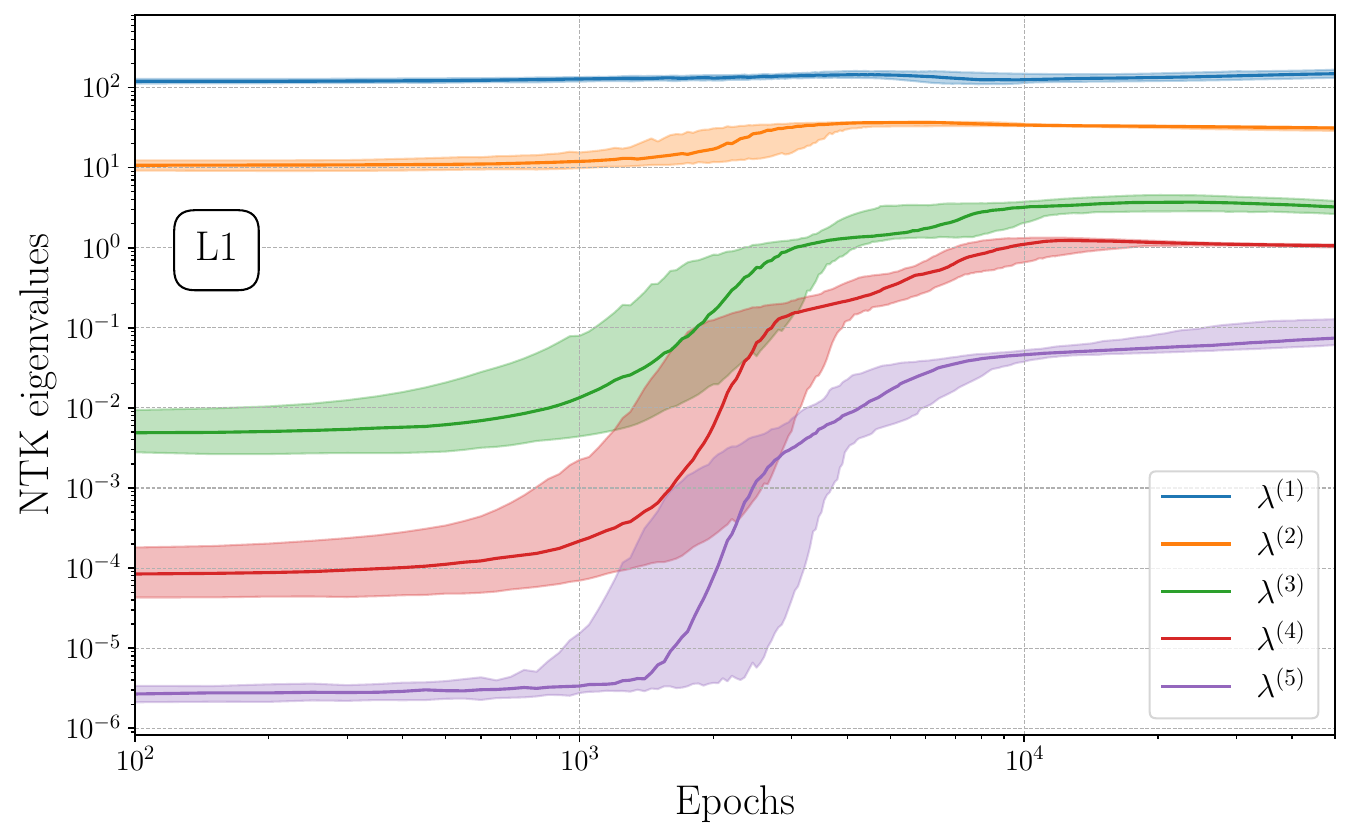}
  \includegraphics[width=0.3\textwidth]{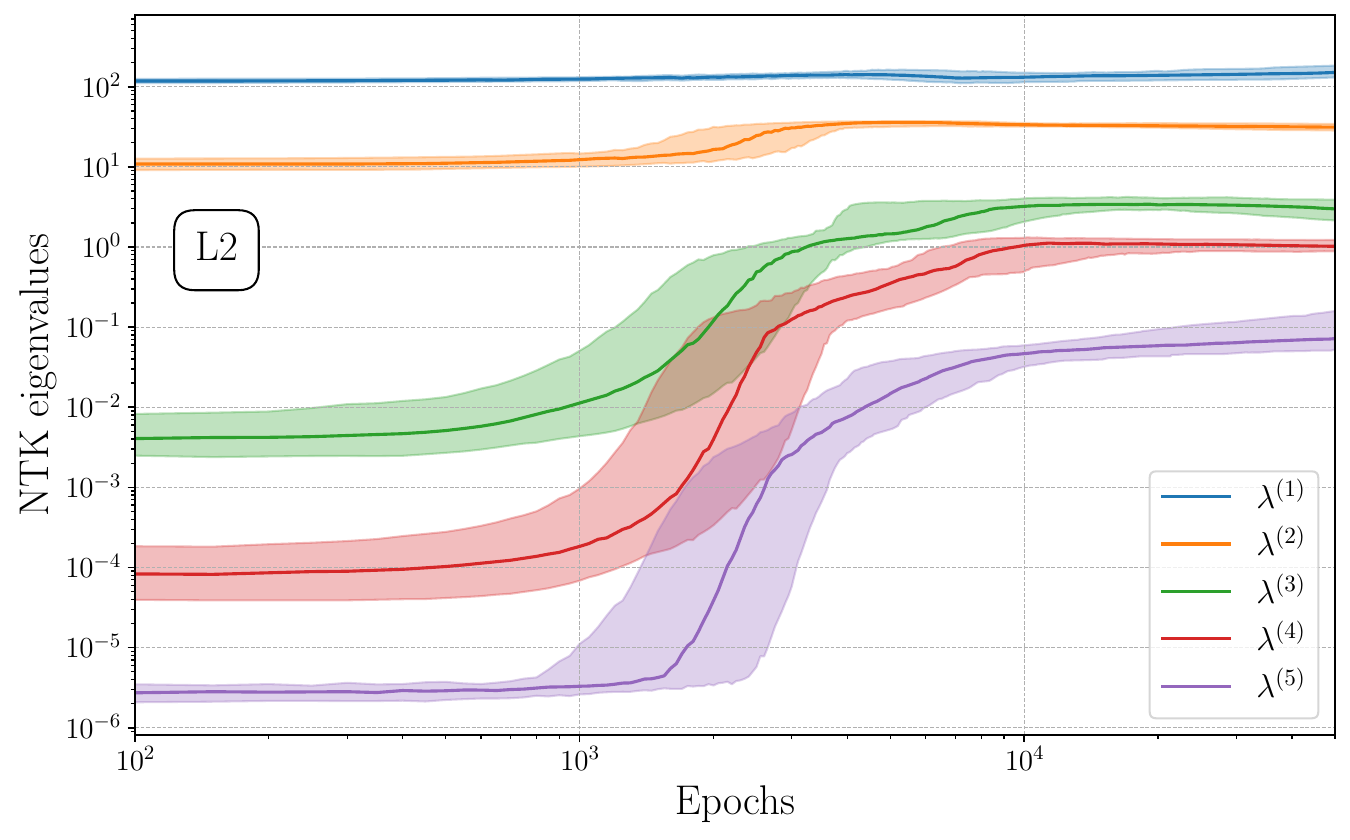} 
  \caption{Evolution during training of the first five eigenvalues of the NTK
  using L0 (left), L1 (center), and L2 (right) data. Solid lines represent the
  median over the ensemble of networks, while solid bands correspond to 68\%
  confidence level. Note that the subdominant eigenvalues $\lambda^{(3)}$, $\lambda^{(4)}$
  and $\lambda^{(5)}$ have increased by one or two orders of magnitude by the end of the rich 
  training phase.}
  \label{fig:NTKEigvalsTime}
\end{figure}

Further insight on the evolution of the NTK can be obtained by studying its
eigensystem as a function of the training time. In Fig.~\ref{fig:NTKEigvalsTime}
we report the variation of the first five eigenvalues of the NTK, using the
standard NNPDF architecture, for L0, L1, and L2 data. We see that the
hierarchical structure observed at initialization is preserved, but the size of
the subdominant eigenvalues increases significantly in the early stages of
training -- by one or two orders of magnitude depending on the specific
eigenvalue. 

\FloatBarrier

\begin{figure}[ht]
  \centering
  \includegraphics[width=0.30\textwidth]{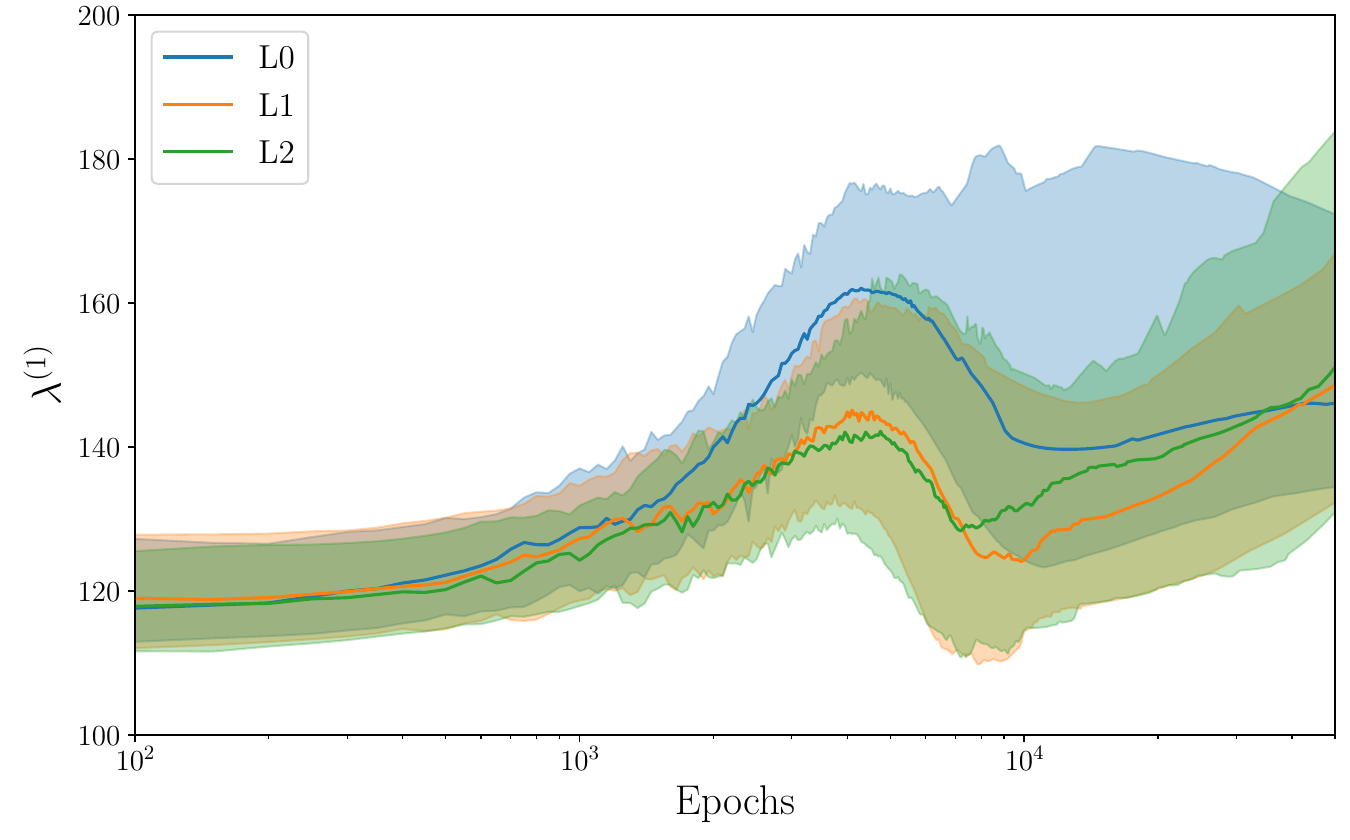}
  \includegraphics[width=0.30\textwidth]{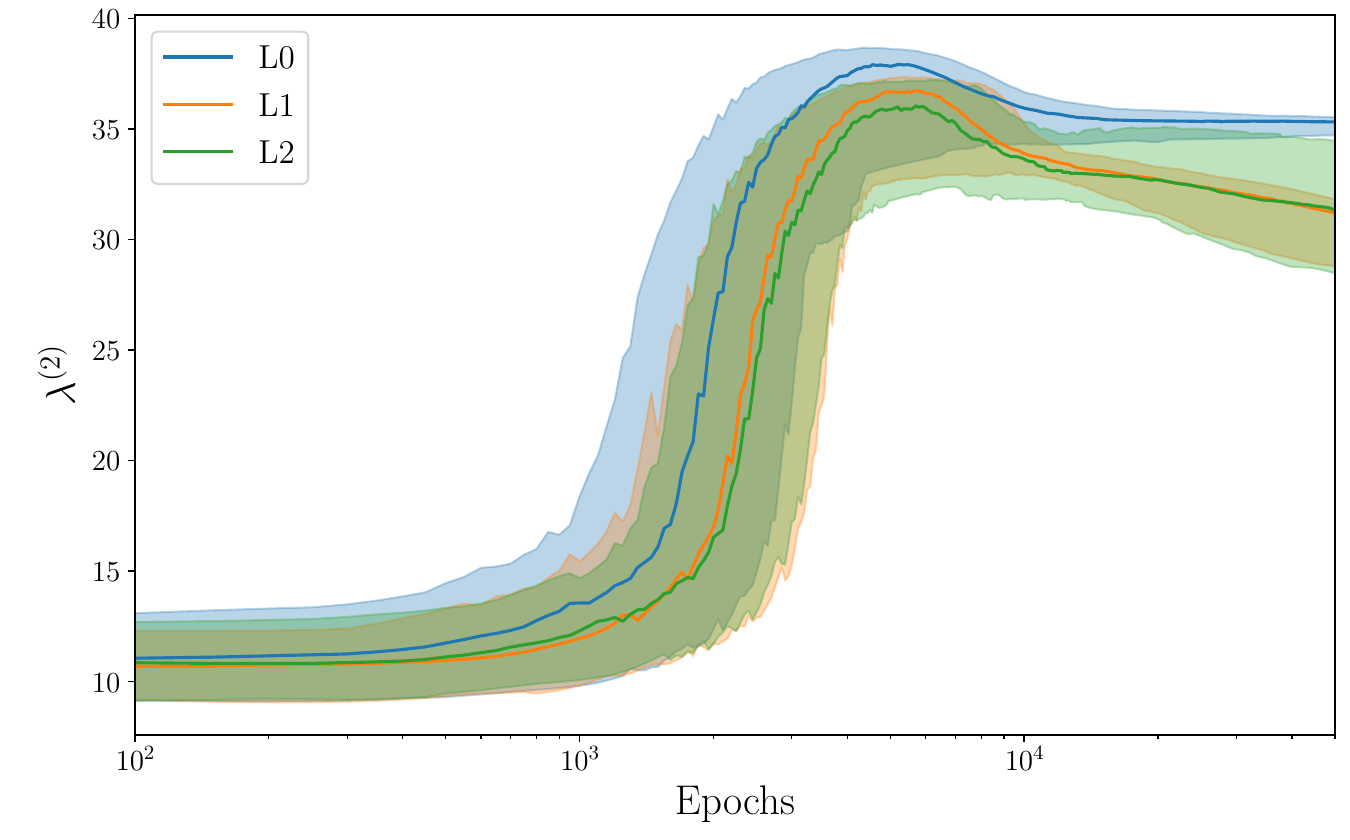}
  \includegraphics[width=0.30\textwidth]{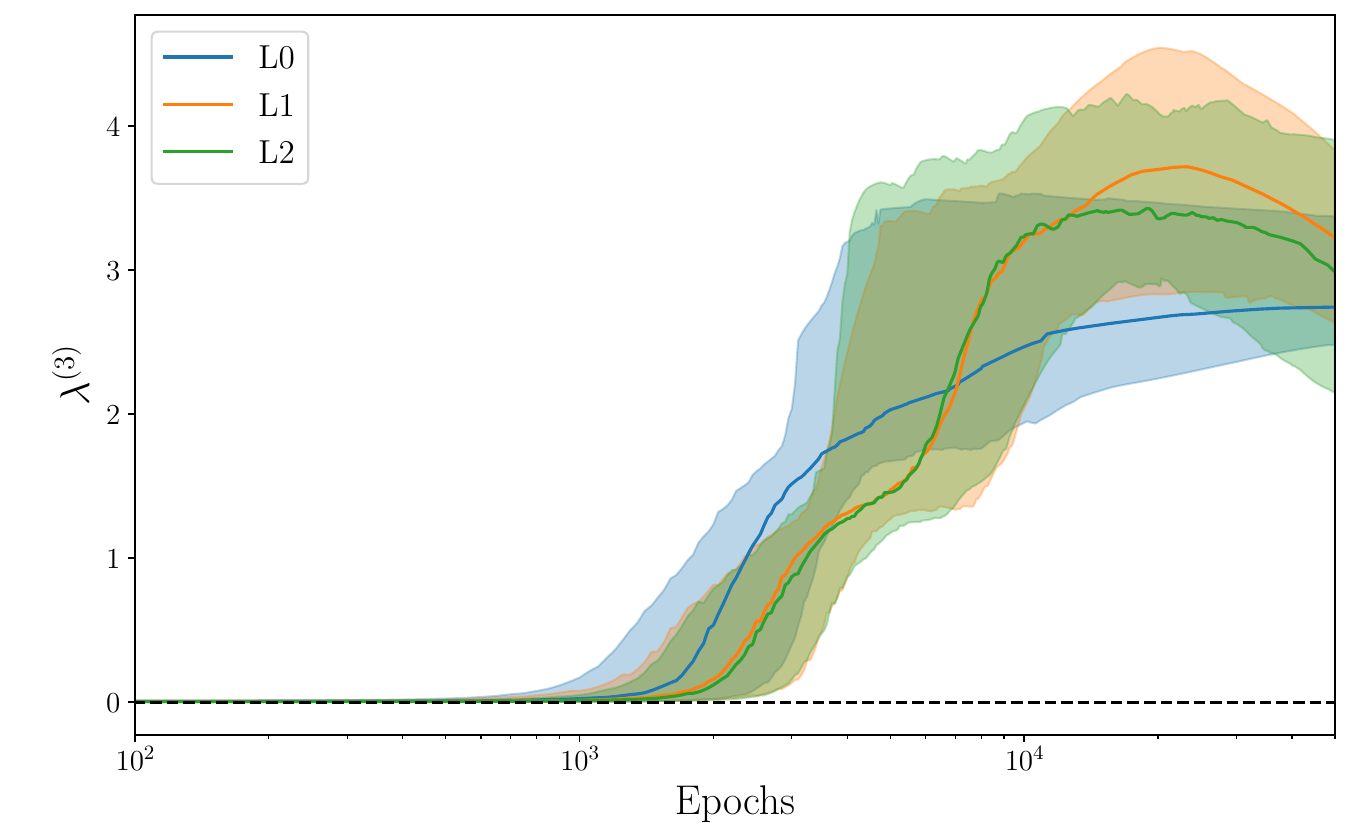}
  \includegraphics[width=0.30\textwidth]{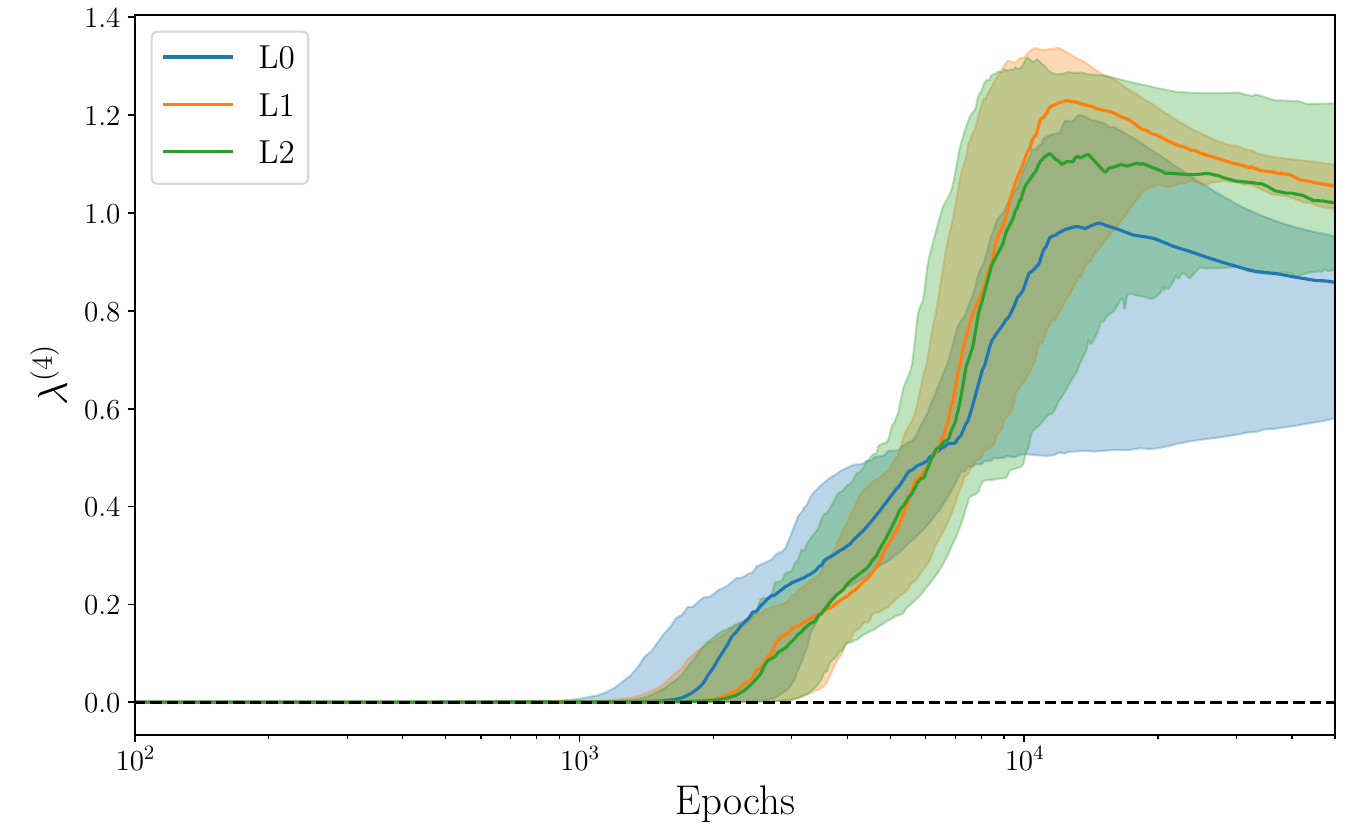}
  \includegraphics[width=0.30\textwidth]{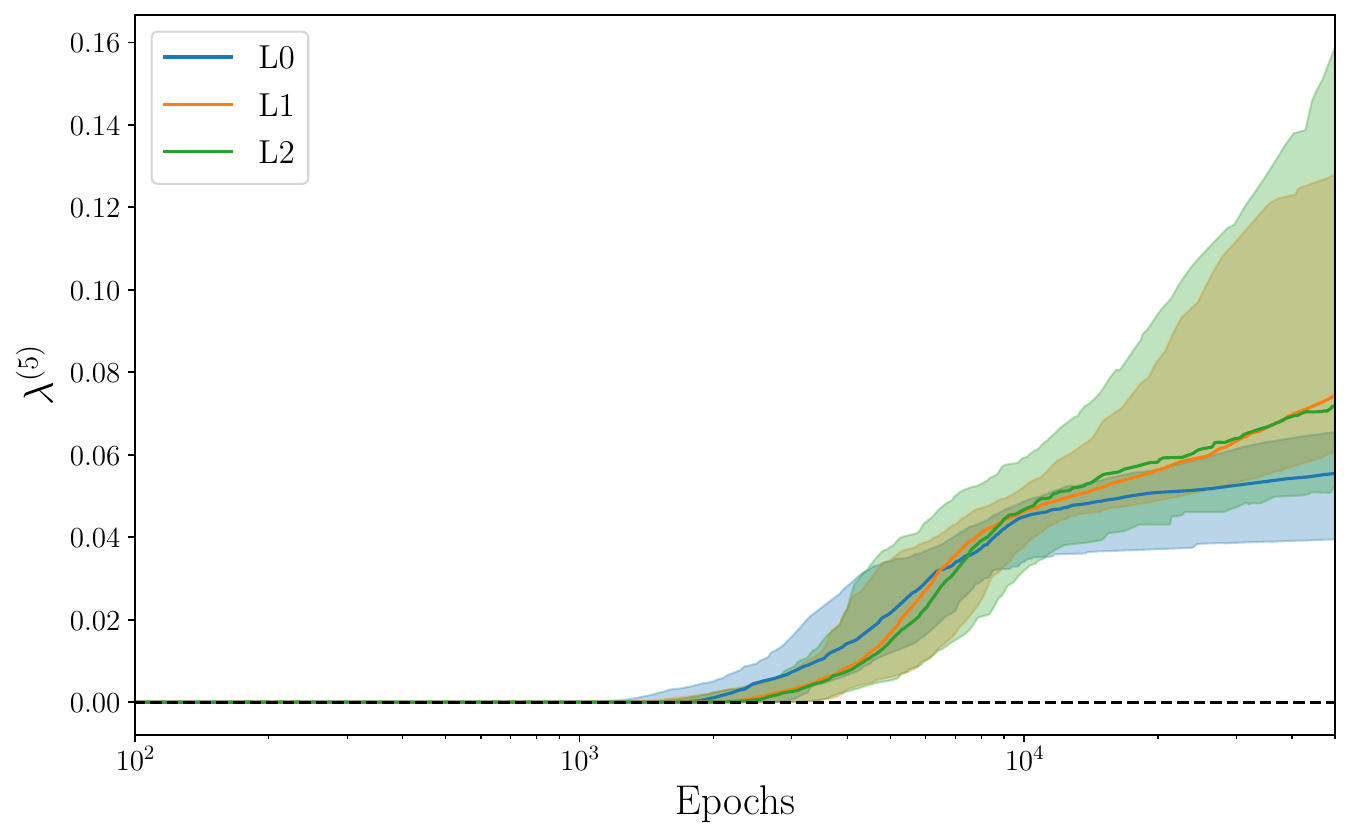}
  \caption{The first five eigenvalues of the NTK for L0, L1, and L2 data. Solid
  lines represent the median over the ensemble of networks, while solid bands
  correspond to 68\% confidence level. Each plot corresponds to a different eigenvalue, 
  as indicated by the label on the vertical axis. Note the different scales on the 
  vertical axes, which reflects the hierarchy of eigenvalues discussed above. Different
  colours correspond to different synthetic data, the agreement between these bands 
  confirms that the evolution of the eigensystem of the NTK does not depend on the 
  level of noise in the data.}
  \label{fig:EigvalsComparison}
\end{figure}

In Fig.~\ref{fig:EigvalsComparison}, the same first five eigenvalues of the NTK
are displayed for L0, L1, and L2 data. We observe a common pattern across all
data types, consistently with the observation made before in
Fig.~\ref{fig:NTKTime}. This indicates the NTK evolution is insensitive to
the noise included in the synthetic data. The increase of the subdominant
eigenvalues, combined with the analysis of Eqs.~\eqref{eq:FlowParallel}
and~\eqref{eq:FlowPerp} in Sect.~\ref{sec:LazyTraining}, suggests that 
more ``physical'' features become learnable before lazy training sets in.

\FloatBarrier

\begin{figure}[ht]
  \centering
  \includegraphics[width=0.30\textwidth]{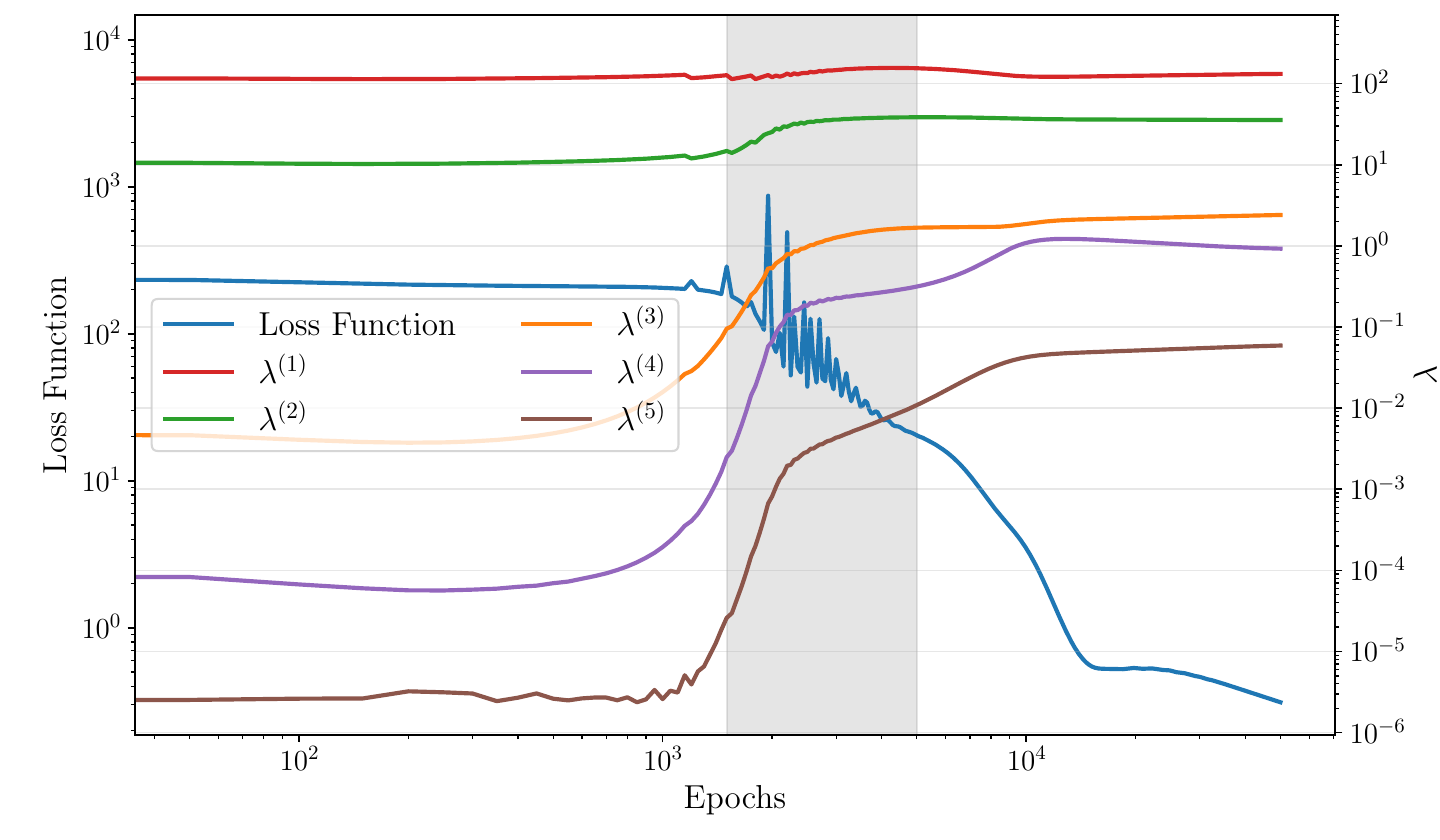}
  \includegraphics[width=0.30\textwidth]{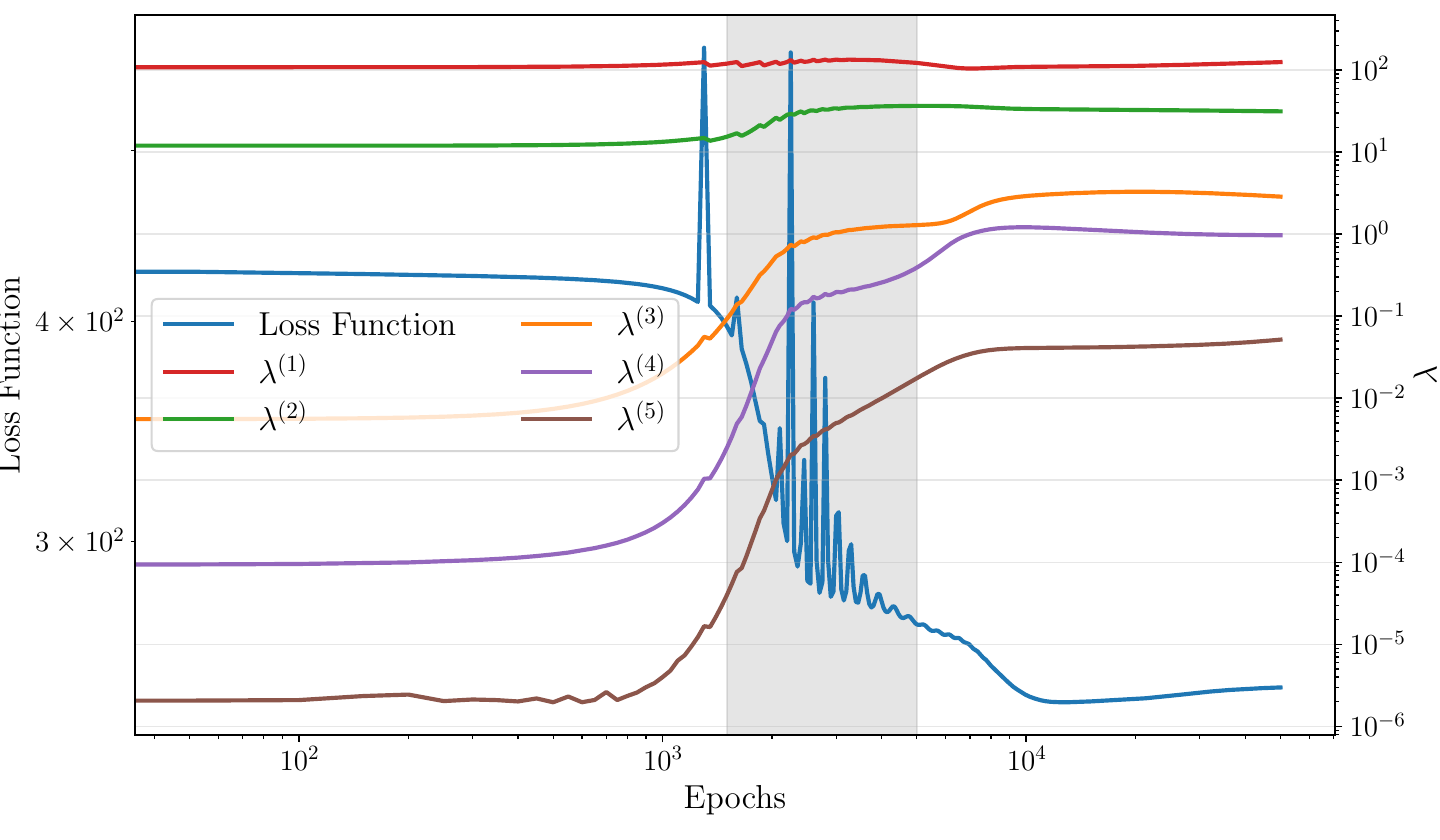}
  \includegraphics[width=0.30\textwidth]{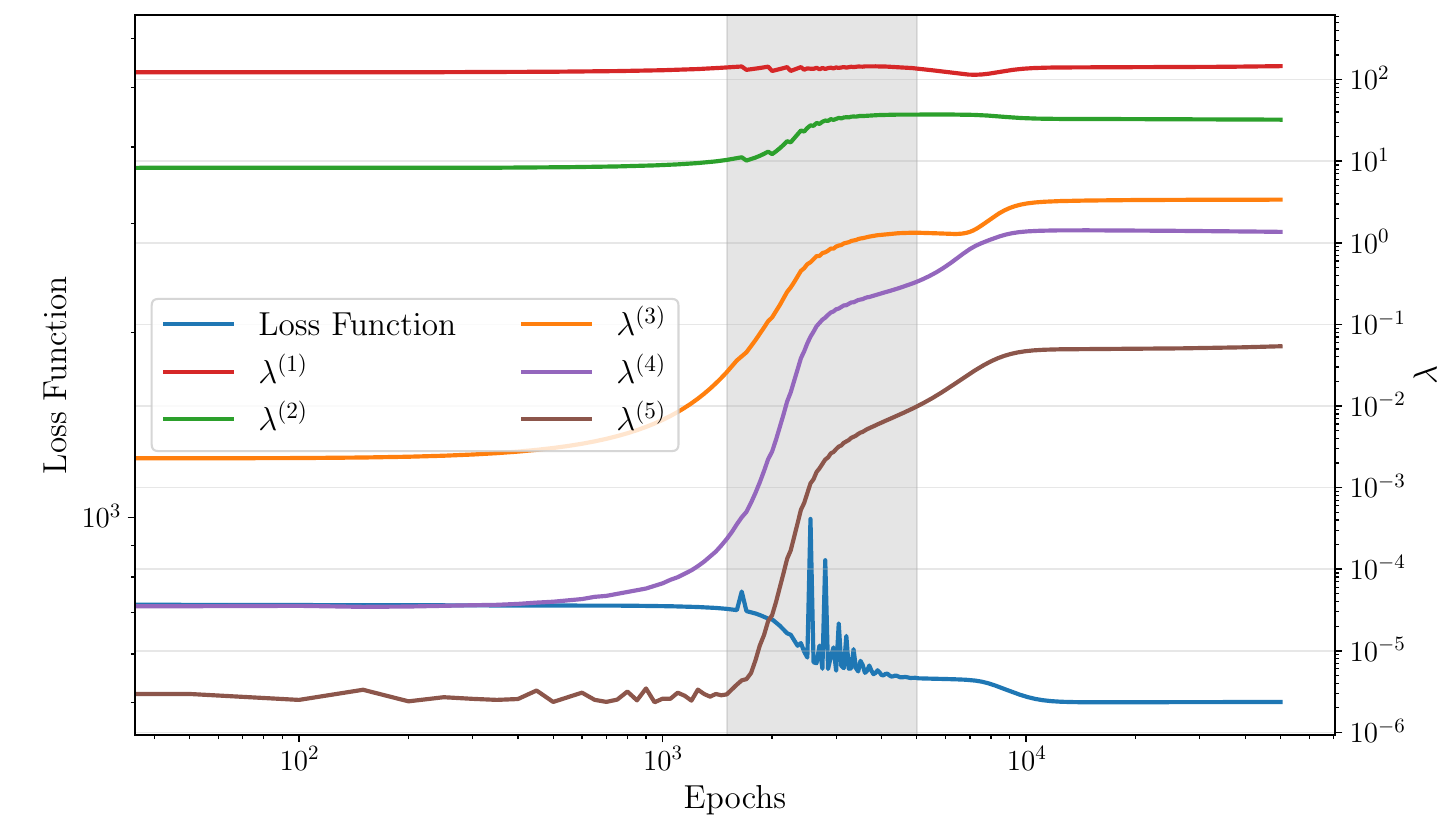}
  \caption{Variation of the loss function overlaid with the first five
  eigenvalues for a selected replica over the ensemble using L0 (left), L1
  (center), and L2 (right) data. Left scale refers to the loss, while the right
  scale refers to the eigenvalues.}
  \label{fig:Loss}
\end{figure}

\paragraph{Connection with the loss function} Finally, in Fig.~\ref{fig:Loss} we
show the variation of the loss function during training, overlaid with the first
five eigenvalues of the NTK, for a selected replica over the ensemble. It is
interesting to see that in correspondence with the sudden variation of the
subdominant eigenvalues, the loss function drops significantly, at the cost of
an instability localized in the descent. The NN is
learning new features, changing its internal representation to accommodate the
new information. After this initial phase, the eigenvalues stabilize and the
loss function decreases smoothly, as expected in the lazy training regime.
The
eigenvectors corresponding to the larger eigenvalues can be interpreted as {\em
learnable}\ features, while the small (or zero) eigenvalues correspond to
directions in which the field $f$ never evolves during training.

\FloatBarrier

\subsubsection{Eigenvectors as Features}
\label{sec:NTKAlign}

\begin{figure}[ht!]
  \centering
  \includegraphics[width=1\textwidth]{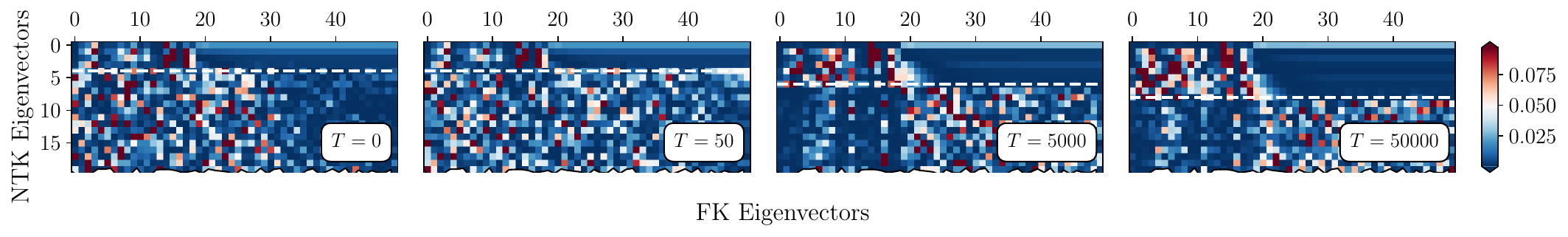}
  \caption{Matrix $A$ as defined in Eq.~\eqref{eq:MatrixA} for L2 data and for a
  single replica of the NTK. The matrix is shown at different epochs of the
  training process, indicated in the top of each panel.}
  \label{fig:NtkMAlign}
\end{figure}

It has been argued above that there is a non-trivial interplay between the
eigenspace of the NTK and that of the matrix $M$. Indeed, the former encodes the
model dependence, while the latter yields physical information. Of course the
two matrices are independent at initialization, and we do not expect any
alignment pattern between the two. However, this picture does change during
training, as the NTK evolves and the model learns the target function. To
quantify this alignment, we define the matrix $A$, 
\begin{equation}
  \label{eq:MatrixA}
  A_{kk'} = \left( z^{(k)}, v^{(k')} \right)^2 = \cos^2(\theta_{kk'}) \;,
\end{equation}
where $z^{(k)}$ and $v^{(k')}$ are the $k$-th and $k'$-th eigenvectors of the
NTK and $M$, respectively. The matrix $A$ is thus a measure of the alignment
between the eigenspaces of the two matrices. The rows of the matrix correspond
to the eigenvectors of the NTK, ordered by the value of the corresponding
eigenvalues, with the eigenvectors corresponding to the larger eigenvalues at
the top of the matrix. The columns correspond to eigenvectors of the matrix $M$,
also ordered by the values of the corresponding eigenvalues, with the largest
eigenvalues to the left in this case. In Fig.~\ref{fig:NtkMAlign}, we show the
matrix $A$ at different epochs of the training for L2 data and a single NTK
replica. 

The blue rectangle in the top right corner of the matrix shows that the
eigenvectors of the NTK corresponding to the largest eigenvalues are orthogonal
to the eigenvectors of $M$ that are in the kernel of $M$, \ie, to the directions
that do not contribute to the observables. It is useful to remember that the
largest eigenvalues of the NTK correspond to the directions that are orthogonal
to $\ker\Theta$, \ie, the directions that are learnable during the training
process. In order to have a robust training process, we expect these learnable
directions to align with the directions that actually contribute to the loss
function, which are the ones corresponding to the largest eigenvalues of $M$.
Consistently with this intuition, we see that the size of this blue rectangle
increases with training time. In particular, it is clear from our plot that it
becomes deeper by the onset of the lazy training regime: more of the learnable
directions -- the {\it features}\ that the network can learn -- are aligned with
the directions that contribute most to the observables.

\begin{figure}[ht!]
  \centering
  \includegraphics[width=0.60\textwidth]{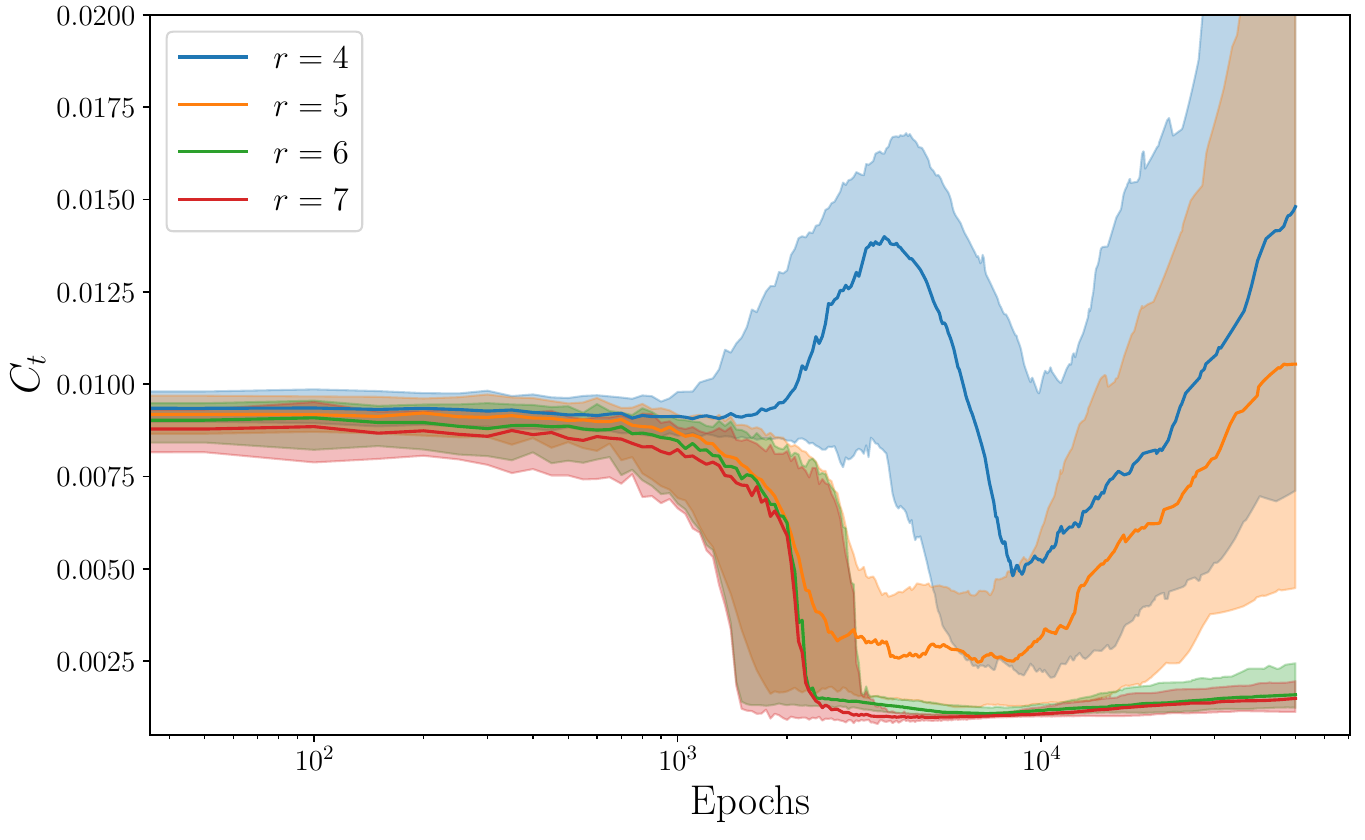}
  \caption{Reconstruction error $C_t(r)$ as defined in Eq.~\eqref{eq:NTKCoverage}
  as a function of the training time and for different numbers of eigenvectors
  $r$. Note that $f_0$ lies entirely in the subspace spanned by the first four
  eigenvectors of the NTK by the onset of the lazy training regime. We see that
  the NTK has aligned its features with the physically relevant directions of
  the problem.}
  \label{fig:NTKAlignFin}
\end{figure}
The eigenvectors of the NTK form an orthonormal basis in
$\mathbb{R}^{\ngrid}$ at any epoch of the training process. However, 
only a subset of these eigenvectors contributes to the training dynamics. The 
{\em expressivity}\  of the vectors belonging to the subspace orthogonal
to the kernel is quantified by measuring how well the internal representation of the neural
network can reconstruct the input function $f_0$ used to generate the data. 
We define a new figure of merit
\begin{equation}
  C_t(r) = 1 - \sum_{k=1}^{r} \frac{(z^{(k)}, f_0)^2}{\Vert f_0 \Vert^2} \,,
  \label{eq:NTKCoverage}
\end{equation}
which measures the reconstruction error of the input function $f_0$ when
projected onto the subspace spanned by the first $r$ eigenvectors of the NTK at
training time $t$. We show $C_t(r)$ as a function of the training time and for
different choices of $r$ in Fig.~\ref{fig:NTKAlignFin}. Inspecting the figure,
we see that in the early stages of training the reconstruction error does not
change significantly with time. This behaviour is shared across all values of
$r$. In fact, in this first phase the NTK has not yet found a suitable internal
representation and the inclusion of more eigenvectors corresponding to yet
undiscovered directions -- those associated with a small eigenvalue as in
Fig.~\ref{fig:NTKEigvalsTime} -- does not result in an improvement of the
reconstruction of $f_0$. Conversely, once the onset of lazy training is
approached, we identify two distinct behaviours depending on the number of
eigenvectors. If we include the eigenvectors up to the modes discovered during
training (\eg, $r=6$ and $r=7$ in the figure), the reconstruction error drops
significantly and remains almost constant throughout the rest of the learning
process. On the other hand, if we consider fewer eigenvectors (\eg, $r=4$ and
$r=5$ in the figure), the reconstruction error becomes larger and grows
indefinitely with training time. The subset of
eigenvectors orthogonal to $\ker \Theta$ is capable of providing an effective
lower dimensional basis, provided that the new modes discovered by the NTK
during training are included. Note that the reconstruction of the input function
improves as long as the eigenvectors do not belong to $\ker \Theta$ --
the eigenvectors that belong to the kernel only add noise to the reconstruction, 
without bringing any new physical information.

\begin{figure}[ht!]
  \centering
  \includegraphics[width=0.90\textwidth]{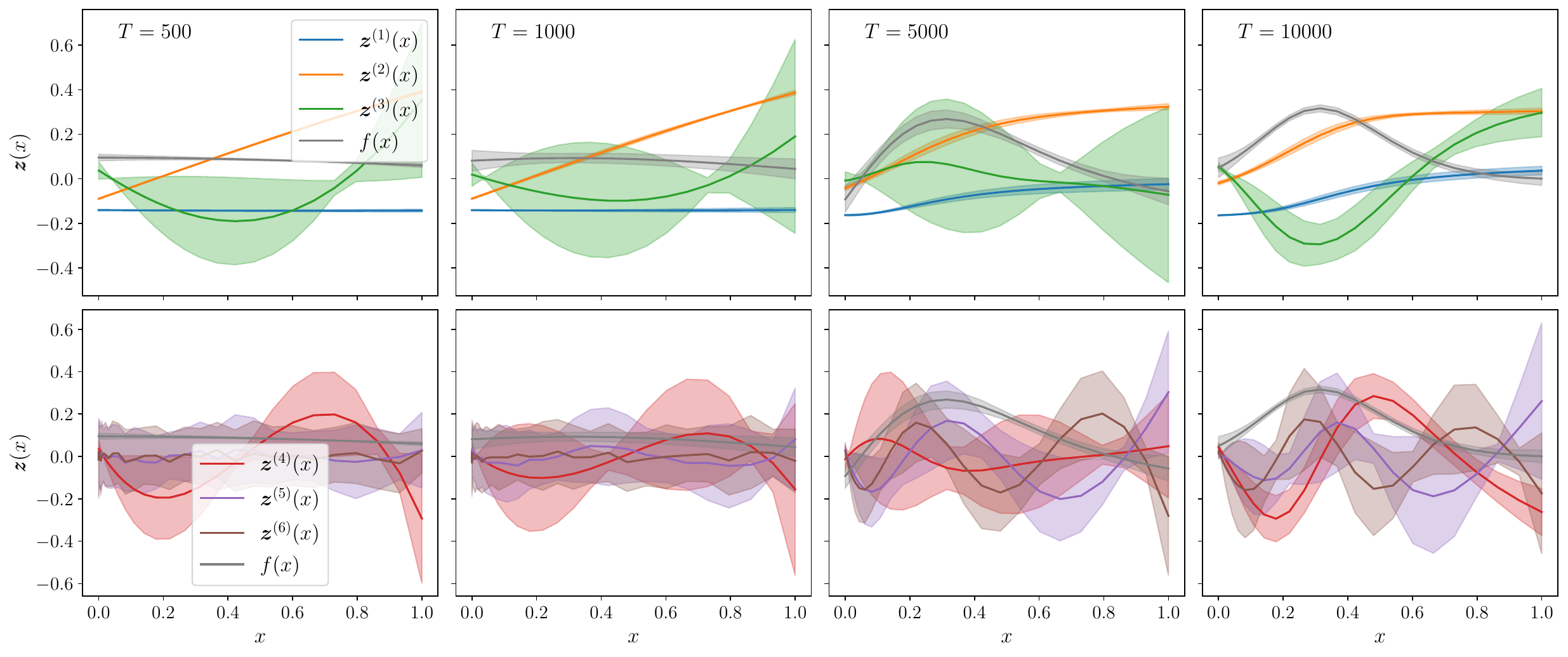}
  \caption{First five eigenvectors of the NTK at different training times and as
  function of the input $x$-grid. We also show the output of the network at the
  same training time, which is displayed in gray. L1 data is used.}
  \label{fig:NTKEigVecs}
\end{figure}
A complementary picture is displayed in Fig.~\ref{fig:NTKEigVecs}. Here, we
show the eigenvectors of the NTK at different training times as functions of the
$x$-grid, denoted by $z^{(i)}$. Together with the eigenvectors, we also show
the output of the trained neural network at the corresponding training time.
From these plots, we see that as the training progresses, the shape of the
eigenvectors becomes more structured in order to reproduce the output function.
Again, this conclusion supports the observations made previously on various
occasions, that during training the neural network is changing its internal
representation and the NTK encodes this information.

\FloatBarrier

\newpage

\section{Lazy Training Regime}
\label{sec:LazyTraining}

In the previous section we presented an empirical study of the training dynamics
through the lens of the NTK. We observed that the NTK is able to capture the
main features of the training process, and that its time evolution is
characterized by a rapid initial transient, followed by a slower evolution
during the rest of the training. We now turn our attention on this last stage of
the training, where the NTK has stabilized and becomes approximately constant.
In doing so, we build upon the results presented in
Refs.~\cite{jacot2018neural,lee2019wide} and derive the analytical solution of the 
flow equation, which
allows us to write an explicit expression for the trained field as a function of
the field at initialization and the data.

\subsection{Analytical Results}
\label{eq:AnlyticalLazySolution}

\subsubsection{Solution of the Flow Equation}
\label{sec:Lazy}

The lazy training regime is characterized by a slow-evolving NTK. We denote as
$t_{\rm ref}$ the time at which the onset of this regime occurs. The NTK is then
\textit{frozen} to its value at $t_{\rm ref}$, and from this time onward the NTK
is taken to be constant
\begin{equation}
  \Theta_t = \Theta_{t_{\rm ref}} \equiv \Theta, \quad \textrm{for } t \geq t_{\rm ref}\, .
\end{equation} 
The flow equation can then be written as
\begin{align}
  \ddt f_t = -\Theta M f_t + b\, ,
  \label{eq:FlowEqTwo}
\end{align}
where $M$ and $b$ are defined as in Eq.~\eqref{eq:MandBDef}. Note that now
neither $\Theta$ nor $b$ depend on the training time $t$. In order to solve this
first-order linear differential equation, we observe that the eigenvectors of
$\Theta$,
\begin{align}
    \label{eq:ThetaEigensystem}
    \Theta z^{(k)} = \lambda^{(k)} z^{(k)}\, ,
\end{align}
provide a basis for expanding Eq.~\eqref{eq:FlowEqTwo}. Furthermore, owing to
the spectrum hierarchy of the NTK (Fig.~\ref{fig:NTKEigvalsTime}), it is
necessary to distinguish the components of $f_t$ that are in the kernel of
$\Theta$ from the ones that are in the orthogonal complement. We introduce the
notation
\begin{align}
    \label{eq:ParallelComponents}
    &f^\parallel_{t,k} = \left(z^{(k)}, f_t\right)\, , \quad \text{if}\ \lambda^{(k)} = 0\, , \\
    \label{eq:OrthogonalComponents}
    &f^\perp_{t,k} = \frac{1}{\sqrt{\lambda^{(k)}}} \left(z^{(k)}, f_t\right)\, , \quad
        \text{if}\ \lambda^{(k)} \neq 0\, ,
\end{align}
where the scalar product has been defined as
\begin{equation}
  \left(f'_{t'}, f_t\right) = \sum_{i,\alpha} f'_{t',i\alpha} f_{t,i\alpha}\,.
\end{equation}
One can readily see that the components in the kernel of $\Theta$, $\text{ker}\
\Theta$, do not evolve during training,\footnote{Despite this result having been
obtained using the frozen NTK, it is worth mentioning that at any time during
training the kernel of the NTK is always defined and in general non-empty.
Hence, even in the initial stages of the training process, there is a component 
that is completely determined by the initial condition, \ie, by the prior distribution in
functional space.}
\begin{align}
    \label{eq:FlowParallel}
    \ddt f^\parallel_{t,k} = 0
        \quad \Longrightarrow \quad f^\parallel_{t,k} = f^\parallel_{0,k}\, .
\end{align}
This means that the final solution is affected by an irreducible noise that
is purely dictated by the initial condition.

The flow equation for the orthogonal components can be written as
\begin{align}
    \label{eq:FlowPerp}
    \ddt f^\perp_{t,k} = - H^\perp_{kk'} f^\perp_{t,k'}
        + B^\perp_{k}\, ,
\end{align}
where  we introduced
\begin{align}
    H^\perp_{kk'} &= \sqrt{\lambda^{(k)}} \left(z^{(k)}, M z^{(k')}\right) \sqrt{\lambda^{(k')}}\, ,\\
    B^\perp_k &= -\sqrt{\lambda^{(k)}} \left[\left(z^{(k)}, M z^{(k')}\right) f^\parallel_{0,k'}
        - \left(z^{(k)}, \FKtabT C_Y^{-1} Y\right)\right]\, .
\end{align}
As discussed above, the indices on quantities that have a $\perp$ suffix only
span the space orthogonal to the kernel of $\Theta$, while the indices on
quantities that have a $\parallel$ suffix span the kernel. We refer to $H^\perp$
as the flow (or training) Hamiltonian, emphasizing that training can only take place in the space
orthogonal to the kernel of $\Theta$; we see explicitly in the definition above
that the flow dynamics is determined by a combination of the architecture of the
NN, encoded in the NTK, and the data, on which $M$ depends. More specifically,
the matrix elements of $M$ can be written as
\begin{align}
    \label{eq:MMatElems}
    \left(z^{(k)}, M z^{(k')}\right) = T^{(k)T} C_Y^{-1} T^{(k')}\, ,
\end{align}
where $T^{(k)} = T[z^{(k)}]$ is the vector of theory predictions for the data
obtained using $z^{(k)}$ as the input PDF. Similarly, we have
\begin{align}
    \label{eq:BMatElems}
    \left(z^{(k)}, \FKtabT C_Y^{-1} Y\right) = T^{(k)T} C_Y^{-1} Y\, .
\end{align}
Denoting by $d^\perp$ the dimension of the subspace orthogonal to $\text{ker}\
\Theta$, $H^\perp$ is a $d^\perp\times d^\perp$ symmetric matrix, whose
eigenvalues and eigenvectors satisfy
\begin{align}
    H^\perp_{kk'} w^{(i)}_{k'} = h^{(i)} w^{(i)}_{k}\, .
\end{align}
\begin{figure}[h!]
  \centering
  \includegraphics[width=0.3\textwidth]{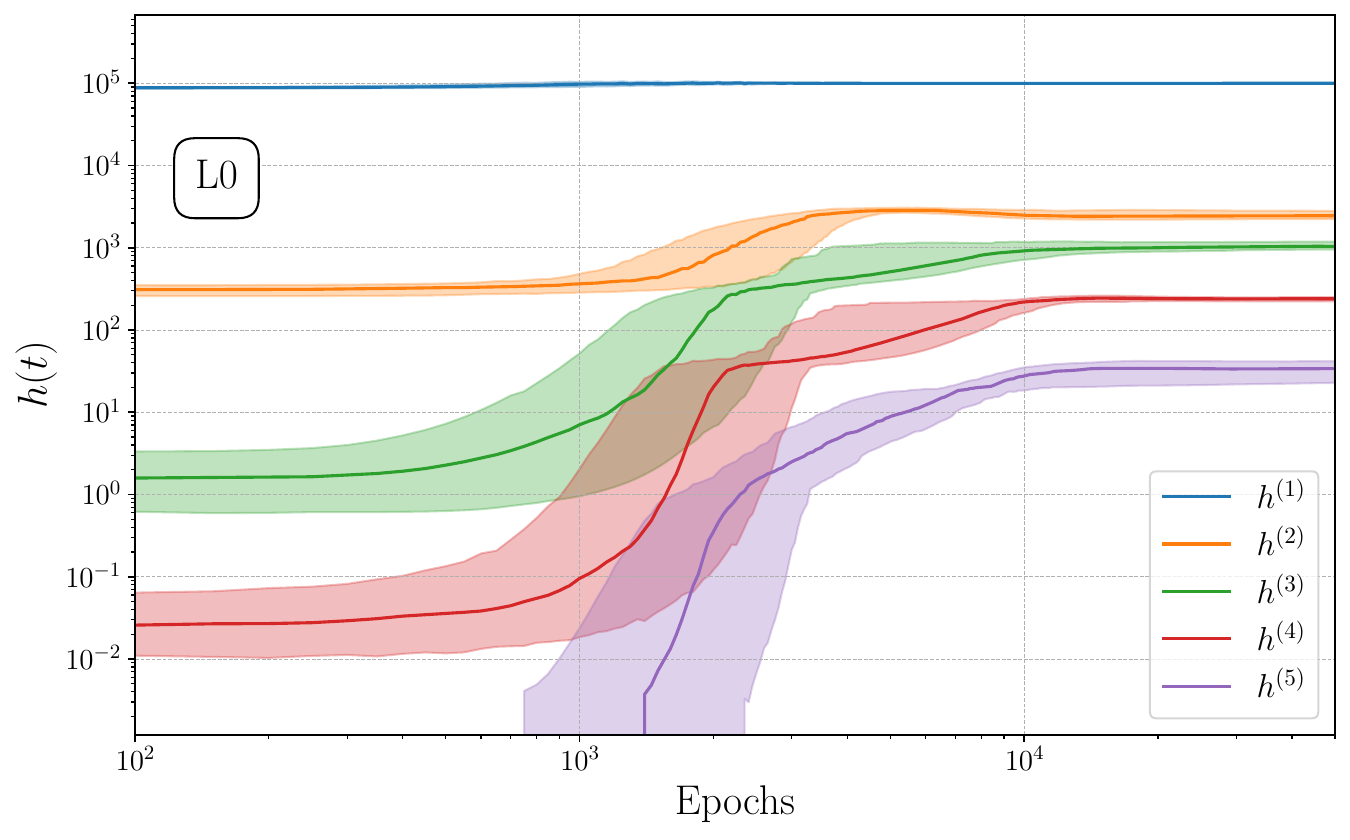}
  \includegraphics[width=0.3\textwidth]{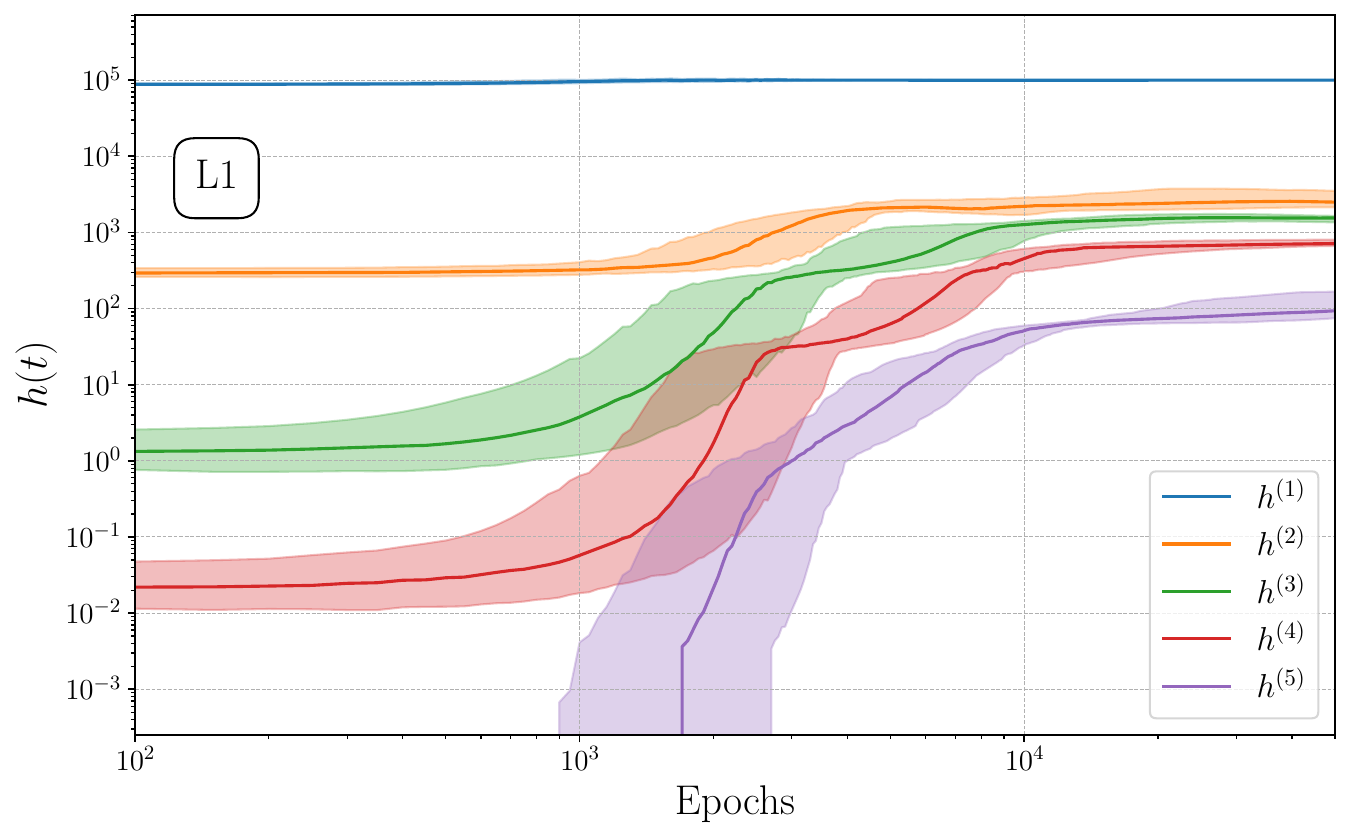}
  \includegraphics[width=0.3\textwidth]{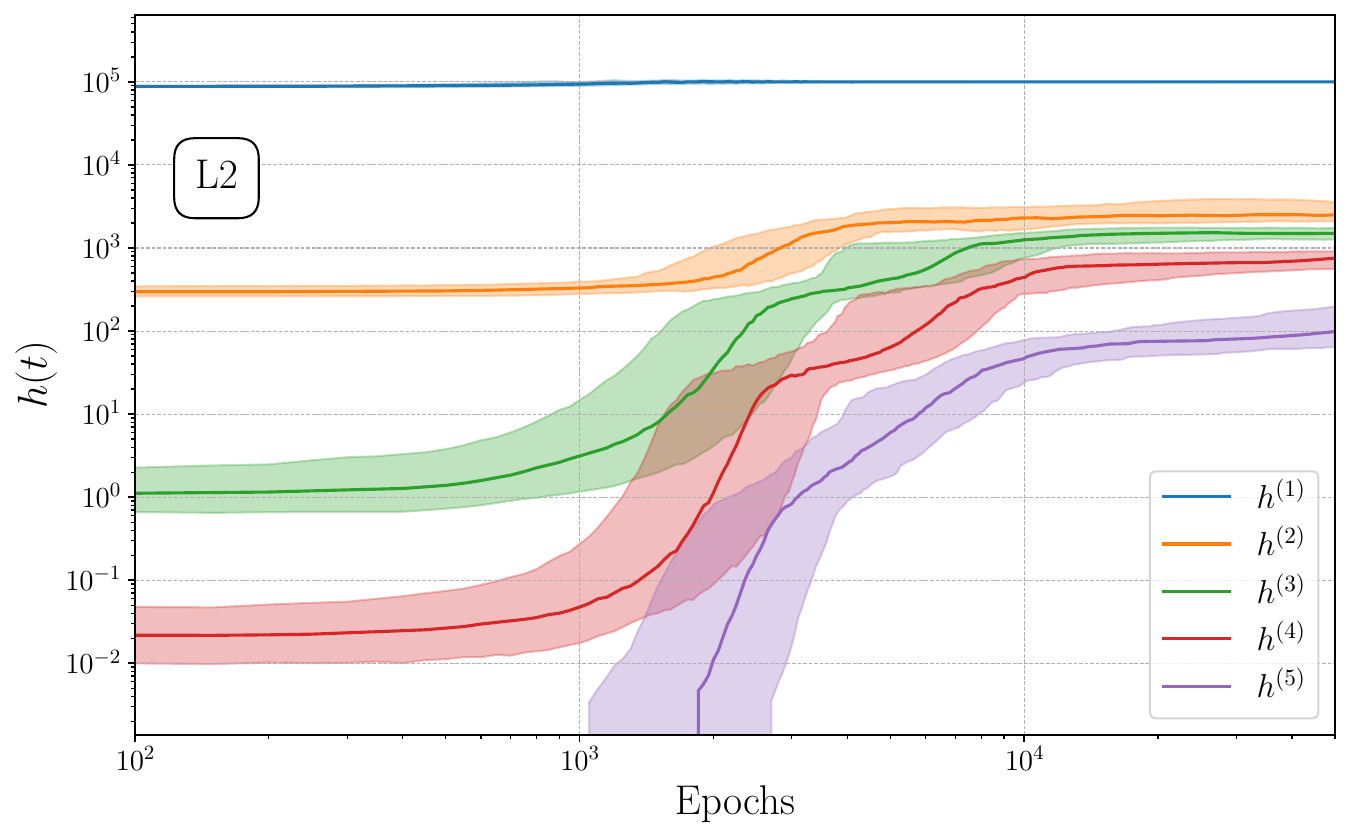}
  \caption{Evolution during training of the first five eigenvalues of
  $H^{\perp}$ using L0 (left), L1 (center), and L2 (right) data. Solid lines
  represent the median over the ensemble of networks, while solid bands
  correspond to 68\% confidence level. Note that the subdominant eigenvalues
  $\lambda^{(3)}$, $\lambda^{(4)}$ and $\lambda^{(5)}$ have increased by one or
  two orders of magnitude by the end of the rich training phase.}
  \label{fig:HPerpEigvalsTime}
\end{figure}
\begin{figure}[h!]
  \centering
  \includegraphics[width=0.90\textwidth]{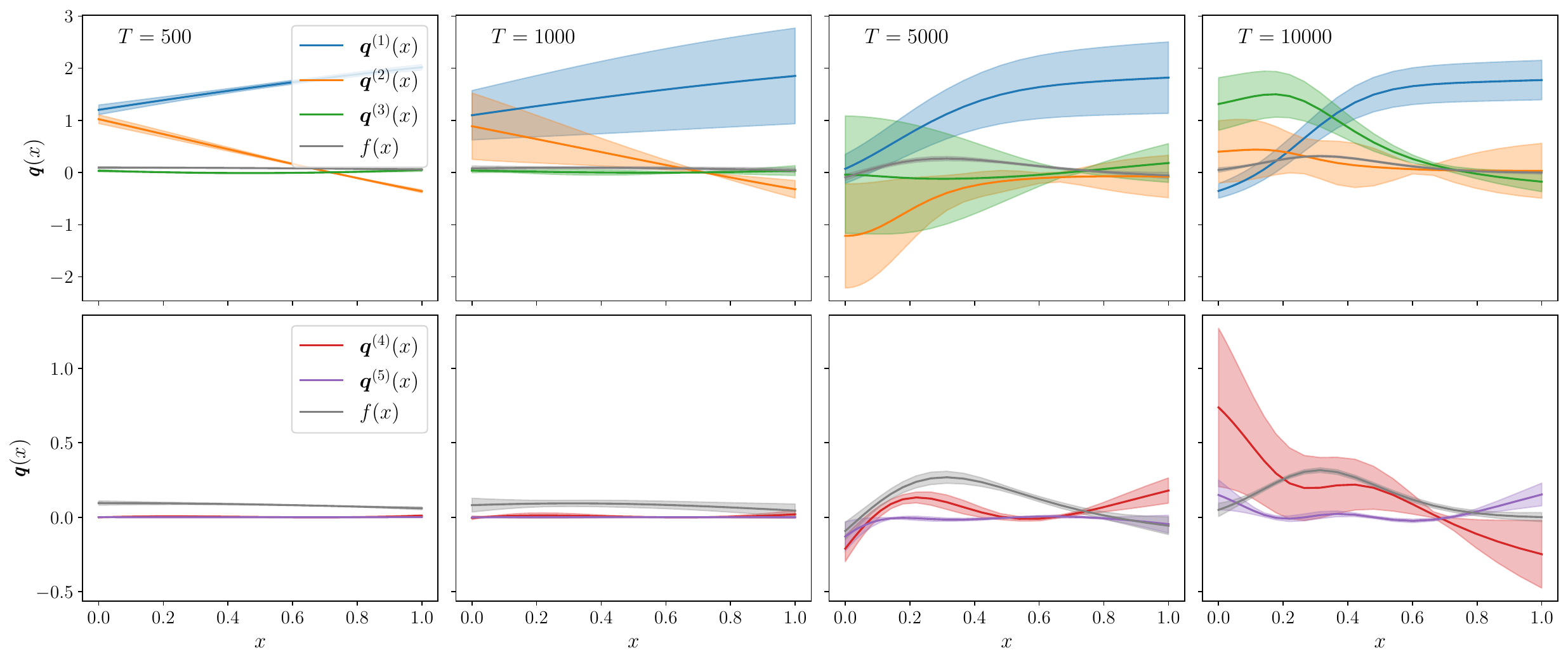}
  \caption{First five eigenvectors of the combined matrix $H=\Theta M$, as in
  Eq.~\eqref{eq:FlowEquationNoIndices}, at different training times and as a
  function of the input $x$-grid. We also show the output of the network at the
  same training time, which is displayed in gray. L1 data is used.}
  \label{fig:HEigVecs}
\end{figure}

In Fig.~\ref{fig:HPerpEigvalsTime} we show the evolution during training of the
first five eigenvalues of $H^{\perp}$ for the three different closure datasets.
In Fig.~\ref{fig:HEigVecs} we show the first five eigenvectors of $H$, denoted
as $q^{(i)}$, at different training times as functions of the $x$-grid. It
should not come as too much of a surprise that the eigenvalues and eigenvectors
of $H^{\perp}$ have a similar behaviour to those of the NTK (see
Figs.~\ref{fig:NTKEigvalsTime} and \ref{fig:NTKEigVecs}), from which they are constructed.
However, we see that $h^{(i)}$ are larger by around three orders of magnitude
than the NTK eigenvalues.

The solution to Eq.~\eqref{eq:FlowPerp} can be written as the sum of the
solution of the homogeneous equation, $\hat{f}^{\perp}_{t,k}$, and a particular
solution of the full equation. The solution of the homogeneous equation is
\begin{align}
    \label{eq:HomoSoln}
    \hat{f}^{\perp}_{t,k} = \sum_{i=1}^{d^\perp} f^{\perp (i)}_{0} e^{-h^{(i)}t} w^{(i)}_k\, ,
\end{align}
where
\begin{align}
    \label{eq:InitialCi}
    f^{\perp (i)}_{0} = \sum_{k=1}^{d_\perp} w^{(i)}_k f^\perp_{0,k}\, ,
\end{align}
guarantees that the initial condition $\hat{f}^\perp_{t,k}=f^\perp_{0,k}$ is
satisfied. Similarly, if we define
\begin{align}
    \label{eq:BiDef}
    \Upsilon^{(i)} = \sum_{k=1}^{d_\perp} w^{(i)}_k B^\perp_{k}\, ,
\end{align}
then
\begin{align}
    \label{eq:PartSol}
    \check{f}^\perp_{t,k} = \sideset{}{'}\sum_{i} \frac{1}{h^{(i)}} \Upsilon^{(i)}
        \left(1 - e^{-h^{(i)}t}\right) w^{(i)}_k\, ,
\end{align}
where the sum only involves the non-zero modes of $H^\perp$, is a particular
solution of the inhomogeneous equation, which satisfies the boundary condition
$\check{f}^{\perp}_{0,k}=0$. Hence, the solution of the flow equation in the
subspace orthogonal to $\text{ker}\ \Theta$ is
\begin{align}
    f^\perp_{t,k}
    \label{eq:FlowSolution}
        &= \hat{f}^\perp_{t,k} + \check{f}^\perp_{t,k}
        \, .
\end{align}
Finally, collecting the parallel contribution, Eq.~\eqref{eq:FlowParallel}, and
the solution of the orthogonal component, Eq.~\eqref{eq:FlowSolution}, yields a
simple expression,

\begin{align}
    \label{eq:AnalyticSol}
    \boxed{f_{t,\alpha}
        = U(t)_{\alpha\alpha'} f_{0,\alpha'} + V(t)_{\alpha I} Y_{I}\, .
    }
\end{align} 

The two evolution operators $U(t)$ and $V(t)$ have lengthy, yet explicit,
expressions, which we summarize here: 
\begin{align}
    \label{eq:UDef}
    U(t)_{\alpha\alpha'} = \hat{U}^\perp(t)_{\alpha\alpha'}
        + \check{U}^\perp(t)_{\alpha\alpha'} + U^\parallel_{\alpha\alpha'}\, ,
\end{align}
where
\begin{align}
    \hat{U}^\perp(t)_{\alpha\alpha'}
        = \sum_{k,k'\in\perp} \sqrt{\lambda^{(k)}} z^{(k)}_\alpha 
            \left[\sum_i w^{(i)}_{k} e^{-h^{(i)}t} w^{(i)}_{k'}\right]
            z^{(k')}_{\alpha'} \frac{1}{\sqrt{\lambda^{(k')}}}\, ,
\end{align}
and
\begin{align}
    U^\parallel_{\alpha\alpha'}
        = \sum_{k''\in\parallel} z^{(k)}_\alpha z^{(k)}_{\alpha'} \, .
\end{align}
The contributions from $\check{U}^\perp(t)$ and $V(t)$ are more easily expressed
by introducing the operator
\begin{align}
    \label{eq:MOperatorDef}
    \mathcal{M}(t)_{\alpha\alpha'} 
        = \sum_{k,k'\in\perp} \sqrt{\lambda^{(k)}} z^{(k)}_\alpha 
            \left[\sideset{}{'}\sum_{i} w^{(i)}_{k} \frac{1}{h^{(i)}}\, 
            \left( 1- e^{-h^{(i)}t}\right) w^{(i)}_{k'}\right]
            z^{(k')}_{\alpha'} \sqrt{\lambda^{(k')}}\,. 
\end{align}
Then, we can write
\begin{align}
    \label{eq:UperpCheck}
    \check{U}^\perp(t)
        = - \mathcal{M}(t)\; \FKtabT C_Y^{-1} \FKtab 
            \left[\sum_{k''\in\parallel} z^{(k'')} z^{(k'') T}\right]\, ,
\end{align}
and
\begin{align}
    \label{eq:VDef}
    V(t) = \mathcal{M}(t)\; \FKtabT C_Y^{-1}\, ,
\end{align}
where we note that the term in the bracket in Eq.~\eqref{eq:UperpCheck} is
simply the projector on the kernel of the NTK. The four terms that appear in the
analytical solution have a clear physical interpretation:
\begin{itemize}
    \item The first term $\hat{U}^\perp(t)$ suppresses the components of the
    initial condition that lie in the subspace orthogonal to the kernel of the
    NTK. These are the components that are learned by the network during
    training. While the trained solution still depends on its value at
    initialization, that dependence is suppressed during training. This
    suppression is exponential in the training time, and the rates are given by
    the eigenvalues of $H^{\perp}$.
    \item The contribution from $U^\parallel$ yields the component of the
    initial condition that lies in the kernel of the NTK. As such, those
    components remain unchanged during training and are part of the trained
    field at all times $t$. 
    \item The two remaining contributions are best understood by combining them
    together,
    \begin{align}
        \label{eq:DataCorrectedInference}
        \check{U}^{\perp}(t) f_{0} + V(t) Y 
            = \mathcal{M}(t)\; \FKtabT C_Y^{-1} \left[Y - \FKtab f_{0}^{\parallel}\right]\, .
    \end{align}
    The parallel component of the initial condition $f_{0}^{\parallel}$ does not
    evolve during training, and therefore it yields a contribution $\FKtab
    f_{0}^{\parallel}$ to the theoretical prediction of the data points at all
    times $t$. This is taken into account by subtracting this contribution from
    the data, before the inference is performed.
\end{itemize}

\medskip
\noindent
\fbox{
    \begin{minipage}{0.9\textwidth}
        \begin{exercise}
            Derive the analytic solution in Eq.~\eqref{eq:AnalyticSol} by solving 
            the flow equation in the subspace orthogonal to the kernel of the NTK, 
            and then adding the contribution from the kernel, as explained in the text. 
            In doing so, you will need to compute the eigenvalues and eigenvectors of $H^\perp$, and use them to write the
             solution of the homogeneous and inhomogeneous equations. Finally, you
             will need to express the solution in terms of the initial condition
             $f_0$ and the data $Y$ by using the definitions of $B^\perp$ and
             $H^\perp$.
        \end{exercise}
    \end{minipage}
}
\medskip

\noindent

\begin{figure}[ht!]
  \centering
  \includegraphics[width=0.9\textwidth]{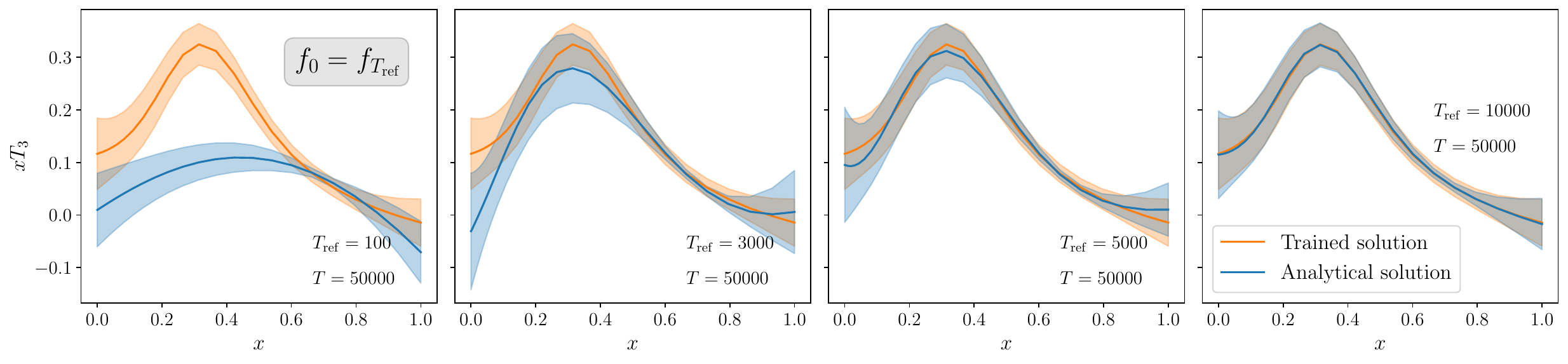} 
  \caption{Comparison of the trained and analytical evolution at the end of
  training. Each panel corresponds to a different frozen NTK, whereby the
  analytical solution is computed starting from $f_{T_{\rm Ref}}$. The orange
  curve represents the final trained function after 50000 iterations of GD, and
  is the same across panels. Error bands represent one-sigma uncertainties
  across replicas. L2 data is used.}
  \label{fig:OnsetLazyL2}
\end{figure}

The solution in Eq.~\eqref{eq:AnalyticSol} is the main result of this section.
It shows that the training process can be described as the sum of a linear
transformation of the initial fields $f_{0,\alpha}$, and a linear transformation
of the data $Y_I$. The two transformations depend on the flow time $t$ and are
given by the evolution operators $U(t)$ and $V(t)$. Fig.~\ref{fig:OnsetLazyL2}
compares the analytical solution with the trained function at the end of
training, for different choices of the frozen NTK. The NN is trained using the
numerical GD until $t_{\rm ref}$, at which point the NTK is frozen. The
evolution time $t$ used in the analytical solution is the difference between the
total training time and $t_{\rm ref}$; the initial condition for the analytical
solution is the trained solution at $t_{\rm ref}$. Central value and uncertainty
bands are obtained by computing the analytical solution for each replica of the
initial condition and frozen NTK.\footnote{Unless stated otherwise, in this
section central values and uncertainties are always computed as ensemble
averages across replicas.} The figures agree with the expectation; the closer
$t_{\rm ref}$ is to the onset of the lazy regime, the better the agreement
between the analytical solution and the trained function.

\FloatBarrier

\begin{figure}[ht]
    \centering
    \includegraphics[width=0.9\textwidth]{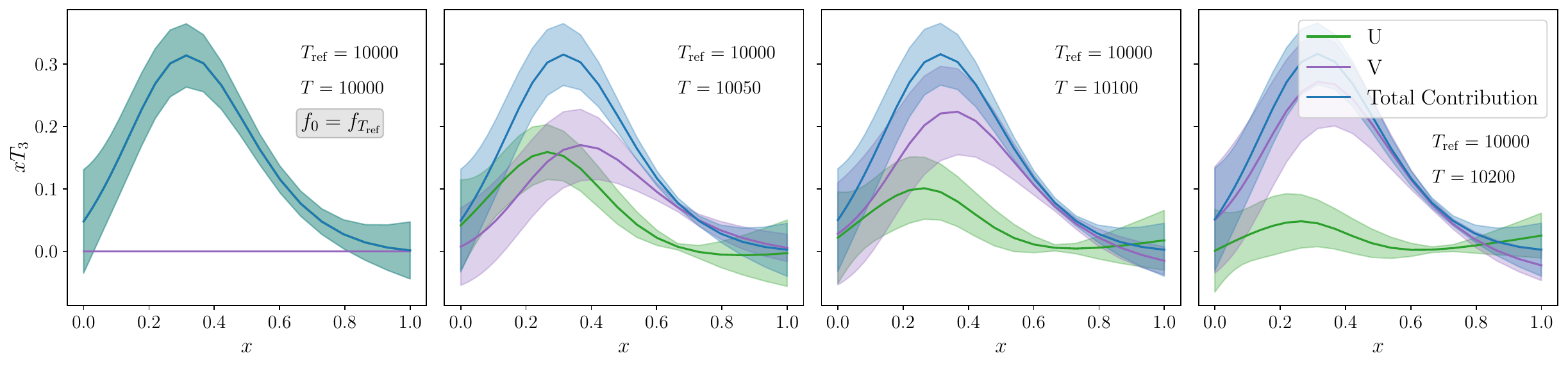} 
    \caption{Decomposition of the analytical solution into the two contributions
    from $U$ and $V$ at different training times. The frozen NTK is fixed across
    panels, and corresponds to the one at $T_{\rm ref} = 10000$. The initial
    condition for the analytical solution is always $f_{T_{\rm ref}}$. 
    As in Fig.~\ref{fig:OnsetLazyL2}, L2 data is used.}
    \label{fig:FrefDecompositionL2}
  \end{figure}

A complementary perspective is provided in Fig.~\ref{fig:FrefDecompositionL2},
where the analytical solution is decomposed into the two contributions from $U$
and $V$. In each panel, the initial condition $f_{t_{\rm ref}}$ is evolved
analytically for different training times by keeping the frozen NTK fixed. We
see that as training proceeds, the contribution from $U$ is progressively
suppressed, in accordance with the observation made above. On the other hand,
the contribution from $V$ grows and becomes dominant at later epochs, indicating
that the trained function is mostly determined by the data, rather than the
initial condition of the network. We also observe that such behaviour happens
quite rapidly -- in a training time interval $\Delta T \approx 200$ --
as a consequence of the fact that the time scales in the analytical
solution are determined by the inverse of the eigenvalues of $H^\perp$, and
the latter are typically large.

The analytical solution in Eq.~\eqref{eq:AnalyticSol} sheds a new light onto the
behaviour of the numerical training of a neural network. Given these results, it
is natural to ask whether the information encoded in the NTK alone can drive
training, independently of the initial condition, \ie, whether the analytical
solution can be used to perform kernel learning. We address this question in 
Sect.~\ref{sec:NTKKernelLearning}, after deriving a few additional analytical results 
in the subsections below. 

\subsubsection{Crosschecks using L0 data}
\label{sec:AnalyticalChecks}

The analytical solution enables rigorous validation of our implementation
through crosschecks with L0 data, where we have complete control over the data
generation process. In this case, the realization of the dataset is completely
determined by the input PDFs
\begin{equation}
    \label{eq:DataL0NoIndices}
    Y = \FKtab \fin\, .
\end{equation}
Note that using L0 data only affects the second term in
Eq.~\eqref{eq:AnalyticSol}.\footnote{To be more precise, since the analytical
solution requires the NTK to be frozen at a certain epoch $T_{\rm ref}$, the NTK
also depends on the data used in the training.} We can then rewrite the combined
term in Eq.~\eqref{eq:DataCorrectedInference} as follows
\begin{align}
  \label{eq:TrainingOnLevelZero}
  \check{U}^\perp(t) f_0 + V(t) Y 
    &= \mathcal{M}(t)\, \FKtabT C_{Y}^{-1} \FKtab\, 
      \left[\fin - f_{0}^\parallel\right]\, .
\end{align}
The subtraction taking place in the square brackets of
Eq.~\eqref{eq:TrainingOnLevelZero} suggests that the effective function that
the neural network actually sees is not the input function $\fin$ used to
generate the data, but rather the difference between $\fin$ and the component of
the initial function $f_0$ that lies in the subspace spanned by the kernel of
the NTK, \ie, $f_0^\parallel$. In other words, the parallel component
$f_0^\parallel$, which, we remind the reader, does not evolve during the analytic training,
acts as a constant ``bias'' in the training process, effectively shifting the 
input function seen by the neural network. Of course the actual magnitude of
this irreducible noise depends both on how $f_0$ and the kernel of the NTK are
distributed over the ensemble. 

\begin{figure}[t]
  \centering
  \includegraphics[width=\textwidth]{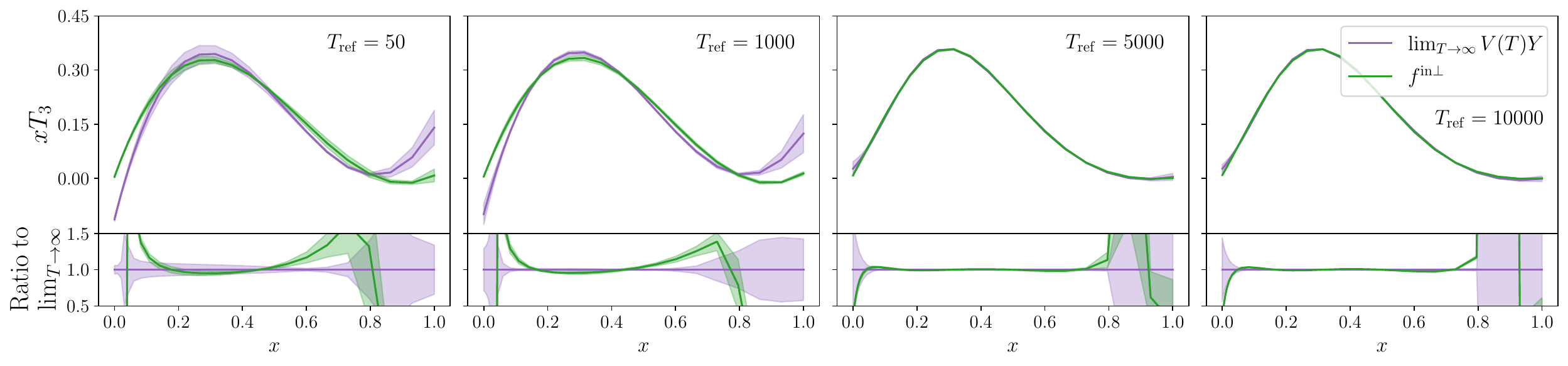}  
  \caption{Test of the $t\to\infty$ limit of the L0 training for different frozen
  NTKs. The green curve represents the projection of the input function $\fin$
  onto the subspace orthogonal to the kernel of the NTK at $t_{\rm ref}$, \ie,
  $\finperp$. The purple curve represents the contribution of the operator $V$,
  computed with the NTK at $T_{\rm ref}$, in the limit of infinite training
  time.}
  \label{fig:InfiniteTimeVterm}
\end{figure}

Note that the observation above remains true even in the limit of infinite
training, 
\begin{align}
    \label{eq:LevelZeroClosureInfiniteTraining}
    \lim_{t\to\infty} V(t) Y = \finperp + \mathcal{M}_{\infty} M \finpar\, ,
\end{align}
which shows that the $V$ component of the trained solution reproduces exactly
the component of the PDF that lies in the subspace orthogonal to the kernel of
$\Theta$. We compare the asymptotic behaviour of $V(t) Y$ and $\finperp$ in
Fig.~\ref{fig:InfiniteTimeVterm}, where we see that the analytical solution at
infinite training time reproduces the expected result, \ie, it coincides 
with $\finperp$, as long as $T_{\mathrm{ref}}> 1000$. The second term in 
the square bracket on the right-hand side of
Eq.~\eqref{eq:TrainingOnLevelZero} is the contribution from the parallel
component at initialization that does not evolve in the training process. Given
that $f_0$ is almost normally distributed around zero, that term does not
contribute to the central value of the fitted PDF, \ie, to the average of the
trained solution over replicas. 

\begin{figure}[h!]
  \centering
  \includegraphics[width=0.95\textwidth]{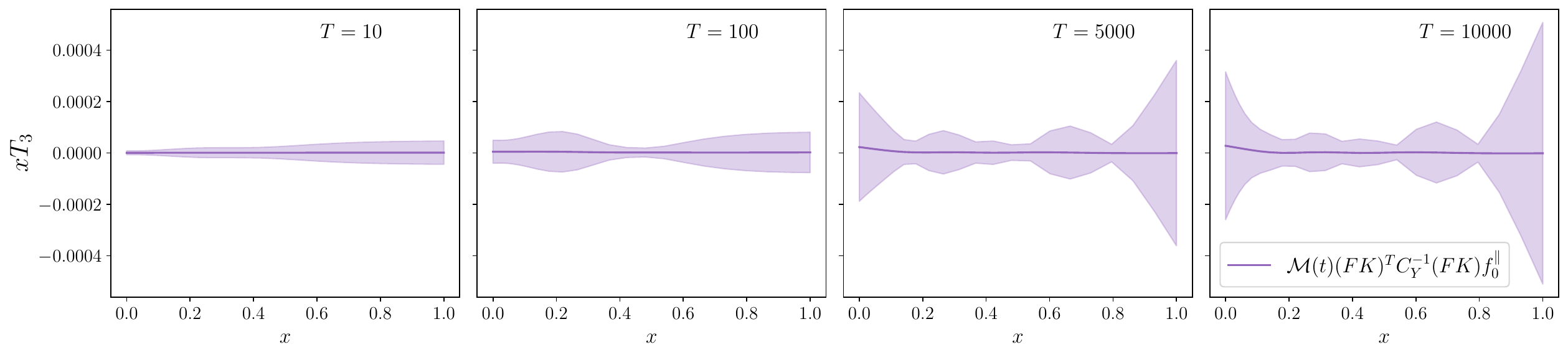} 
  \caption{Test of the average of the parallel contribution for different
  epochs. The reference epoch at which the frozen NTK is chosen is $T_{\rm ref}
  = 10000$. L2 data is used in the plot. Note that the scale on the vertical 
  axis is three orders of magnitude smaller than in Fig.~\ref{fig:InfiniteTimeVterm}.}
  \label{fig:AverageLevelZeroUcheck}
\end{figure}

The time evolution of 
\begin{align}
  \label{eq:AverageLevelZeroUcheck}
  \mathbb{E}\left[\mathcal{M}(t)\, \FKtabT C_{Y}^{-1} \FKtab\, 
    f_{0}^\parallel\right]\, ,
\end{align}
is shown in Fig.~\ref{fig:AverageLevelZeroUcheck}.

\subsubsection{Infinite Training Time}
In the limit of infinite training time, the evolution operators $U(t)$ and
$V(t)$ simplify and yield an elegant interpretation of the minimum of the cost
function. For large training times, we have
\begin{align}
    \label{eq:UhatInfty}
    \hat{U}^\perp_{\infty, \alpha\alpha'}
        &= \lim_{t\to\infty}\hat{U}^\perp(t)_{\alpha\alpha'} = 0\, \\
    \label{eq:MOperatorInfty}
    \mathcal{M}_{\infty, \alpha\alpha'} 
        &= \lim_{t\to\infty}\mathcal{M}(t)_{\alpha\alpha'} = \sum_{k,k'\in\perp} \sqrt{\lambda^{(k)}} z^{(k)}_\alpha 
        \left[\sideset{}{'}\sum_{i} w^{(i)}_{k} \frac{1}{h^{(i)}}\, 
        w^{(i)}_{k'}\right] z^{(k')}_{\alpha'} \sqrt{\lambda^{(k')}}\, ,
\end{align}
and explicit expressions for $\check{U}^\perp_{\infty}$ and $V_{\infty}$ are
obtained from $\mathcal{M}_{\infty}$. The term in the square bracket in
Eq.~\eqref{eq:MOperatorInfty} is the spectral decomposition of the pseudo-inverse
of $H^\perp$ in $d_\perp$-dimensional orthogonal subspace. So, the operator
$\mathcal{M}_{\infty}$ acts as follows on a field $f_{\alpha}$:
\begin{enumerate}
    \item The term on the right of the square bracket computes the coordinate
    $f_{k'}$ introduced in Eq.~\eqref{eq:OrthogonalComponents}. The $f_k$ are 
    the coordinates of the component $f^\perp$ that evolves during
    training, 
    \begin{align}
        \label{eq:RightOfTheBracket}
        f^\perp = \sum_{k\in\perp} \sqrt{\lambda^{(k)}} f_k\, z^{(k)}\,  .
    \end{align}
    \item The term in the square bracket applies the pseudo-inverse to the
    coordinates $f_k$, 
    \begin{align}
        \label{eq:ApplyPseudoInv}
        f'_k = \left(H^\perp\right)^+_{kk'} f_{k'}\, .
    \end{align}
    \item The final term on the left of the square bracket reconstructs the full
    field corresponding to the coordinates $f'_{k}$,
    \begin{align}
        \label{eq:LeftOfTheBracket}
        f^{'\perp} = \sum_{k\in\perp} \sqrt{\lambda^{(k)}} f'_{k}\, z^{(k)}\, .
    \end{align}   
\end{enumerate}

As discussed at the end of Sect.~\ref{sec:Lazy} it is convenient to combine the
contributions of $\check{U}^\perp_{\infty}$ and $V_{\infty}$,
\begin{align}
    \label{eq:DataCorrectedInferenceAtInfty}
    \check{U}^{\perp}_{\infty} f_{0} + V_{\infty} Y 
        = \mathcal{M}_{\infty}\; \FKtabT C_Y^{-1} \left[Y - \FKtab f_{0}^{\parallel}\right]\, .
\end{align}
The contribution to the observables from the parallel components of $f$ does not
change during training, therefore that contribution is subtracted from the data
and the orthogonal components of $f$ are adjusted to minimize the $\chi^2$ of
the corrected data. The minimum of the $\chi^2$ in the orthogonal subspace is
found applying $\mathcal{M}_{\infty}$, \ie, by projecting in the orthogonal
subspace, applying the pseudo-inverse and finally recomputing the full field as
detailed above.

\FloatBarrier

\subsection{Numerical Results}
\label{sec:NTKKernelLearning}

The results shown in Sec.~\ref{sec:Lazy} and Sec.~\ref{sec:NTKDuringTraining}
support the idea that the NTK is capable to encode in its eigenvectors the
physical features learned during training. 
Let us now study the analytical solution that we obtain by choosing the initial 
condition to be an ensemble of
networks at initialization as in Fig.~\ref{fig:prior}. The analytical solution 
is computed using the NTK frozen at $t_{\rm ref}$.

\subsubsection{Central Value and Covariance of the Trained Fields}
\label{sec:CentralAndCovariance}

The analytical solution in Eq.~\eqref{eq:AnalyticSol} is inherently stochastic,
since the frozen NTK at $t_{\rm ref}$ is actually obtained from an ensemble of
networks. The central value of the analytical solution is 
defined as
\begin{align}
    \label{eq:MeanValAtT}
    \bar{f}_{t,\alpha} = \mathbb{E}\left[f_{t,\alpha}\right]
        = \mathbb{E}\left[U(t)_{\alpha\alpha'} f_{0,\alpha'}\right]
            + \mathbb{E}\left[V(t)_{\alpha I} Y_I\right] \, .
\end{align}
More interestingly, the covariance matrix at any time $t$ is given by
\begin{align}
    \cov[f_t,f_t^T]
        &= \mathbb{E}\left[U(t) f_0 f_0^T U(t)^T\right] 
            - \mathbb{E}\left[U(t) f_0\right] \mathbb{E}\left[f_0^T U(t)^T\right]  \nonumber \\
        &\quad + \mathbb{E}\left[U(t) f_0 Y^T V(t)^T\right] 
            - \mathbb{E}\left[U(t) f_0\right] \mathbb{E}\left[Y^T V(t)^T\right] \nonumber \\
        &\quad + \mathbb{E}\left[V(t) Y f_0^T U(t)^T\right]
            - \mathbb{E}\left[V(t) Y\right] \mathbb{E}\left[f_0^T U(t)^T\right] \nonumber \\
    \label{eq:CovAtT}
        &\quad + \mathbb{E}\left[V(t) Y Y^T V(t)^T\right]
            - \mathbb{E}\left[V(t) Y\right] \mathbb{E}\left[Y^T V(t)^T\right] \, .
\end{align}
Note that the first and the fourth lines above yield symmetric matrices, while
the third line is just the transpose of the second, thereby ensuring that the
whole covariance matrix is the sum of three symmetric matrices and therefore is
symmetric, 
\begin{align}
    \label{eq:SumOfCovariances}
    \cov[f_t,f_t^T] = C_t^{(00)} + C_t^{(0Y)} + C_t^{(YY)}\, ,
\end{align}
where
\begin{align}
    \label{eq:C00term}
    C_t^{(00)} 
        &= \mathbb{E}\left[U(t) f_0 f_0^T U(t)^T\right] 
        - \mathbb{E}\left[U(t) f_0\right] \mathbb{E}\left[f_0^T U(t)^T\right]\, ,\\
    C_t^{(0Y)}
        &= \mathbb{E}\left[U(t) f_0 Y^T V(t)^T\right] 
        - \mathbb{E}\left[U(t) f_0\right] \mathbb{E}\left[Y^T V(t)^T\right] \nonumber \\
        \label{eq:C0Yterm}
        &\quad + \mathbb{E}\left[V(t) Y f_0^T U(t)^T\right]
            - \mathbb{E}\left[V(t) Y\right] \mathbb{E}\left[f_0^T U(t)^T\right] \, ,\\
    C_t^{(YY)}
        &= \mathbb{E}\left[V(t) Y Y^T V(t)^T\right]
        - \mathbb{E}\left[V(t) Y\right] \mathbb{E}\left[Y^T V(t)^T\right]\, .
        \label{eq:CYYterm}
\end{align}
Eq.~\eqref{eq:SumOfCovariances} shows explicitly the various contributions to 
the covariance matrix. Indeed, $C_t^{(00)}$ quantifies the
contribution to the covariance matrix that is purely due to the fluctuations of
the initial condition, while $C_t^{(YY)}$ quantifies the contribution that is
purely due to the statistical fluctuations of the data. The mixed term
$C_t^{(0Y)}$ accounts for the correlations between the two sources of
uncertainty.

\subsubsection{Convergence of the Analytical Solution}
\begin{figure}[t]
  \centering
  \includegraphics[width=0.9\textwidth]{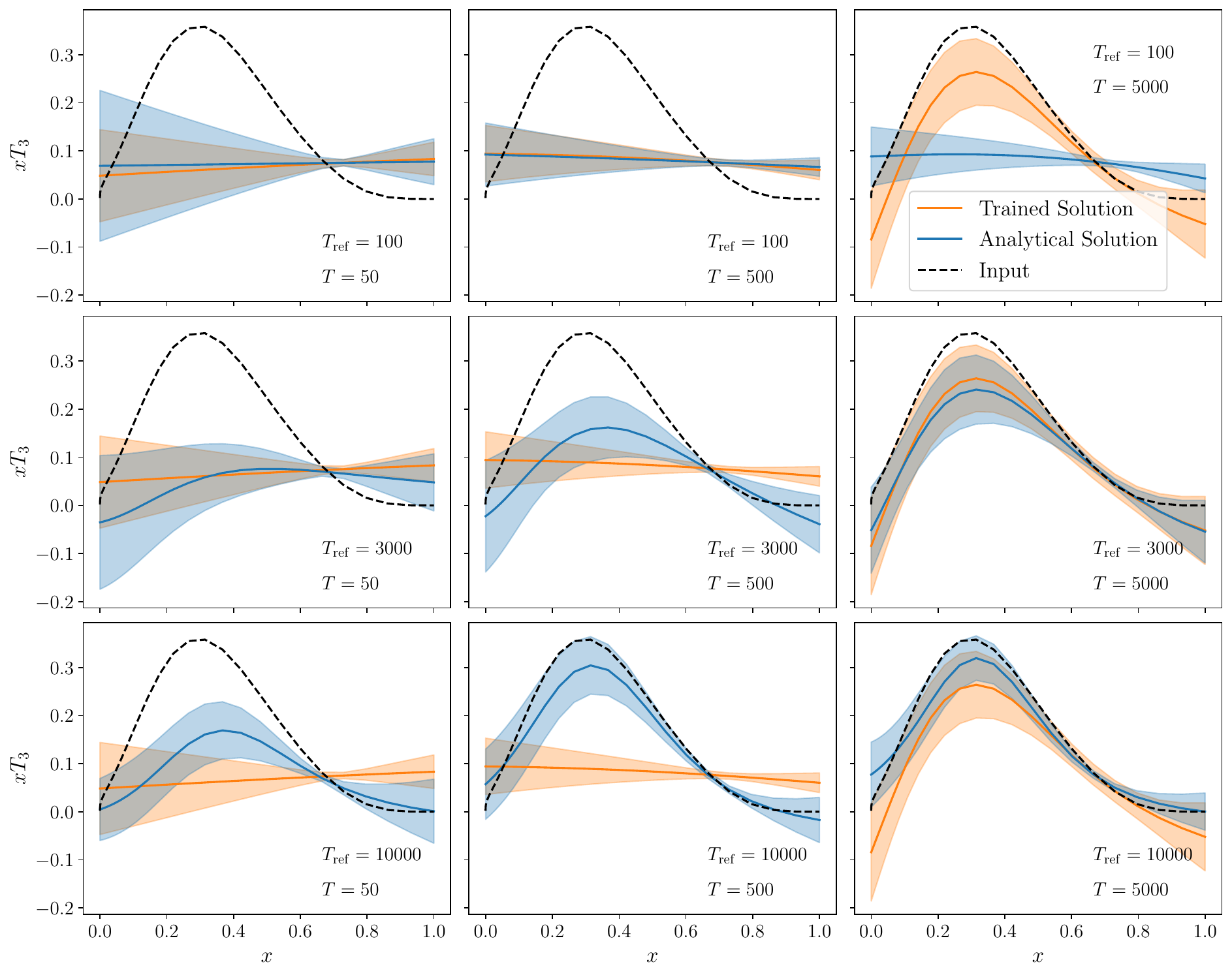} 
  \caption{Comparison of the trained (orange) and analytical (blue) evolution
  starting from an ensemble of networks at initialization as the initial
  condition. Each row corresponds to a different frozen NTK, while the columns
  represent different training times. The dashed line represents the input 
  function used to generate the synthetic data, \ie, the {\em true}\ result. 
  L2 data is used.}
  \label{fig:EvolutionGridF0L2}
\end{figure}

In order to study the convergence of the analytical solution, we compare
\begin{itemize}
    \item [$\bullet$] the analytical solution (AS), obtained using an ensemble of
    networks at initialization as the initial condition; 
    \item [$\bullet$] the trained solution (TS), obtained by training another   
    ensemble of networks, drawn from the same prior distribution, using GD.
\end{itemize}
This comparison is shown in
Fig.~\ref{fig:EvolutionGridF0L2} for L2 data, where the rows in the grid
correspond to different frozen NTKs, while the columns represent numerical and
analytical evolution after $T=50, 500$ and 5000 epochs. These results deserve a
few comments.

The first observation is that the NTK at early stages of training is not able to
drive the prior towards the true function, as shown in the first and, though
less dramatically, in the second row of Fig.~\ref{fig:EvolutionGridF0L2}. This
is expected, as we extensively discussed in Sec.~\ref{sec:NTKDuringTraining},
since at early stages of training by GD the NTK has not yet aligned its 
internal representation with the data.

More significantly, we observe a significant discrepancy between the AS and the
TS even at $T=5000$. At the beginning of
training, there is clearly no difference between AS and TS since both represent
the neural network output at initialization, with
variations due only to different initialization seeds. During early training
stages, AS and TS differ as expected. Indeed, the analytical solution is
computed using the frozen NTK at $t_{\rm ref}$, while the trained solution
evolves with an NTK that is still changing as shown in Sect.~\ref{sec:training}.
Crucially, if the NTK at $t_{\rm ref}$ has already learned from the data and
aligned with the solution, the AS converges faster to the target, while the TS
requires additional epochs before evolving in the correct direction.

\subsubsection{Connection with Linear Methods}
\label{sec:ConnectWithLinear}
We can consider a simplifying limit of Eq.~\eqref{eq:MeanValAtT}, where the
initial condition $f_0$ and the data $Y$ are statistically independent from the
respective evolution operators $U(t)$ and $V(t)$. Note that the first term on
the right-hand side of Eq.~\eqref{eq:MeanValAtT} can only be non-zero because of
the correlations between $U(t)$ and $f_0$. In the absence of such correlations,
the first term would be given by the product of the expectation values and
therefore would vanish if $f_0$ is an ensemble of networks at initialization.
Under these assumptions, we have
\begin{align}
    \label{eq:MeanUt}
    \bar{U}(t)
        &= \mathbb{E}\left[U(t)\right]\, , \\
    \label{eq:MeanVt}
    \bar{V}(t)
        &= \mathbb{E}\left[V(t)\right]\, ,
\end{align}
and
\begin{equation}
    \label{eq:MeanValAtTNoCorr}
    \bar{f}_{t,\alpha} = \bar{U}(t)_{\alpha\alpha'} \bar{f}_{0,\alpha'}
        + \bar{V}(t)_{\alpha I} Y_I = \bar{V}(t)_{\alpha I} Y_I \, .
\end{equation}
The second term in Eq.~\eqref{eq:MeanValAtT}, or equivalently
Eq.~\eqref{eq:MeanValAtTNoCorr}, explicitly shows the contribution of each data
point to the central value of the trained fields at each value of $x_{\alpha}$.
It is worthwhile remarking that in this limit, the central value from the set of
trained networks is a linear combination of the data points, with coefficients
given by the evolution operator $V(t)_{\alpha I}$.

\begin{figure}[ht!]
  \centering
  \includegraphics[width=0.95\textwidth]{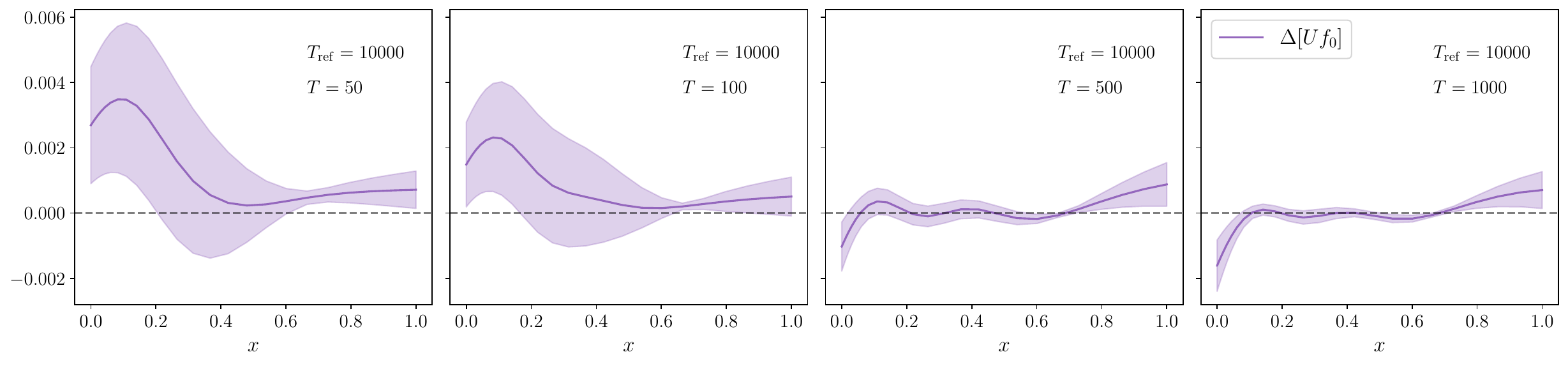}
  \includegraphics[width=0.95\textwidth]{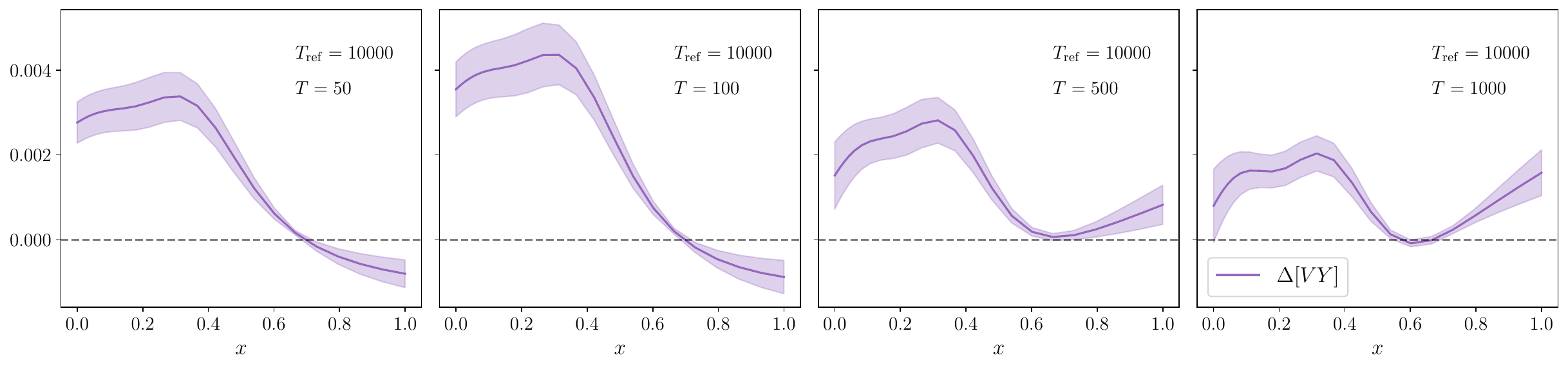}
  \caption{Behaviour of $\Delta [U(t)f_0]$ and $\Delta [V(t)Y]$, as defined in
  Eqs.~\eqref{eq:DeltaExpValUtF0} and~\eqref{eq:DeltaExpValVtY}, as functions of
  the training time. The operators $U(T)$ and $V(T)$ are constructed by taking
  the NTK at $T_{\rm ref} = 10000$, which is fixed across panels. The
  uncertainties are extracted from the bootstrap ensemble as discussed in the
  text.}
    \label{fig:xT3_exp_val}
\end{figure}

In the absence of general theorems, we verify this assumption empirically. From
the ensemble of replicas, we generate bootstrap samples and compute the
following two estimators,
\begin{align}
    \label{eq:DeltaExpValUtF0}
    \Delta[U(t)f_0] &= \mathbb{E}\left[U(t) f_{0}\right] 
      - \mathbb{E}\left[U(t) \right] \mathbb{E}\left[f_{0}\right]\, , \\
    \label{eq:DeltaExpValVtY}
    \Delta[V(t)Y] &= \mathbb{E}\left[V(t) Y\right] 
      - \mathbb{E}\left[V(t) \right] \mathbb{E}\left[Y\right]\, .
\end{align}
for different training times, using the same frozen NTK and L2 data. The results
are shown in Fig.~\ref{fig:xT3_exp_val} for the $U$ (upper panel) and $V$ (lower
panel) contributions. The error bands are computed using bootstrap error. By
inspecting the figures, we see two distinct patterns emerging. For the operator
$U$, $\Delta[U(t) f_0]$ is different from zero for small training times, and thus
the correlations between $U(t)$ and $f_0$ are non-negligible. However, as
training proceeds, $\Delta[U(t) f_0]$ becomes compatible with zero within the error
bars, suggesting that the correlations are progressively suppressed. The case of
the $V$ operator is even more striking, as $\Delta[V(t) Y]$ is clearly
non-negligible across all training times, although it also shows a decreasing
trend as training proceeds. This suggests that the correlations between $V(t)$
and $Y$ cannot be neglected. 

Let us conclude this brief discussion by noting that Eq.~\eqref{eq:MeanValAtTNoCorr} 
resembles the structure of a linear method, like Backus-Gilbert or 
Gaussian Processes. 

\subsubsection{Error decomposition}
\label{sec:ErrorDec}
\begin{figure}[ht!]
  \centering
  \includegraphics[width=0.80\textwidth]{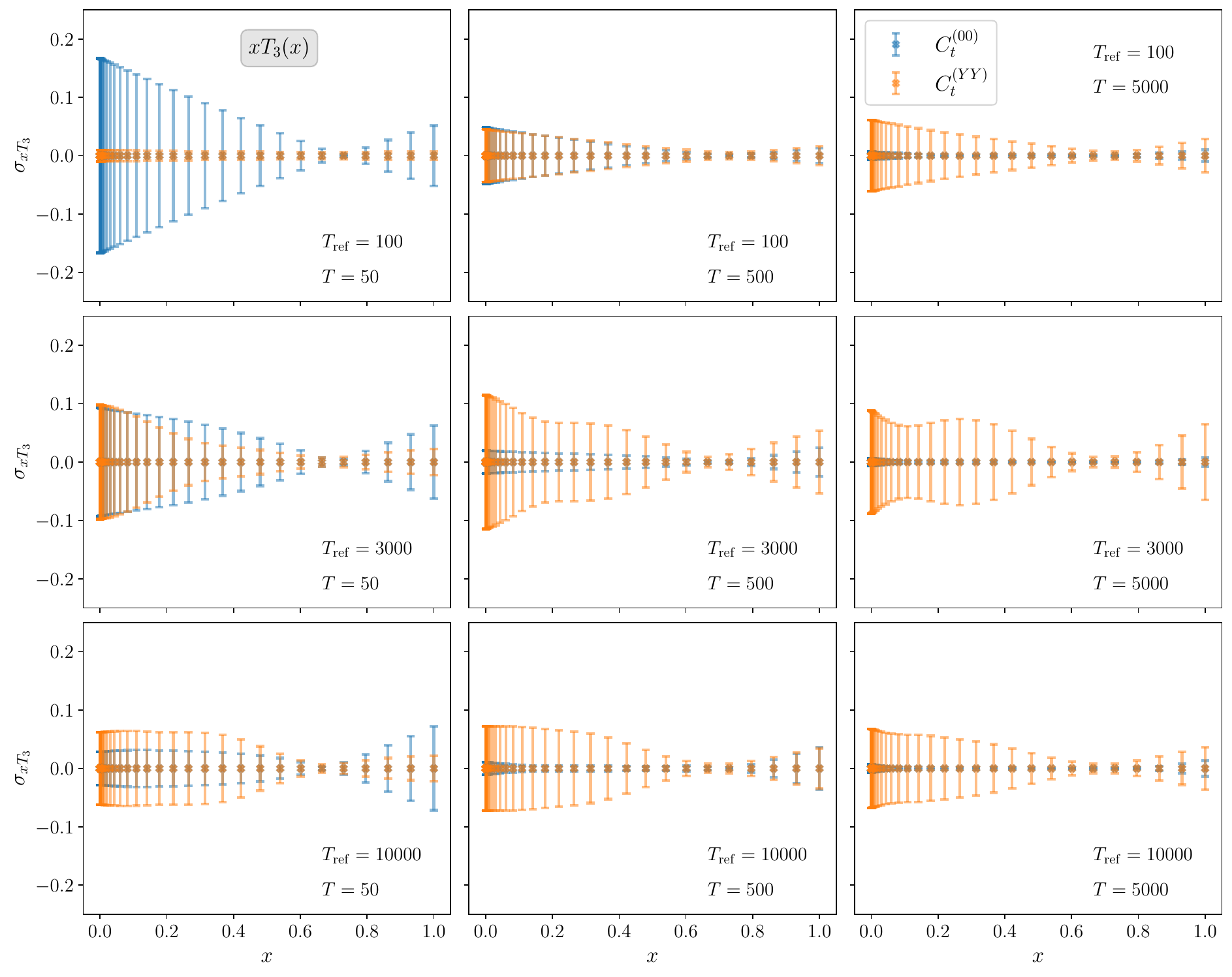}
  \caption{Decomposition of the error budget of the trained fields into the two
  components from the initial condition (blue) and from the data (orange), as
  defined in Eqs.~\eqref{eq:C00term} and~\eqref{eq:CYYterm}. Each row
  corresponds to a different frozen NTK, while the columns represent different
  training epochs. L2 data is used. We see that if the NTK is taken at later
  stages of training, the contribution from the initial condition is severely
  suppressed towards the end of training. }
  \label{fig:ErrorBudgetL2}
\end{figure}

The analytical expression for the covariance matrix,
Eq.~\eqref{eq:SumOfCovariances}, allows us to monitor the relative size of the
three contributions as training proceeds. For a properly trained ensemble of
networks, the covariance of the trained fields should be dominated by the
statistical error on the data. We show the diagonal entries of the two
contributions $C_t^{(00)}$ (blue band) and $C_t^{(YY)}$ (orange band) to the
error budget in Fig.~\ref{fig:ErrorBudgetL2}, for different frozen NTKs (rows)
and different training epochs (columns), using L2 data as before. We do not show
the mixed term $C_t^{(0Y)}$ since it is negligible with respect to the other two
contributions -- the two sources of uncertainty are largely uncorrelated. In
general, we observe that towards the end of the training process the
contribution from the data $C_t^{(YY)}$ becomes dominant with respect to the
contribution coming from the initial condition $C_t^{(00)}$. We also see that
the suppression of the initial condition is more severe and happens earlier when
the frozen NTK is taken at later stages of training.

\begin{figure}[ht!]
  \centering
  \includegraphics[width=0.80\textwidth]{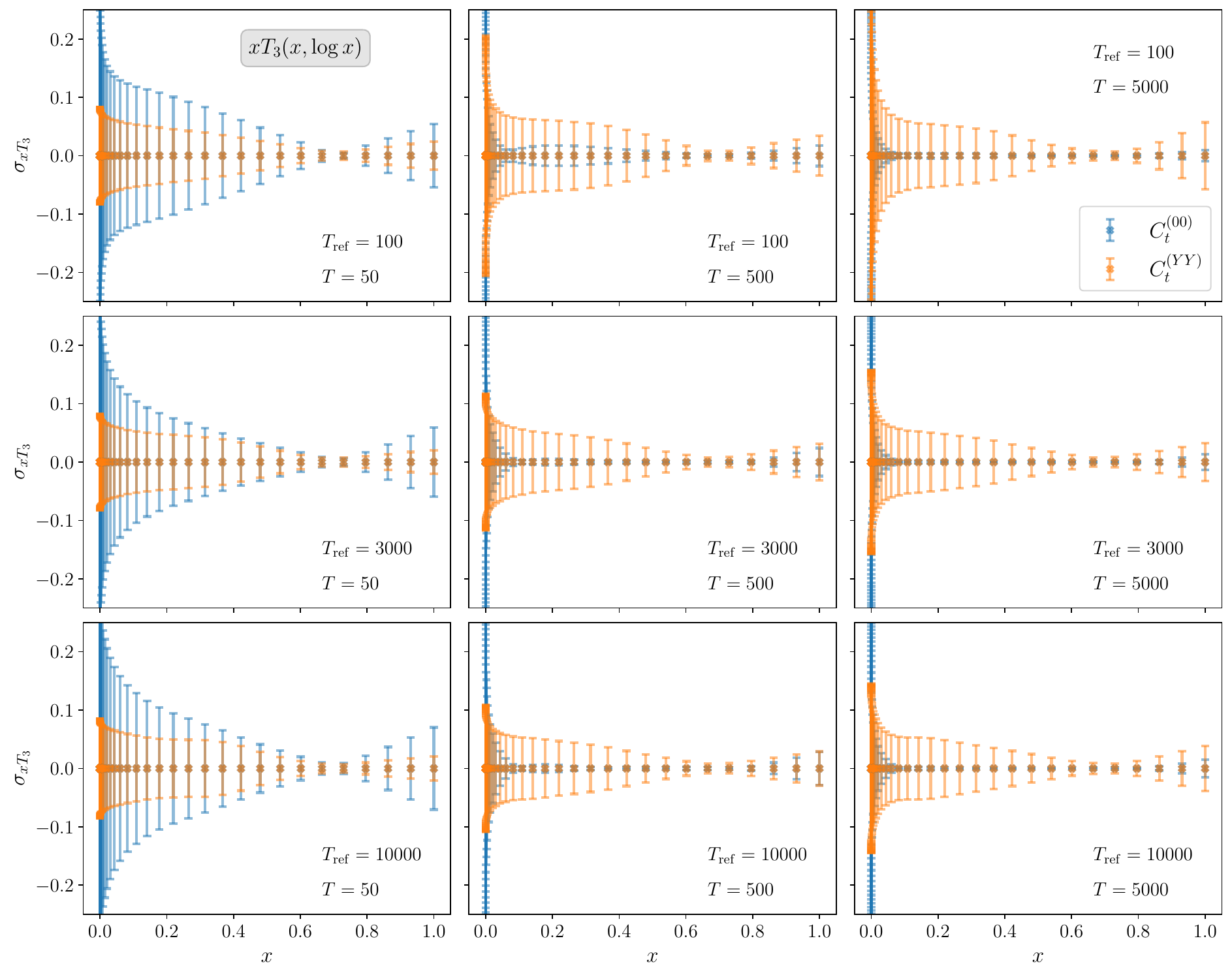}
  \caption{Similar to Fig.~\ref{fig:ErrorBudgetL2}, but now for the case of scaled
  input $f(x, \log x)$.}
  \label{fig:ErrorBudgetL2Logx}
\end{figure}

In order to study the dependence of the error decomposition on the initial
condition, we repeat the same analysis for the case of scaled input $f(x, \log
x)$, \ie\ with two neurons in the input layer. 
The initial condition $f_0$ of the analytical solution is drawn from the same
prior distribution as the trained solution, and corresponds to the orange curve
in Fig.~\ref{fig:prior}. The resulting error decomposition is shown in
Fig.~\ref{fig:ErrorBudgetL2Logx} for L2 data, where panels are ordered as in
Fig.~\ref{fig:ErrorBudgetL2}. Inspecting the figures, we observe that now the
contribution of the initial condition becomes dominant in the small-$x$ region,
even for large training times and irrespective of the epoch at which the NTK is
frozen. This result reflects the behaviour of the prior distribution at
small-$x$, where indeed error bands are significantly enlarged with respect to
the case of linear input. Interestingly, even the contribution from the data,
$C^{(YY)}_t$, increases at small-$x$ towards larger training times. This can be
explained by observing that, despite not being explicitly dependent on the
initial condition $f_0$, the evolution operator $V(t)$ is constructed from a
frozen NTK that has encoded the dependence on the architecture through the
training process (see Eq.~\eqref{eq:VDef}). That the difference between
Figs.~\ref{fig:ErrorBudgetL2} and \ref{fig:ErrorBudgetL2Logx} is primarily
localized at small-$x$ showcases that, for the region left uncovered by the
data, the methodology is not able to suppress the dependence on the initial
condition. In fact, where the information from the data is available
(corresponding roughly to $x \gtrsim 0.01$), the dependence on the initial
condition lessens as the analytical solution evolves (left to right). This
reduction occurs more rapidly for frozen NTKs taken at later stages of training
(top to bottom), showing that there is a non-trivial interplay between the
information provided by the data and that acquired by the NTK.

These studies reveal the intricate connection between the prior distribution and
the uncertainties of PDFs. In the region constrained by data, the error is
dominated by the statistical error on the data, rather than by the fluctuations
of the initial fields. This is an important step in our study of the error
estimates. It guarantees that the error bars computed from the ensemble of
trained PDFs are not biased by the choice of prior, which depends on the selected
architecture, activation function, and probability distributions for the biases
and weights at initialization.


\FloatBarrier

\clearpage

\section{Attempts to a Synthesis}
\label{sec:Synth}

Let us try to conclude these lectures by summarising some lessons we learned along the way. 

First of all, it is worthwhile emphasizing that the discussion of the training based on the NTK can
be adapted to any parametrization and minimization process, including
fixed functional forms~\cite{Ablat:2024hbm,Bailey:2020ooq,Alekhin:2017kpj} or
kernels~\cite{Costantini:2025wxp}. 
What information can be extracted from the NTK in those cases is an open question for future investigations, 
aiming at comparing different procedures within a unified framework. 

From a more general perspective, we can trace the difficulties in solving inverse problems to 
the fact that the initial question is ill-posed. A stable solution can only be obtained by regularizing
the problem in one way or another. 

The probabilistic approach that we have advocated here has the advantage of casting different approaches as 
alternative ways to regularise the problem and define a posterior distribution for the solution. 
Having analysed in detail Backus-Gilbert, Gaussian Processes and Neural Network solutions, 
we have been able to expose some relations between these different approaches. 
In particular, we found that the BG solution can be understood
as special case of the GP solution. A careful analysis of the systematic errors in these two frameworks 
will allow for more robust results. 

It is interesting to remark that the GP solution is based on Bayesian inference, while the other two 
rely on minimizing some figure of merit. In particular, for the case of Neural Networks, the solution 
is obtained by finding an {\em optimal}\ stopping time, rather than letting the minimization reach 
an absolute minimum that would correspond to an overfitted solution. Understanding the connection between 
the stopping condition and the results obtained with BG/GP methods is an open question, which will shed light 
on the errors quoted with these different methods. Understanding the training dynamics of NNs is 
a fascinating problem, and we have only started to scratch the surface of the problem. 

\clearpage

\paragraph{Acknowledgements.} I would like to thank my collaborators, who made my work on 
inverse problems so much more enjoyable. In particular, I benefitted from discussions with 
Marc Costantini, Richard Kenway, Marco Panero, Alberto Ramos, Nazario Tantalo and Maria Ubiali. 
My collaborators in 
the NNPDF collaboration have provided a constant stream of phenomenological results on 
the usage of NNs, prompting me to understand better the training process. None of this 
would have been possible without the work of a cohort of brilliant PhD students,
who have been sharing the daily challenges of these problems with me, 
and who have been a constant source of excellent discussions. 
So, huge thanks to Alessandro Candido, Amedeo Chiefa, Tommaso Giani, Alessandro Lupo,
Giacomo Petrillo and Michael Wilson. A great thank you also to the organisers of the CERN 
summer school, who gave me the opportunity to give these lectures, and to the students 
who attended them, providing further motivation to try to summarise some of the recent 
results we obtained in this field. I hope that these lectures will be useful by providing 
a unifying perspective on the different methods currently in use. LDD is supported by an STFC 
Consolidated Grant (ST/T000600/1, ST/X000494/1).

\newpage

\bibliographystyle{unsrt}
\bibliography{cern24.bib}
% \bibliography{sn-bibliography}% common bib file
%% if required, the content of .bbl file can be included here once bbl is generated
%\input InverseDelDebbio.bbl

\bigskip

{\bf Data Availability Statement}: No new data were created or analysed in this study.

\end{document}